\documentclass[a4paper,fleqn,usenatbib,useAMS,times]{mnras}
\usepackage[T1]{fontenc}
\usepackage{ae,aecompl}
\usepackage{newtxtext}
\usepackage{newtxmath}
\usepackage{graphicx}	
\usepackage{amsmath}	
\usepackage{amssymb}	
\usepackage{rotfloat} 
\usepackage{float} 
\usepackage{graphicx}
\usepackage{longtable}
\usepackage{url}
\usepackage{dcolumn}
\usepackage{placeins}
\usepackage{multirow} 
\usepackage{hyperref}
\usepackage{bm}		
\usepackage{ragged2e}
\usepackage[T1]{fontenc}
\usepackage[none]{hyphenat}
\usepackage{color,soul}
\usepackage{adjustbox}
\usepackage{multicol}
\usepackage{caption}
\usepackage{subcaption}
\usepackage{supertabular}
\usepackage{longtable}
\usepackage[english]{babel}
\usepackage{blindtext}
\usepackage{tabu}
\usepackage{booktabs,caption,fixltx2e}
\usepackage{threeparttable}
\usepackage{array}
\usepackage{afterpage}
\usepackage{capt-of}
\usepackage[stable]{footmisc} 
\usepackage{booktabs}
\usepackage{natbib} 
\usepackage{threeparttable}
\usepackage{Custom_Functions}

\newcommand{\lya}{Ly$\alpha$}

\newcommand{\HI}{\mbox{H\,{\sc i}}}

\newcommand{\HeII}{\mbox{He\,{\sc ii}}}

\newcommand{\OII}{\mbox{O\,{\sc ii}}}

\newcommand{\CII}{\mbox{C\,{\sc ii}}}
\newcommand{\CIII}{\mbox{C\,{\sc iii}}}
\newcommand{\CIV}{\mbox{C\,{\sc iv}}}

\newcommand{\SiII}{\mbox{Si\,{\sc ii}}}
\newcommand{\SiIII}{\mbox{Si\,{\sc iii}}}
\newcommand{\SiIV}{\mbox{Si\,{\sc iv}}}

\defcitealias{Schaye2007}{S07}
\defcitealias{Kim2016}{KIM16}
\defcitealias{Haardt2012}{HM12}
\defcitealias{Khaire2016}{KS16a}
\defcitealias{Khaire2017}{KS16}
\defcitealias{Khaire2018}{KS18}

\newcommand{\be}{\begin{equation}}
\newcommand{\en}{\end{equation}}

\def\kms{km~s$^{-1}$}

\def\cmsq{cm$^{-2}$}
\def\cmcb{cm$^{-3}$}

\def\nce{$N({\CIII})$}
\def\nc4{$N({\CIV})$}
\def\nhi{$N({\HI})$}
\def\si{${\sim}$}

\def\HM12{\citetalias{Haardt2012}}
\def\ks18{\citetalias{Khaire2018}}
\def\KIM16{\citetalias{Kim2016}}
\def\KS18{\citetalias{Khaire2018}}

\newcommand{\CLOUDY}{\mbox{\scriptsize{CLOUDY}}}

\newcommand{\pirate}{$\Gamma_{\rm HI}$}

\newcommand{\nh}{$n_{\rm \tiny H}$}

\def\rat{$N({\CIII})/N({\CIV})$}

\def\ratb{$N({\SiIII})/N({\SiIV})$}
\newcommand{\sag}{$S_1$}
\newcommand{\sah}{$S_2$}
\newcommand{\sai}{$S_3$}
\newcommand{\may}{$M1$}
\newcommand{\mah}{$M2$}
\newcommand{\mai}{$M3$}
\def\Mo{{\textcolor{blue}{\may}}}
\def\Mt{{\textcolor{blue}{\mah}}}
\def\Mth{{\textcolor{blue}{\mai}}}
\def\sam{{\textcolor{blue}{\sag}}}
\def\san{{\textcolor{blue}{\sah}}}
\def\sao{{\textcolor{blue}{\sai}}}
\newcommand{\nhio}{$N({\HI})_{obs}$}
\newcommand{\nhij}{$N({\HI})_{J}$}

\title[High$-z$ C~{\sc iii} absorbers]{Physical conditions in high$-z$ optically thin C~{\sc iii} absorbers: Origin of cloud sizes and associated correlations.
}

\author[Mohapatra, A. et al.]{Abhisek Mohapatra$^{1}$\thanks{Contact e-mail: \href{abhisekphy@gmail.com}{abhisekphy@gmail.com}}, R. Srianand$^{2}$, Vikram Khaire$^{3}$ \& Ananta C. Pradhan$^{1}$\
\\
$^{1}$Department of Physics and Astronomy, National Institute of Technology, Rourkela,  Odisha 769\ 008, India\\
$^{2}$Inter-University Centre for Astronomy \& Astrophysics, Postbag 4, Ganeshkhind, Pune 411\ 007, India\\
$^{3}$Department of Physics, Broida Hall, University of California, Santa Barbara, CA 93106-9530, USA}
\date{Accepted\dotfill. Received\dotfill; in original form\dotfill}

\pubyear{2019}

\begin{document}
\label{firstpage}
\pagerange{\pageref{firstpage}--\pageref{lastpage1}}
\maketitle

\begin{abstract}
We present detailed photoionization models of well aligned optically thin \CIII\
absorption components at $2.1 < z < 3.4$. Using our
models we estimate density (\nh),
metallicity ({\it $[C/H]$}),
total hydrogen column density and line-of-sight thickness ($L$)
in each \CIII\ components. We estimate the systematic
errors in these quantities contributed by the allowed range of the quasar
spectral index used in the ultraviolet
background radiation calculations. 
Our inferred \nh\ and overdensity ($\Delta$)
are much  higher than the measurements available in the literature and favor the absorption
originating from gas associated with circumgalactic medium and
probably not in hydrostatic equilibrium. 
{We also notice
\nh, {\it L} and {\it $[C/H]$} associated with \CIII\ components show statistically significant
redshift evolution. To some extent, these redshift evolutions are driven by the appearance of compact, high \nh\
and high {\it $[C/H]$} components only in the low$-z$ end. 
We find more than 5$\sigma$ level
correlation between {\it $[C/H]$} and $L$, {\it L} and neutral hydrogen column density (\nhi), \nhi\ and {\it $[C/H]$}.
We show {\it L} versus {\it $[C/H]$} correlation
can be well reproduced if {\it L} is
{governed} by the product of gas cooling time and sound
speed as expected in the case of cloud fragmentation under thermal instabilities.
This allows us to explain other observed correlations by simple photoionization considerations. 
Studying the optically thin
\CIII\ absorbers over a large $z$ range and probably correlating their $z$ evolution with global
star formation rate density evolution can shed light into the 
physics of cold clump formation and their evolution in
the circumgalactic medium.}
\end{abstract}

\noindent \section{Introduction}
Absorption lines seen in the spectra of distant quasars are used to probe the
physical and chemical state of gas associated either with circum-galactic medium (CGM) of galaxies
or intergalactic medium (IGM). In particular, under the assumption of the
absorbing gas in ionization and thermal equilibrium with the metagalactic
ionizing ultraviolet background (UVB), one can derive the physical and chemical
properties
of the absorbing gas.
On the other hand, detection of absorption produced by
a range of ions from the same atom can allow one to probe the
shape of the UVB \citep[see][]{Fechner2011}. 
The UVB cannot be measured directly but obtained from the synthesis models which 
perform radiative transfer of UV photons emitted by quasars and galaxies through the IGM across different
redshifts \citep[e.g.,][]{Miralda1990, Shapiro1994, Shull1999,Haardt1996,Faucher2009,Haardt2012, Khaire2018}.

The UVB plays a crucial role in the absorption line studies.
For a given set of absorption lines, photoionization
models for an assumed UVB allow us to constrain
properties of the absorbing gas such as density, total column
density, metallicity and line-of-sight thickness. Nevertheless, the inferred 
properties of the absorbing gas will be accurate
when many metal absorption lines are originating from the same absorbing gas (i.e. co-spatial) having
a temperature of few $10^4$ K,
where collisions are sub-dominant. These inferred properties can 
provide a powerful tool to study the origin of the absorbing clouds, for example,
by distinguishing the inflow of pristine IGM gas from outflow of the metal enriched gas.
Moreover, the correlations in 
these inferred parameters can further answer important questions such as stability 
and fate of these clouds and provide essential constraints for hydrodynamic simulations of galaxy formation, 
the CGM and IGM.

There are many theoretical explanations on the formation and the stability of IGM and CGM clouds. The IGM 
clouds are thought to be in hydrostatic equilibrium
\citep{Ikeuchi1986, Rees1986, Schaye2001, Dedikov2004} or supported by ambient  
pressure \citep{Sargent1980, Williger1992, Schaye2007}.
However, there are intervening absorbers with sizes of  tens to hundreds of
kiloparsecs which are much larger to be 
confined by external pressure from a confining medium or dark matter halos \citep{Bechtold1994, Dinshaw1997, Smette1992}.
Similarly, the metal-enriched cold (\si $10^4$K) gas which is supposed to originate from the vicinity of galaxies or CGM
and need not to be stable due to either of the above two main confinements.  
Despite several observations, the physical and chemical properties of this cold gas
is still uncertain and an important
challenge to track. Previous work suggests that hot winds can sweep up
metal-rich
cold interstellar gas to the CGM by means of radiation/ram pressure
\citep[e.g.][]{Mccourt2015, Heckman2017} or these clouds can form in-situ
due to condensation of the hot wind via
thermal instabilities \citep{Field1965, Thompson2016}. 
Recently, hydrodynamic simulations show that the fragmentations and isobaric
condensations in thermal instability of such cold gas produce very small parsec size cloudlets in the galactic hot halo environment \citep{Mccourt2018, Liang2018}.   
Our goal is to understand this in observations since the current state of the art cosmo-hydrodynamical
simulations
can not resolve the relevant physics in more details and, in particular, over
the scales involved in the problem \citep[see,][]{Gronke2018, Mccourt2018, Sparre2018,Voort2018}.

In order to derive the properties of the IGM or CGM gas clouds and 
understand their physical origin, we need a large sample of absorption systems 
especially having well aligned metal transitions. 
Such a large 
sample of the optically thin C~{\sc iii} components at $2.1 < z < 3.4$ is provided by
\citet[][hereafter \citetalias{Kim2016}]{Kim2016}.  
\citetalias{Kim2016} have
identified  the well aligned absorbers and provided H~{\sc i}, C~{\sc iii} and
C~{\sc iv} column densities and their kinetic temperatures for the components. 
We perform the photoionization modeling for this sample
using our
updated \citep[][hereafter \citetalias{Khaire2018}]{Khaire2018} UVB along with 
the \citet[][hereafter \citetalias{Haardt2012}] {Haardt2012} UVB for comparison.
Note that the UVB is an essential part of  such analysis however it is quite uncertain mainly because of the available choices for
different input parameters during the synthesis models.
For such purpose, we have been studying the synthesis models of the UVB 
\citep[see,][]{Khaire2013,Khaire2015, UVKhaire2015, Khaire2016, Khaire2017,Khaire2018},
the measurements of the UVB \citep[see,][]{Gaikwad2017, UVGaikwad2017}
and its implications on the absorption line studies 
\citep[see,][]{Pachat2016,Hussain2017,Pachat2017a,Muzahid2018}
to address above mentioned issues. 
Here, we take UVB uncertainties due to the variations in quasar spectral energy distributions (SEDs) into account and derive 
various physical parameters
such as density, metallicity and line-of-sight thickness
of the gas clouds along with the associated uncertainties in these
parameters.  
We study associated correlation between the derived parameters and explain these 
correlations with a toy model where the size of the clouds follows the gas cooling length.
We show that, this toy model
hold clues to understand the origin and fate of these clouds.

This paper is organized as follows: We describe the details of the
data in Section~2. We give
brief information about our photoionization models
and present the re-analysis of the optically thin \CIII\ components
in Section~3.
In our photoionization models we consider two stopping criteria and wide range of UVBs
resulting from the variations in quasar SEDs.
In Section~4 and Section~5,
we explore the redshift evolution of the
derived parameters and any possible correlations between these
parameters, respectively. We construct a simple toy model in
Section~6 using which 
using which we explain the observed correlations.
We summarize our results in Section~7. 
Throughout this article, we adopt flat
$\Lambda$CDM cosmology with
$H_0 =70$ {\kms} $Mpc^{-1}$ , $\Omega_\Lambda$ = 0.7, and $\Omega_m$ = 0.3 \citep{Plank2016}.
We use the notation [X/Y] = ${\rm log}$~(X/Y) - ${\rm log}$~(X/Y$)_{\odot}$ for abundances of heavy elements with solar relative 
abundances taken from \citet{Grevesse2010}.
\noindent \section{Data}
\label{s1}
In this work, we model the intervening optically thin
\CIII\ components and systems in the redshift range $2.1 < z < 3.4$ along 19 quasar sightlines
analyzed by \citetalias{Kim2016}\footnote{\url{http://vizier.cfa.harvard.edu/viz-bin/VizieR?-source=
J/MNRAS/456/3509}}.
We choose this sample because it provides {\CIII} and {\CIV} column densities in individual
components where the basic idea of homogeneous cloud is most probably valid. Also the
ionization potential of these two ions lie
on either side of He~{\sc ii} Lyman break so that they are sensitive to
the changes in the quasar SEDs used to generate the meta-galactic UVB.

In this work, we mainly focus on 53 absorption systems where {\CIII} and {\CIV} column
densities are measured using Voigt profile fitting. There are 132 {\CIII} components in these
systems with {\CIV} component association. Out of these,
there are 104 clean {\CIII} components identified by \citetalias{Kim2016}
(i.e unsaturated and no upper limits) with
well-aligned\footnote{Well-aligned components refers to those {\CIII} and {\CIV} absorptions
which have same $z$
and Doppler parameter ($b$) during
the Voigt profile decomposition.} {\CIV} absorption component.
For 24 {\CIV} components only upper limits can be obtained for $N$(\CIII). In the remaining 4 components
{\CIII} is either saturated or blended with other strong absorption lines. We take lower limits on
$N$(\CIII) for these components.

We perform ionization modeling of these above mentioned 132 components, under the assumption of
constant density gas, to
derive physical conditions.
Since the  \CIII\ and \CIV\ absorption are well aligned the 
ratio of their column densities can be used to constrain ionization parameter (U)
or {\nh}
of gas for a given ionizing UVB.
We estimate the range of {\nh} for such components for each UVB considered in this study. 
Moreover, measuring the column density of {\HI} associated with the components
allow us to constrain the metallicity and derive the cloud properties.

We define two sub-samples \sam\ and \san\ out of these 132 intervening components on
basis of associated {\HI} components as following. The sub-sample
\sam\ consists of 32 intervening {\CIII} components
which have a co-aligned\footnote{Co-aligned components refers to the aligned {\CIII+\CIV} components which have a {\HI} absorption component
at the same velocity centroid as that of {\CIII} absorption.} {\HI} components 
which is required to probe the temperature, turbulent broadening and also the gas phase metallicity. 
The sub-sample \san\ consists of a total 50 \CIII\ components out of which there are 
33 {\CIII} components which have a moderately-aligned\footnote{Moderately-aligned components are the ones which have a {\HI} component
within a maximum velocity difference of 6.5 km/s without perfect alignment.} {\HI}
components.
In such cases the assumption of {\CIII} and {\HI} originating from a single phase may not be valid.
While density measurements based on \rat\ 
ratio for these components will be robust, metallicity measurements
should be considered as limits as \HI\ association is uncertain.
In addition,
there are 17 components in sub-sample \san\ where {\CIV} and {\HI} are well aligned but we have only
upper limits for {$N$(\CIII)}.
In this case the derived density and {\it $[C/H]$} are upper limits.

While Voigt profile decomposition is historically used for the absorption line analysis, it may not be appropriate if the absorption
profile originates from a complex velocity and density field of the gas as one sees in IGM simulations. In such cases obtaining column density
weighted average
properties of the whole profile may be interesting. So, we construct \textit{{\textcolor{blue}{\sao}}} which includes all
the 53 {\CIII} systems where we
consider the total column density of ions
(i.e sum of the column densities in individual Voigt profile components)
to originate from single cloud in our models.

\noindent \section{Photoionization Models}
In this section we describe our photoionization models. We use {\CLOUDY} version c13.03 \citep{Ferland2013} for our calculations and
assume an absorption component to be a single-phase plane parallel slab of
constant density in thermal
and ionization equilibrium with the assumed UVB. 
The thickness of the absorbing gas in {\CLOUDY} is defined through the stopping criteria. We use following two criteria: (1) total column density of hydrogen
inferred assuming the H~{\sc i} absorption that is produced by a cloud in hydrostatic
equilibrium with the associated dark matter potential or (2) {\nhi} predicted
by the models that is equal to the observed {\nhi} as the stopping criteria.
We also consider models with constant temperature
to incorporate the effect of additional heating sources. The metal composition is assumed to follow the solar composition given by \citet{Grevesse2010}.

\noindent \subsection{The UVBs}
\begin{figure}
  \centering
  \includegraphics[totalheight=0.26\textheight, trim=.1cm 2.1cm 0cm 0.6cm, clip=true, angle=0]{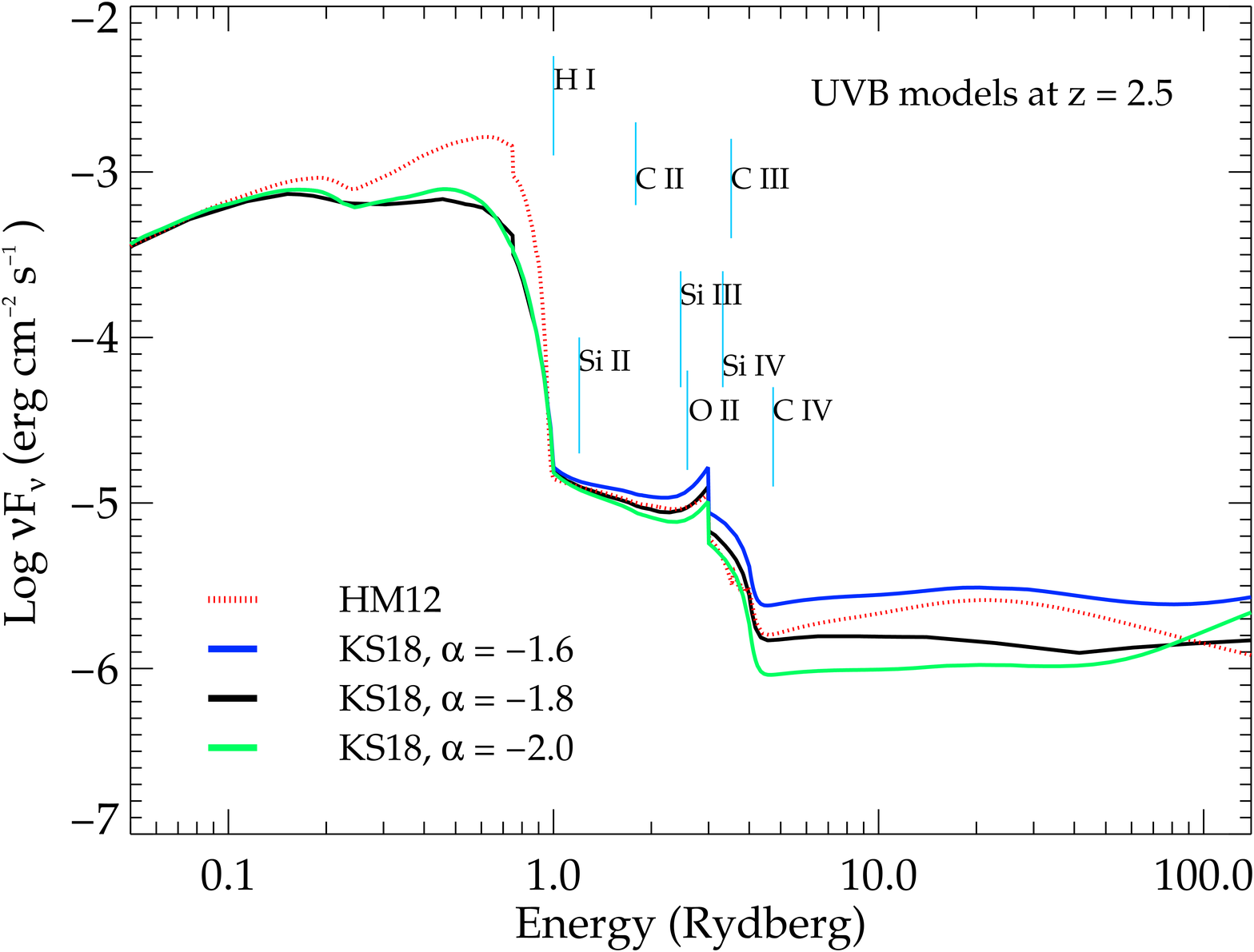}
\caption{Different UVB models at $z$ {\si}  2.5 used in this study. The blue, black and green solid lines are 
used for \citetalias{Khaire2018} UVB with $\alpha$ = $-1.6, -1.8$ and $-2.0$, respectively. The red dotted line shows
the \citetalias{Haardt2012} UVB. 
The vertical lines are used to show the ionization
energy of different ions (ion name is marked right next to the line).}
\label{fig1}
\end{figure}
In our analysis, we use updated UVBs computed by \citetalias{Khaire2018} and 
\citetalias{Haardt2012} considering both quasars and galaxies as ionizing sources.
In Fig.~\ref{fig1}, we compare different UVB spectra at $z=2.5$.
The quasar contribution to the integrated UVB intensity at high energies mainly depends on the assumed
average SED of quasars which is used in the calculation of quasar emissivity. 
Quasar SEDs over the whole range of wavelength is usually approximated by a power law 
of $ f_{\nu} \propto \nu^\alpha $ \citep{Francis1991, Vanden2001, Scott2004, Shull2012,
Telfer2002}.  
The reported values 
of power-law index ($\alpha$) varies from $-0.72$ to $-1.96$ \citep[e.g see Table.~1 of][]{Khaire2017}.
The power law index is observationally
measured up to 2 Ryd \citep{Stevans2014}, 
which is then extrapolated to higher energies to obtain the complete spectrum.
The \citetalias{Haardt2012} UVB  uses $\alpha$ = $-$1.57  consistent with \citet{Telfer2002}. 
\citet{Khaire2017} has recently shown that  the UVB estimated using $\alpha$ = $-1.6$ to $-2.0$ also reproduces the
      {\HeII} {\lya} optical depth as a function of $z$ very well \citep[see also,][]{Gaikwad2018a,Gaikwad2018b}.
      \citetalias{Khaire2018} provides UVB for a range of $\alpha$ values, following their paper we use
      UVB with $\alpha = -1.8$ as our fiducial model.
In addition, \citetalias{Khaire2018} UVB uses
the updated column density
distribution of neutral
hydrogen from \citet{Inoue2014} while calculating the IGM opacity to the UVB. 
\begin{figure*}
  \centering
  \includegraphics[totalheight=0.5\textheight, trim=0.5cm 9cm 0cm 0.559cm, clip=true, angle=0]{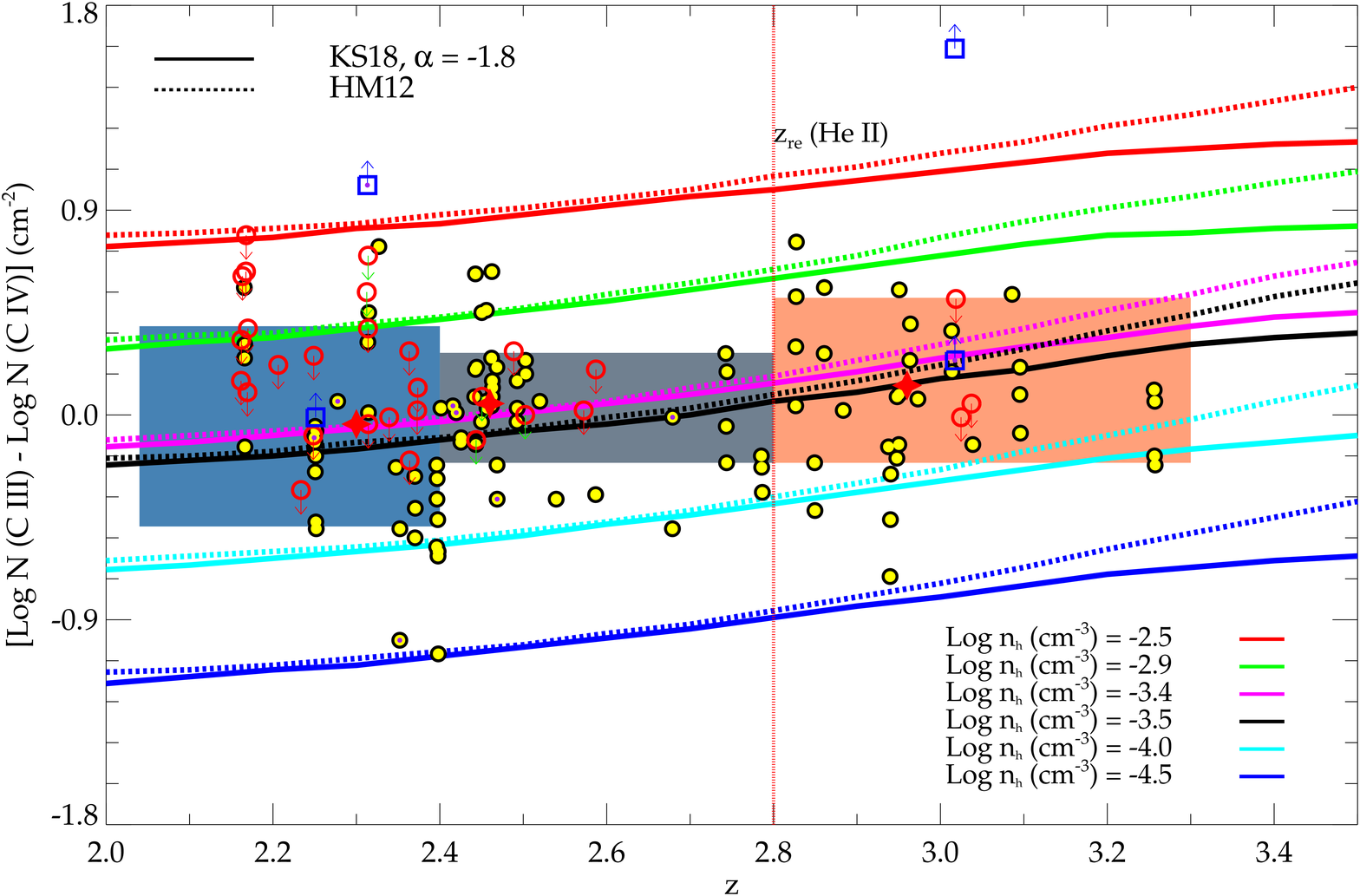}
\caption{\small{The observed ratio of $N$({\CIII}) to $N$({\CIV}) for 132 {\CIII} components as a function
of $z$ [104 clean detections (scattered yellow filled black circles), 24 components with
upper limit on {\CIII} (red open circles) and 4 components with
lower limit on {\CIII} (blue open squares)]. 
Predicted column density ratio of {\CIII} to {\CIV} for two different UVBs for model \Mo\ are overplotted on top of the observed data.
The solid and dashed lines are used to show the model predictions of \citetalias{Khaire2018} and \citetalias{Haardt2012} UVB, respectively.
From top to bottom  
\nh\ is increasing as shown in the legends. The shaded blue, gray and orange
regions show the range for redshift 
2.1 < $z$ $\le$ 2.4, 2.4 < $z$ $\le$ 2.8 and 2.8 < $z$ $\le$ 3.4
where 68\% of observed data lies as obtained from the cumulative probability distribution of the sample data.
The red filled stars are used to mark the median values of the column
density ratios at each median redshift of the above three bins. The vertical red dotted line is used to
mark the reionization redshift of \HeII\ ($z_{re}$(\HeII))
for our fiducial \KS18\ UVB model.
}}
\label{fig2}
\end{figure*}

While these two UVBs (i.e our fiducial and \citetalias{Haardt2012} UVBs) differ in spectral shape at low$-z$
($z < 1.0$) and high$-z$ ($z > 3.5$), in the redshift range of our
interest ($2.1 < z < 3.4$) the difference in {\HI} photoionization rates (\pirate) predicted by the two models is minimum
(see Fig.~4 of \citetalias{Khaire2018}). However, our fiducial \KS18\ UVB spectrum shows
different shape at E $>$ 4 Ryd and E $<$ 1 Ryd compared to that of \citetalias{Haardt2012} because
of the differences in the quasar 
emissivity and SED. At $z = 2.5$ both UVBs provide {\pirate} consistent with the available
measurements \citep{Becker2013,Bolton2007} and 
the difference in {\pirate} between \citetalias{Haardt2012} and \citetalias{Khaire2018} is very small i.e {\si} $5 \times 10^{-13} s^{-1}$.

The ionization energies
of different ions are marked with vertical lines in Fig.~\ref{fig1}. 
For the ions
(for e.g.  {\OII}, {\SiII}, {\SiIII}, {\CII} and {\CIII}) whose ionization energies are in the range
E = 1 $-$ 4 Ryd,
the UVB is contributed by the radiation coming from quasars as well as galaxies. 
However, for ions such as {\CIV} the ionizing background radiation is dominated by quasars with no or negligible
contributions from galaxies.
\noindent \subsection{Modeling {\CIII} absorbers}
\label{s3}
In photoionization models, ratio of
column densities of two successive ionization stages of any element 
(for e.g. $\CIII/\CIV$, $\SiIII/\SiIV$, etc.) usually depends weakly on metallicity and depends mainly
on the intensity and shape of the incident radiation field, density and 
temperature of the gas.
Therefore, the basic idea of these models is to match the observed column density ratios 
of relative ions with the model predictions to determine {\nh} of the absorber for a given ionizing UVB.
The metallicity of the gas is then adjusted to match the individual 
observed column densities of the heavy ions and {\nhi}.
The specified initial conditions for each cloud are: (1) the UVB or ionizing
continuum intensity and shape, (2) assumed \nh\ and the stopping criteria to terminate the calculation and
(3) the chemical composition of the gas.
\subsubsection[]{{\CIII} absorbers as Jean's stable clouds}
In the first set of photoionization models (hereafter, \Mo\ model), we assume
the size of the cloud (or line-of-sight thickness) to be Jean's length 
as suggested for the {\lya}
forest absorption by \citet{Schaye2001}. This is a reasonable approximation if {\CIII} absorbers are
predominantly originating
from the IGM. Under this approximation the stopping total hydrogen column density
($N(H)$) in {\CLOUDY} models for an assumed {\nh}  (\cmcb) is given by \citep{Schaye2001}, 
\begin{equation}
 N(H)_{J} = n_HL_J \sim 1.6 \times 10^{21} cm^{-2} {n_H}^{1/2}T_4^{1/2}\bigg({\frac{f_g}{0.16}}\bigg)^{1/2} 
\label{a15}
 \end{equation}
where, $L_J$ is Jean's length,
$T_4(K)=T(K)/10^4$ and $f_g$ is the gas mass fraction. For the above calculations we use $T$ {\si}$10^4$K, initial
metallicity to be $10^{-2}$ Z$_\odot$ and $f_g$ $=$ 0.16 close to cosmic baryonic mass fraction.
We varied $n_H$ (in a logarithmic scale) and redshift with step size of 0.1
to generate a grid of models and obtain the column densities for several ions of our interest.

In Fig.~\ref{fig2}, we plot the observed ratio of {\nce} to {\nc4} for 132 components
[104 clean detections (yellow filled black
circles), 24 components with upper limit on \CIII\ (red open circles) and 4 components with lower 
limit on \CIII\ (blue open squares)]
in our sample
as a function of $z$.
We divide the entire redshift range into three bins, $[2.1, 2.4]$ (blue shaded region), $[2.4, 2.8]$ (gray shaded region)
and $[2.8, 3.4]$ (orange shaded region). 
In these redshift bins the measured column density ratio of \CIII\ to \CIV\ ranges (in logarithmic units) are
$-0.49$ to $0.39$, $-0.21$ to $0.27$ and $-0.21$ to $0.52$, respectively
where 68\%
of the observed data in each bin are distributed around the observed median.
\begin{figure*}
\centering
\includegraphics[totalheight=0.31\textheight, trim=0.1cm 1cm 0.1cm .1cm, clip=true, angle=0]{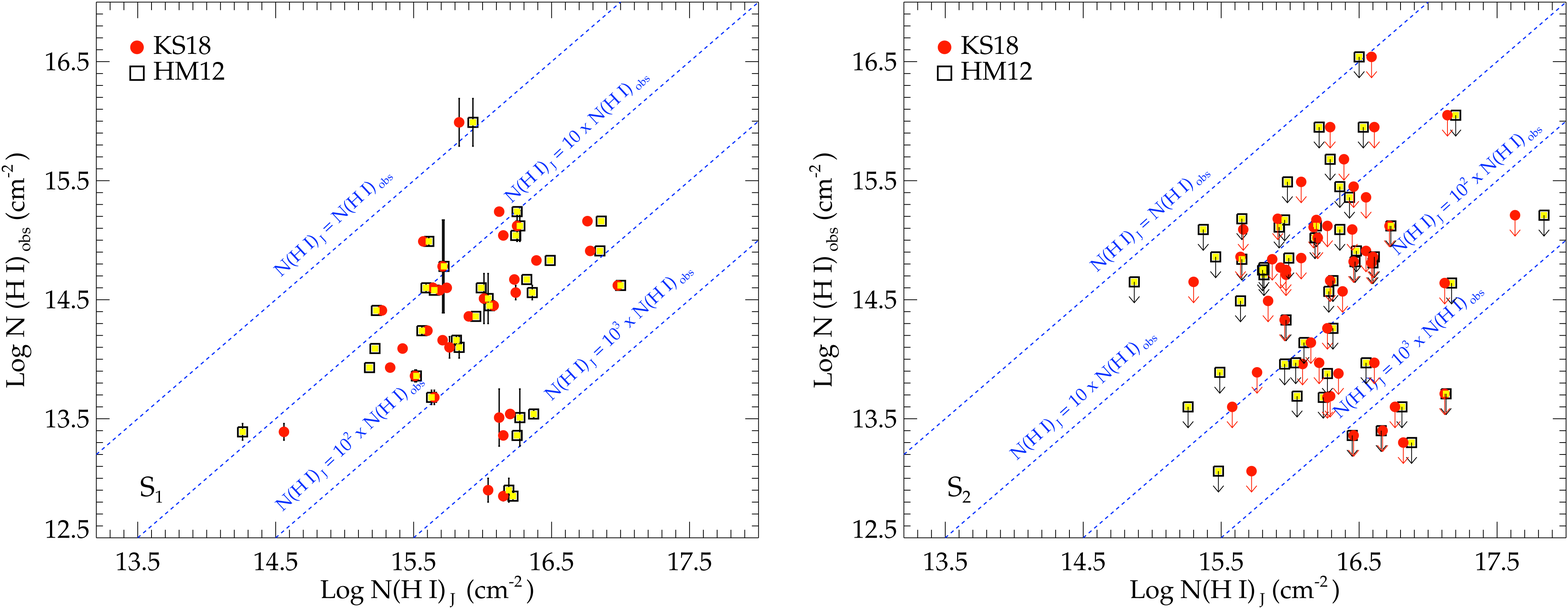}
\caption{The observed neutral hydrogen column density, ${\rm log}$~$N$({\HI}), is plotted against the model \Mo\ predicted neutral hydrogen column density
${\rm log}$~$N$({\HI})$_{J}$ for the two UVBs. Left-hand panel and right-hand panel show data from \sam\ and \san, respectively.
Solid red circles and yellow filled squares
show the results from \citetalias{Khaire2018} and \citetalias{Haardt2012} UVB, respectively.
Black vertical line is used to 
represent error associated with the quantity where error is larger than the symbol size. The downward arrows represent the upper limits.
The dashed blue lines represent different scaling of $N(\HI)_J$ with $N(\HI)_{obs}$. The text next to these line shows the linear scaling relation. }
\label{fig3}
\end{figure*}
The red filled stars are used to mark the median values (-0.04, 0.05 and 0.13) of the observed column
density ratios at each median values of redshift (2.30, 2.46 and 2.96) of the three redshift bins, respectively. 
We see a mild evolution in the median values of observed column density ratio with increasing redshift. As pointed out by
\citetalias{Kim2016}, the apparent deficiency of {\CIII} components at $z\sim2.6$ is not real and caused
by the lower number of quasar sightlines covered in this $z$ in the sample used.
\begin{figure*}
\centering
\begin{subfigure}{.5\textwidth}
  \centering
  \includegraphics[totalheight=0.35\textheight, trim=.1cm 0.5cm 0cm 0.6cm, clip=true, angle=0]{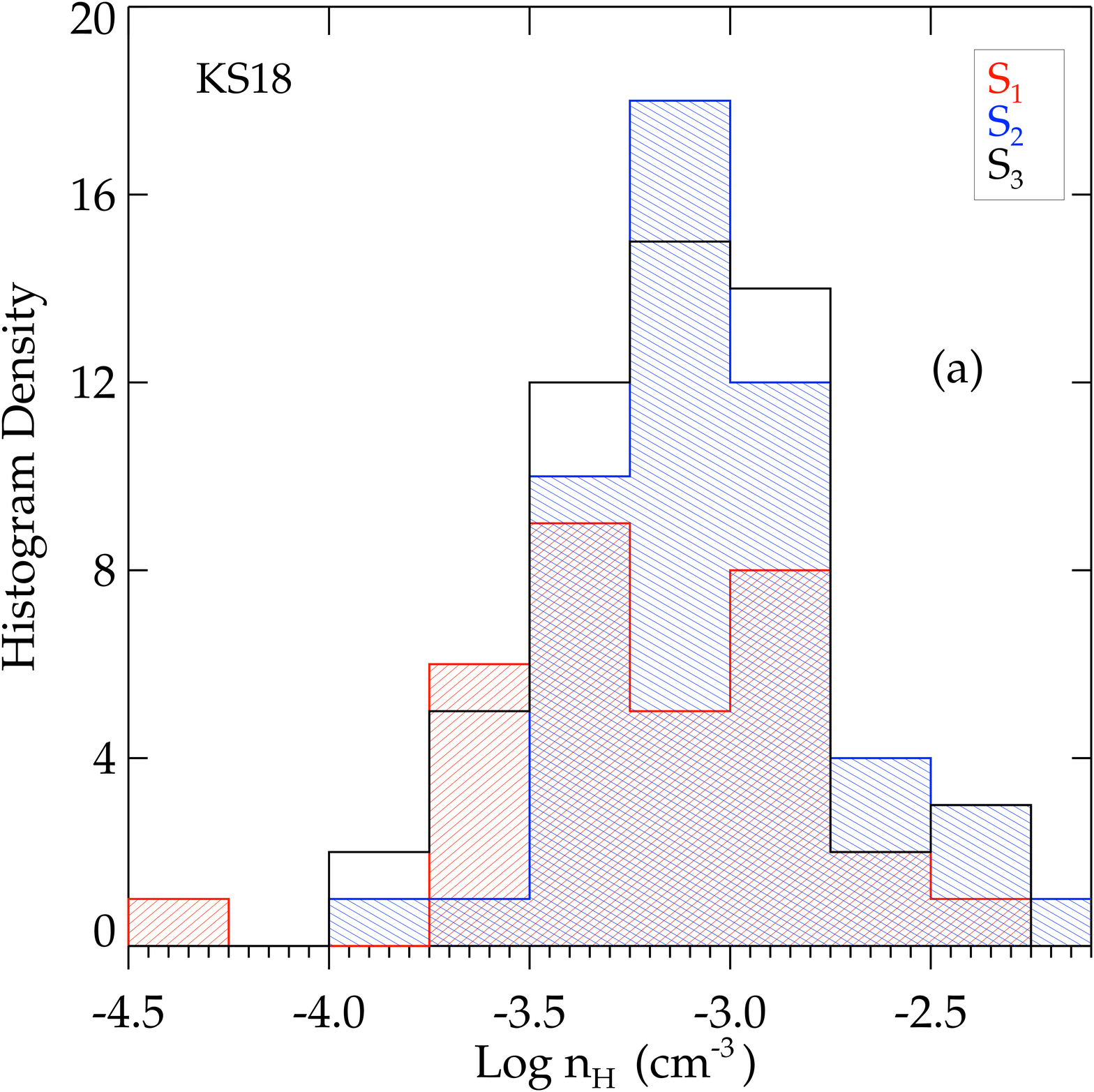}
\end{subfigure}%
\begin{subfigure}{.5\textwidth}
  \centering
 \includegraphics[totalheight=0.35\textheight, trim=.1cm 0.5cm 0cm 0.6cm, clip=true, angle=0]{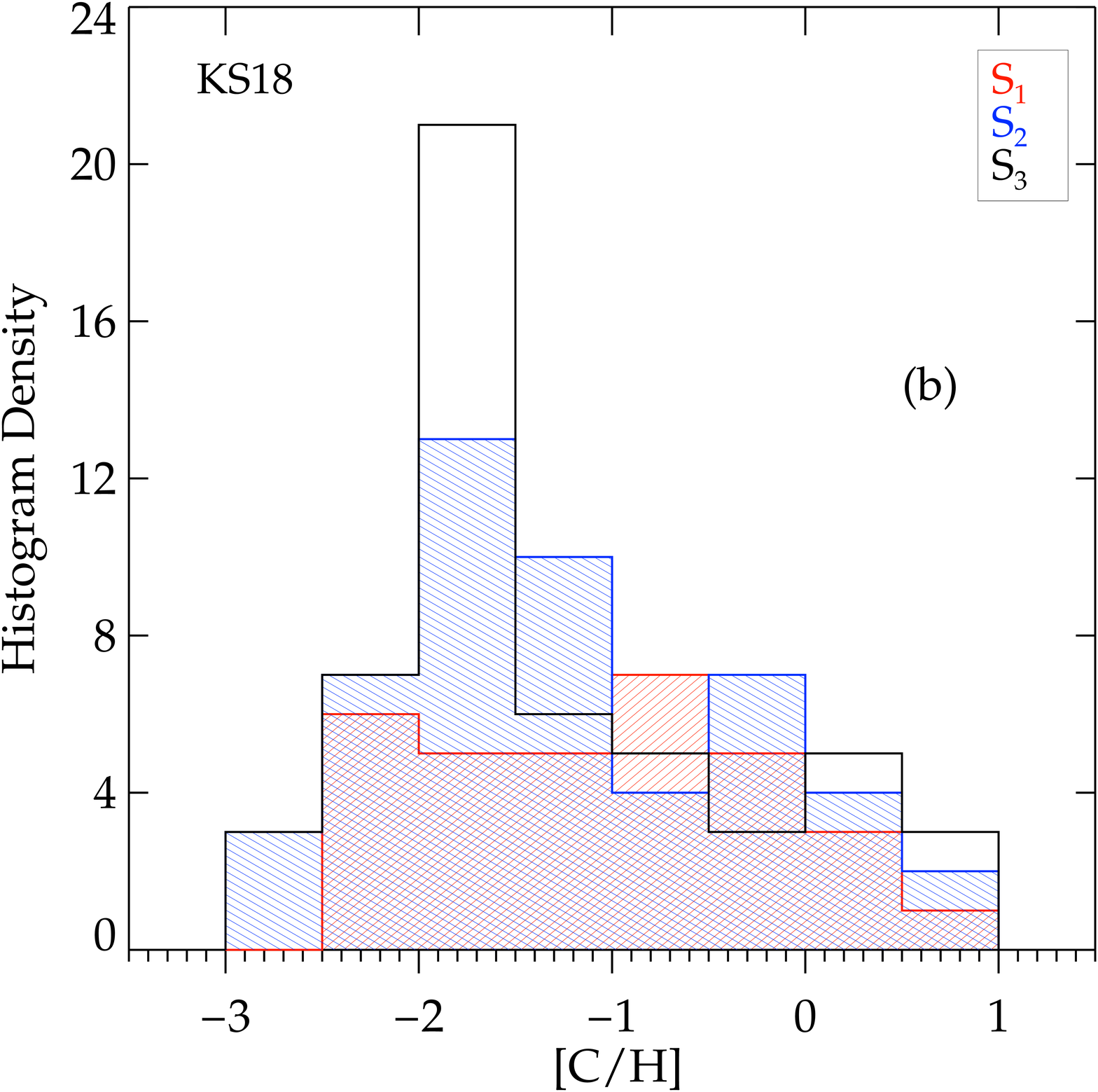}
\end{subfigure}
\begin{subfigure}{.5\textwidth}
  \centering
  \includegraphics[totalheight=0.35\textheight, trim=.1cm 0.5cm 0cm 0.6cm, clip=true, angle=0]{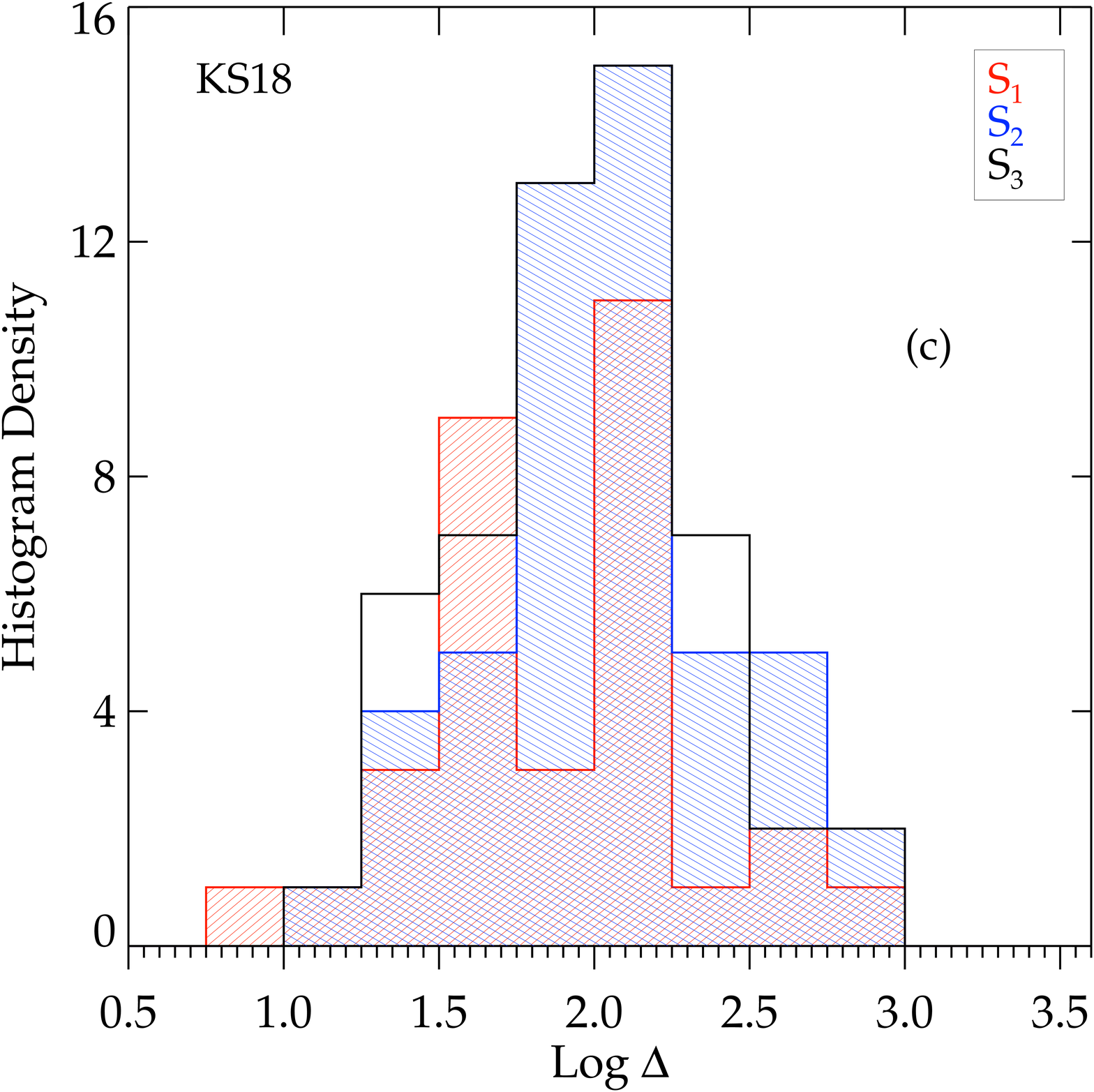}
\end{subfigure}%
\begin{subfigure}{.5\textwidth}
  \centering
 \includegraphics[totalheight=0.35\textheight, trim=.1cm 0.5cm 0cm 0.6cm, clip=true, angle=0]{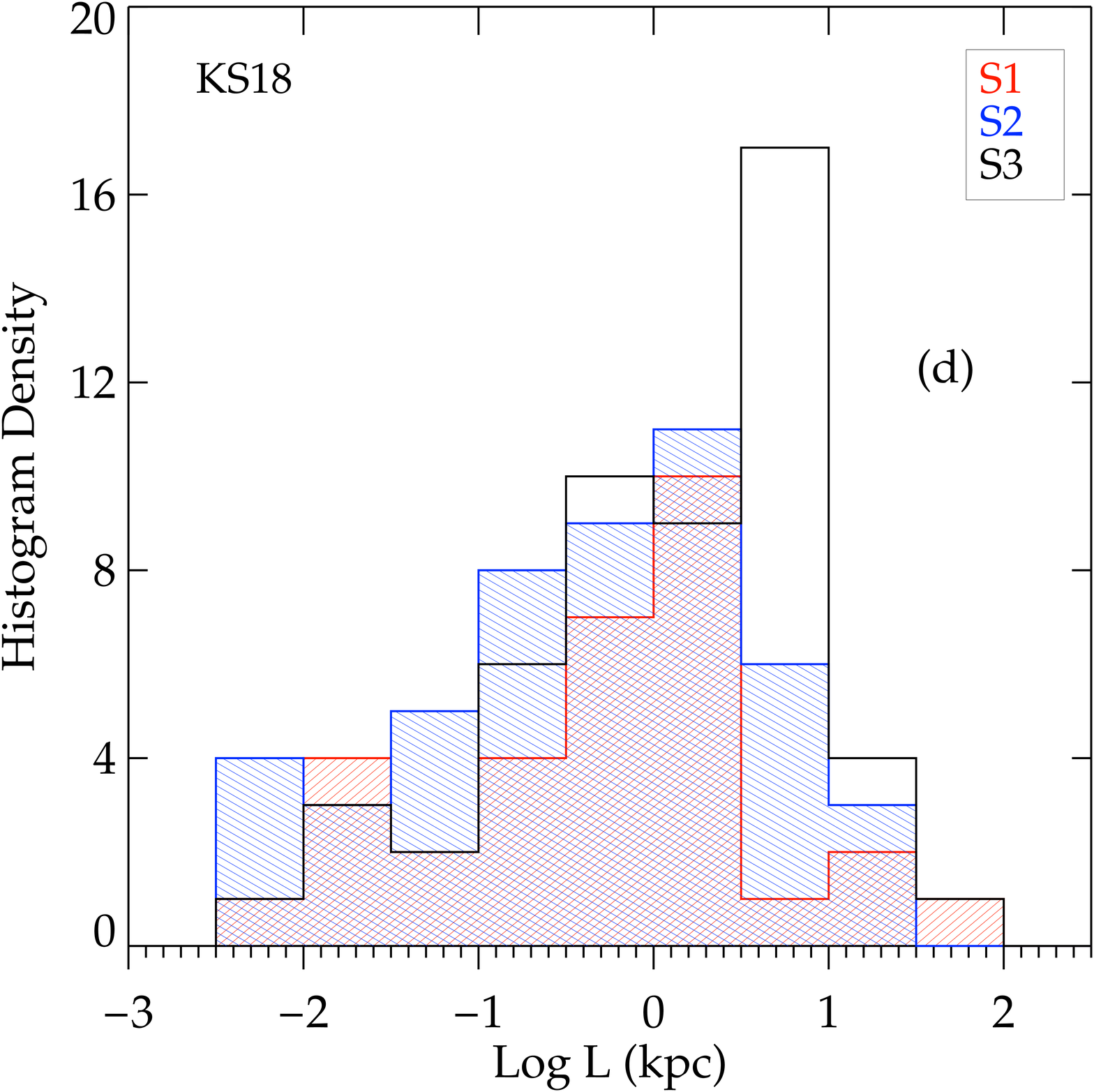}
\end{subfigure}
\caption{Histograms of derived parameters for model \Mt. Panel (a): hydrogen density (\nh), 
panel (b): carbon abundances ({\it $[C/H]$}), panel (c): overdensity ($\Delta$) and panel (d):
line-of-sight thickness, {\it L} of the components in logarithmic scale.
Red, blue and black histograms are used to show the results for {\textcolor{blue}{\sam}}, \san\ and
\sao\ data for our fiducial \KS18\ UVB, respectively.}
\label{fig45}
\end{figure*}

Fig.~\ref{fig2} also shows the redshift evolution of column density ratio of {\CIII} to {\CIV} predicted by our models for two UVBs
(Solid line for our fiducial \KS18\ UVB and dashed line for
\citetalias{Haardt2012} UVB). The model predictions
show that this column density ratio increases mildly with increasing $z$ in the range 2 to 3.4 for a fixed hydrogen
density of the absorber.
Next, we use the observed {\rat} to constrain {\nh} in the three redshift bins
identified above. For the redshift bins [2.1, 2.4] and [2.8, 3.4],
density range between $-$4.0 $<$ ${\rm log}$~{\nh} (in \cmcb) $< -2.9$ consistently reproduce the observed ratios (the blue and orange shaded regions) whereas for
the mid redshift bin we require a density
range from $-$3.6 $<$ ${\rm log}$~{\nh} (in \cmcb) $<$ $-$3.1 for the two UVBs.
As discussed before this redshift range also shows
slight deficit of systems.
The overall range in density for all the absorber is
$-$2.2 $<$ ${\rm log}$~{\nh} (in \cmcb) $<$ $-$4.5, for both the UVBs used in this analysis.

The mild redshift evolution shown by the median values of observed column density ratio in the three redshift bins
are well reproduced by the model with ${\rm log}$~{\nh} (in \cmcb) {{\si}} $-$3.4.
It is also clear from Fig.~\ref{fig2} that model curves based on our fiducial \KS18\ UVB and that
from \citetalias{Haardt2012} differ from each other only at $z > 2.5$. So the {\nh}
required for models with \citetalias{Haardt2012} UVB is slightly lower than that for our fiducial model
in this redshift range. 
The observation as noted before in the text show larger range 
in quasar spectral index ($\alpha$)
in the redshift of our interest  \citep{Khaire2017}. So, we also consider two UVBs with quasar SEDs having
$\alpha$ value $-1.6$ and $-2.0$. The results are shown
in Fig.~\ref{figA111} (in the Appendix~\ref{Ap1}). It is clear from Fig.~\ref{figA111} that a given $N$(\CIII)/$N$(\CIV) can be produced
with lower {\nh} when we use UVBs generated using higher values of
$\alpha$. It is also evident that an uncertainty in $\alpha$ of $\pm0.2$
(with respect to our fiducial value $\alpha = -1.8$) translates to an uncertainty of
$\pm$0.17 dex in the inferred {\nh} over the redshift range of interest in this study. This we will treat as a typical
systematic uncertainty in the inferred {\nh} arising from  allowed uncertainties in the quasar extreme-UV spectral index.
It is also clear that the model predicted curves in $z$ $-$ \nh\ plane are parallel to each other so inferred redshift evolution
of the derived parameters like {\nh} will not depend on our choice of $\alpha$ (unless otherwise there is a
redshift evolution in $\alpha$).

As explained in Section.~\ref{s1}, not  all  {\CIII} components shown in Fig.~\ref{fig2} have an aligned
{\HI} component which makes it difficult for us to constrain
the physical conditions and chemical composition of such absorbers using photoionization models.
However, detailed analysis is possible for the components in the sub-sample {\sam} as these
components have co-aligned {\HI} component association. 
As our stopping criteria in the model \Mo\ is the total hydrogen column density, the
models predicted {\nhi} (i.e., {\nhij})
can be compared with the observed {\nhi} for these components denoted by {\nhio}. 

In Fig.~\ref{fig3}  we compare {\nhij} from both the models with the {\nhio}. The result using our fiducial \KS18\ UVB and
the \citetalias{Haardt2012} UVB are plotted with solid red color circles and yellow filled squares,
respectively. The dotted lines shows {\nhio} $\propto$ {\nhij} for different proportionality  constants (indicated
above each line). It is clear from the figure that apart from one component the {\nhio} is systematically
lower than the one predicted from our models.
The observed {\nhi} in the
sample \sam\ (left-hand panel of Fig.~\ref{fig3}) shows a range,
{\nhi$_{obs}$}{\si} $7 \times 10^{12}$ $-$ $1 \times 10^{16}$  {\cmsq} whereas, the model generated
{\nhi}$_J$ has a range from $2 \times 10^{14}$ $-$ $1 \times 10^{17}$  (in {\cmsq}) . The median value
of the observed neutral hydrogen
column density is, ${\rm log}$~\nhi\ = 14.51 {\cmsq}
whereas our model generated median values of column densities are ${\rm log}$~\nhi\ $=$ 16.04 {\cmsq} and
= 16.01 {\cmsq} for \citetalias{Haardt2012} and \KS18\ UVBs, respectively.
{\it Hence the median values of {\nhi} predicted by model \Mo\ is almost factor {\si} 32
higher than the observed column density for sample \sam\ for the set of UVBs considered here.
Moreover, there are {\si}
78\% components in {\sam} for which {\nhij} is almost factor of 10 to 1000 or more times higher than {\nhio}. 
This implies that if {\CIII} components are in hydrostatic equilibrium then $f_g$ should be less than
$1.6 \times 10^{-3}$ in most of these components.}
Note the density range allowed in these clouds, even considering changes in $\alpha$, is not sufficient to alter this conclusion.
Also most of the cases,
assumed gas temperature is close to  the inferred kinetic temperature from the Doppler parameter ($b$) values by \citetalias{Kim2016}.

In order to check whether this is the case for sample \san\ as well,
we plot {\nhi} predicted by our models with the observed {\nhi} in the right-hand panel of Fig.~\ref{fig3}. Here also we find median
{\nhij} is a factor of \si24 higher than
the median value of \nhi$_{obs}$ with a similar trend as in sample \sam.
Note that in this case
the observed {\nhi} could be an upper limit as \HI\ absorption is not well aligned.
However, the trend clearly supports that the
conclusion derived for the sample \sam\ may not be specific to the components with a well aligned
\HI\ absorption alone and
may be more generic to the {\CIII} components. 
This implies either $f_g$ is very
small in these components or these {\CIII} components are not in hydrostatic
equilibrium with the dark matter
or its self-gravity.
In what follows
we explore models with {\nhi} as stopping criteria without imposing 
hydrostatic equilibrium condition given in Eq.~\ref{a15}.
\subsubsection[]{Models with stopping criteria {\nhi} $\approx$ {\nhio}}
As an alternative approach we construct another set of photoionization models
(hereafter, model \Mt) where the  observed value of
$N({\HI})$ is used as the 
stopping criteria and the gas temperature is self-consistently
calculated under photoionization equilibrium. Following similar approach as \Mo\ we constrain $n_H$ for each component separately
from the 
observed column density ratio of {\CIII} to {\CIV}.
Once $n_H$ is constrained, we generate another grid of model outputs
varying carbon abundances ({\it $[C/H]$}) in logarithmic scale of 0.1 to match
with the individual 
observed column densities of {\CIII} and {\CIV}. This allows us to infer {\nh} and metallicity of the individual
{\CIII} components that have aligned
{\CIV} and {\HI} absorption. We show histograms of different cloud properties obtained from
\Mt\ ({\it which we quote as our fiducial model}) in Fig.~\ref{fig45} for our fiducial \KS18\ UVB. In Table~\ref{tab1} we summarize the observed and model predicted column
densities as well as the {\nh} and {\it $[C/H]$} required by the models for {\sam} components with clear detections.\\

\begin{figure*}
\centering
 \includegraphics[totalheight=0.31\textheight, trim=.1cm .1cm .1cm .1cm, clip=true, angle=0]{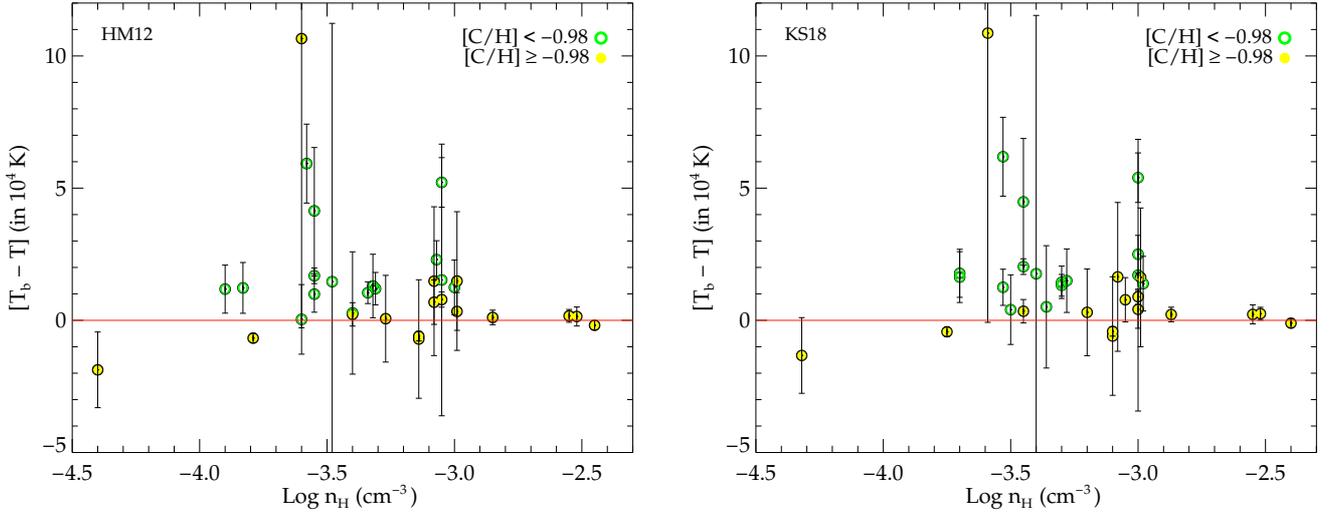}
 \caption{Difference in the temperature predicted by the photoionization model (T) for \citetalias{Haardt2012} (left-hand panel) and our fiducial \KS18\ UVB (right-hand panel)
and the inferred temperature (${\rm T_b}$)
from the Voigt profile fitting. High metallicity (i.e $[C/H]>-0.98$)
components are plotted in yellow color circles and the low metallicity components (i.e $[C/H]<-0.98$)
are plotted with green open circles.
The error bars show the 3$\sigma$ error range in
$T_b$.
It seems that $T_b - T$ deviations from zero occur at low metallicity and low density components.
}
\label{fig_temp}
\end{figure*}

\begin{figure*}
\centering
 \includegraphics[totalheight=0.31\textheight, trim=.1cm .1cm .1cm .1cm, clip=true, angle=0]{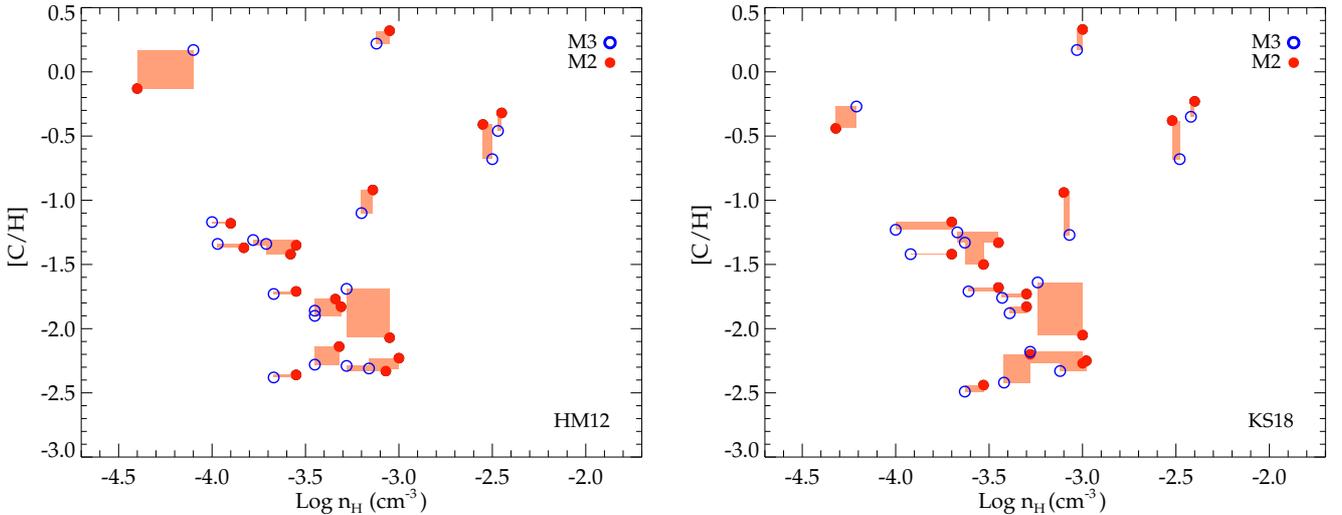}
 \caption{Comparison of hydrogen density
 and metallicity between two models \Mt\ and \Mth for the 17 absorbers which are outside of $3\sigma$ range from $T_b$.
 for \citetalias{Haardt2012} (left-hand panel) and \citetalias{Khaire2018} (right-hand panel) UVBs.
The red solid circles and blue open circles show the \Mt\ and \Mth\
outputs, respectively. 
The orange color patches are used to relate \Mt\ to \Mth\ values.}
\label{fig5}
\end{figure*}

\noindent{\bf Hydrogen density:} The hydrogen density range for the $32$ components in {\sam}
is, 
${\rm log}$~{\nh} (in \cmcb) $\in$ [$-4.3$, $-2.4$], with a median ${\rm log}$~{\nh} (in \cmcb) of $-3.2$
for our fiducial \KS18\ UVB model. As expected this is similar
to the {\nh} in model \Mo\ that was required to fit median {\rat} as a function of $z$.
Left-hand top panel of Fig.~\ref{fig45} shows the histogram distribution of the 
hydrogen density derived in model \Mt\ for all there samples.
The required density distribution for the sub-sample {\san} and the sample {\sao} are also shown in Fig.~\ref{fig45}.
The median ${\rm log}$~{\nh} (in \cmcb) $=$ $-$3.1 is found for both {\san} and {\sao}.
As discussed before changing $\alpha$ between $-1.6$ and $-2.0$ introduces a systematic uncertainty in the inferred {\nh} by $\pm$0.17 dex. 
Assuming the uncertainty due to UVB in-addition to the slight offset 
due to allowed range in {\nh} to match the column densities, we notice that
{\nh} distribution of well aligned components (sample \sam) is almost consistent with other two samples (\san\ and \sao).
{\it However, we note that the density we have obtained
for individual components in {\sam} are systematically higher by 1.2 to 1.3 dex compared to the values quoted by KIM16
(see their table A1 and A2) for the same components.}
As can be seen from figure~12 of \citetalias{Kim2016}, $\nu J_\nu$, values for \HM12\ shown are much lower than other UVB shown in that figure.
In comparison to what we plot in Fig.~\ref{fig1}, the \HM12\ UVB shown in \citetalias{Kim2016}
  are off by a factor 1.1 dex. This could be accounted if \KIM16\ have missed a factor of 4$\pi$. If one uses the spectrum shown by \citetalias{Kim2016}
  in the cloudy models then the derived density is expected to be smaller by at least one order of magnitude.  We provide some details on
  this issue in Appendix~\ref{Ap6}.\\

\noindent{\bf Carbon abundance:} In panel (b) in Fig.~\ref{fig45}, we show the  {\it $[C/H]$}
distribution for the three samples. For {\textcolor{blue}{\sam}}, the derived range for {\it $[C/H]$}
is [$-2.44$, $0.52$] with a median of $-0.98$. We find similar range of {\it $[C/H]$}
for other two samples irrespective of upper limits of column densities. However the median values of {\it $[C/H]$}
for {\san} and {\sao} are $-1.27$ and $-1.64$, respectively which are moderately lower compared that of {\sam}.
There are 4 components in 
\sam\ and in total 8 components in the 53 absorption systems which show super solar metallicity.
We find change in $\alpha$ between $-1.6$ and $-2.0$  leads to an uncertainty of 0.2 dex in the metallicity
measurement based on our fiducial model. As can be seen from Table.~\ref{tab1}, our metallicity measurements
for most of the components in \sam\ are slightly higher (i.e upto 0.4 dex with a median of 0.1 dex)
than those derived by \citetalias{Kim2016}. This differences in density and metallicity motivate us to revisit different
correlations found by \KIM16\ that were instrumental in drawing basic properties of the optically thin
\CIII\ components which we do in the following sections. \\

\noindent{\bf Overdensity ($\Delta$) and cloud thickness:} In panel (c) of Fig.~\ref{fig45} we
show the histogram distribution for overdensity ($\Delta$ = \nh/<$n_H$> with <$n_H$> being the mean hydrogen density
(<$n_H$> $=$ $1.719 \times 10^{-7}$ \cmcb~$(1+z)^3$) of the
uniform IGM that contains all the baryons). Simple analytic models of IGM \lya\ forest at $z\sim2.6$
suggests that most of the absorptions with ${\rm log}$ $N$(H~{\sc i}) $\le$ 14 originate from regions with
$\Delta \le 4.0$ (i.e ${\rm log}$ $\Delta\le -0.60$) \citep[see Eq.~10 and Eq.~11 of][]{Gaikwad2017}.
The data shows a range of overdensities, ${\rm log}$ $\Delta$ $\approx$ $0.75 - 3.00$, with a
median value, ${\rm log}$~$\Delta$ $=$ $2.0$ for all the components in our samples. As can be seen from
Table~\ref{tab1}, that most of the components in sample \sam\ have ${\rm log}$~$N$(H~{\sc i}) $<$ 14.0.
Our derived densities for these components clearly confirm that they are not like typical IGM
clouds but they rather have higher
densities and hence are more compact compared to the IGM clouds having similar $N$(H~{\sc i}). 
Since we have the information of {\nh} (in {\cmcb}) and $N$(H) (in {\cmsq}) for individual components
we calculate the
line-of-sight thickness defined as, {\it L} (in cm) = $N$(H)/{\nh}. Fig.~\ref{fig45} panel (d) shows the histogram
distribution of
line-of-sight length, {\it L} in logarithmic scale. We find {\it L} 
for the 32 components in {\sam} varying over a wide range from 7 pc to 35 kpc which is several times smaller than showed by
\KIM16\ (20 kpc to 480 kpc).
Similar range in {\it L} (\san: 2 pc to 15 kpc, \sao: 8 pc to 35 kpc) is also seen for other two sub-samples
where the line-of-sight thickness could be upper limits.
Note that there are 
3 components (component number 14, 15 and 30 in Table~\ref{tab1}) with super-solar metallicity which have
$L_J$ greater than 1000 times of $L$. 
We provide more detailed discussions on the possible origin for their
line-of-sight thickness (or cloud size) in the following section.

\noindent{\bf Gas temperature:}
In the IGM (at say $z<5$), as the recombination time-scales are larger,
the gas temperature at any epoch is decided by the
energy injection by photoheating during the epoch of H~{\sc i} and He~{\sc ii} reionization followed by adiabatic
cooling due to cosmological expansion \citep{Hui1997}. In the case of CGM, the temperature can be related to various local heating and cooling sources.
However, 
the gas temperature (T) for our fiducial model is self-consistently calculated in {\CLOUDY} under thermal equilibrium
where recombination cooling is equated to the photoheating.
Therefore, the temperature computed by {\CLOUDY} need not be the correct kinetic temperature of the gas even when
our estimated densities are much higher than the mean IGM density (<$n_H$>). In the case
of sample \textcolor{blue}{\sam}, \citetalias{Kim2016} have obtained kinetic temperature (${\rm T_b}$) by decomposing the
thermal and non-thermal contributions to the $b$ values. In Fig.~\ref{fig_temp}, we plot the (${\rm T_b/T}$) as
a function of \nh\ for \citetalias{Haardt2012} and \citetalias{Khaire2018} UVBs. High metallicity (i.e $[C/H]>-0.98$)
components are plotted in yellow color circles and the low metallicity components (i.e $[C/H]<-0.98$) are plotted with
green open circles.
The error bar shows the 3$\sigma$ error range in
$T_b$, where $\sigma$ is the error in $T_b$ obtained from the $b$ error.
About {\si} 53 percent data points (component number 1, 5, 6, 7, 8, 9,
11, 12, 13, 15, 16, 17, 21, 23, 25, 27 and 28 in Table~\ref{tab1}) in \sam\  have
temperature predicted by {\CLOUDY} outside the range of
$T_b \pm 3\sigma$
found by \citetalias{Kim2016}.
It is clear from the Fig.~\ref{fig_temp} that points where  $T$ is consistent with  $T_b$ are the ones
that also have high metallicity (i.e $[C/H] > -0.98$) and high density (${\rm log}$~\nh\ $\ge$ $-3.5$).

To quantify the effect of this discrepancy on the derived values of {\nh} and metallicity 
we setup another set of photoionization models (which we denote as \Mth).
The basic \CLOUDY\ modeling is same as \Mt, however we fix the gas temperature to be  $T_b$
(i.e constant temperature models) given
by \citetalias{Kim2016} obtained from the $b$ value. In Fig.~\ref{fig5}, we show
the comparison of {\nh} and {\it $[C/H]$} obtained
from \Mt\ and \Mth\ for the 17 absorbers which are outside of $3\sigma$ range from $T_b$. The red and blue
solid circles
are used to denote the results for \Mt\ and
\Mth, respectively. The steel patches are used to show the differences between derived quantities from \Mt\ and \Mth.
The mean difference in density between \Mt\ and \Mth\ for the two UVBs are
 ($\Delta$ ${\rm log}$ ${n_{H}}$) =  $0.1$ (resp. 0.12)
and $(\Delta{[C/H]}) = 0.05$ (resp. 0.02) for our fiducial \KS18\ UVB (resp. \citetalias{Haardt2012}).
Therefore, our models with only photoionization considerations do not introduce
notable off-sets in the derived \nh\ and [{\it C/H}].
\subsubsection[]{Model predictions for other ions}
\begin{figure}
\centering
\includegraphics[totalheight=0.287\textheight, trim=0cm 0.0cm 0cm .0cm, clip=true, angle=0]{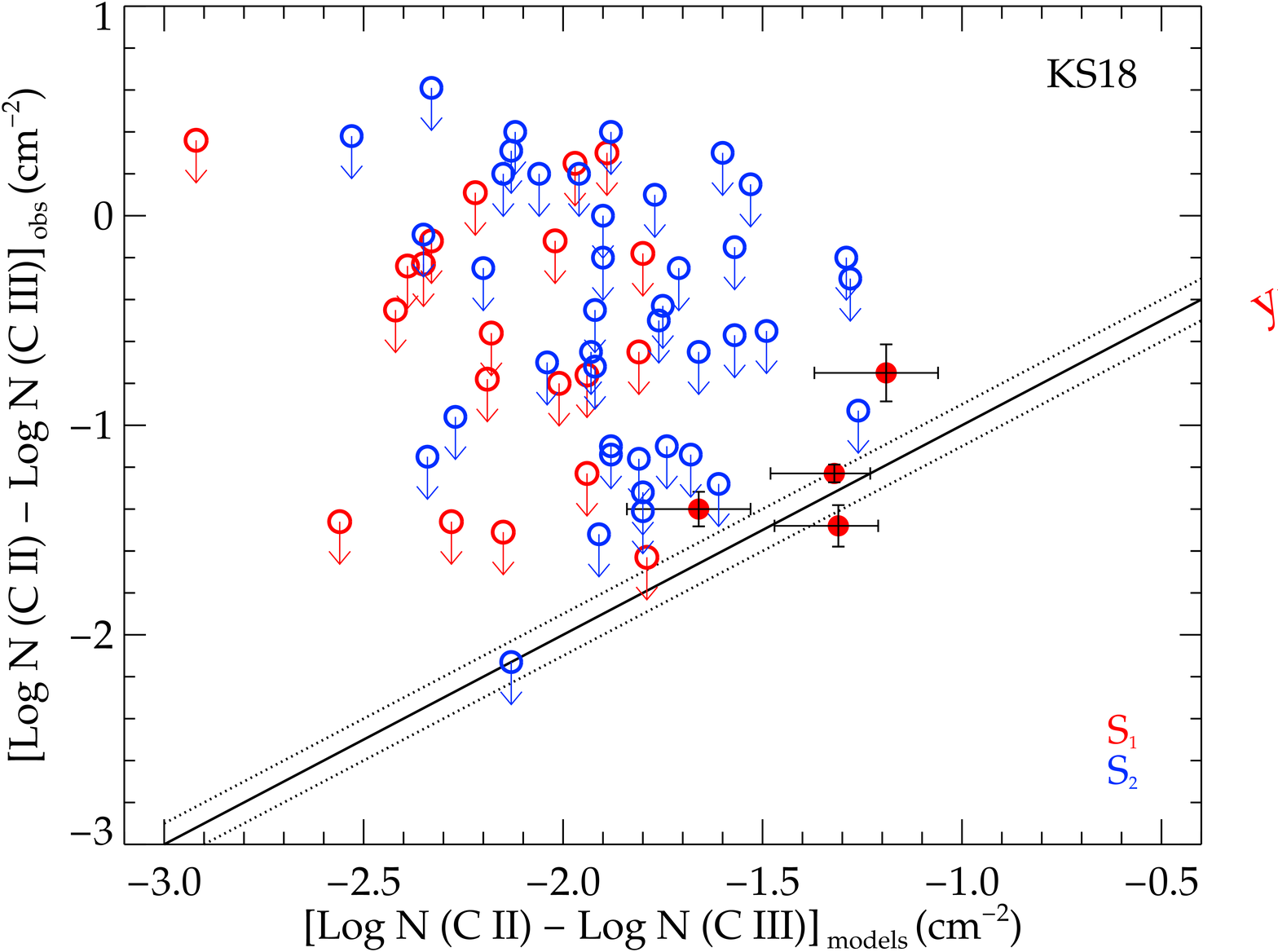}
\caption{Comparison of observed column density ratio of {\CII} to {{\CIII}} with the model predicted ratio for our fiducial
\KS18\ UVB.
Black horizontal bar is used to show
error associated with the observed column density. 
 Red and blue circles are used for {\sam} and {\san}, respectively. Filled circles are used
  to represent measurements and
  open symbols are used for
 upper limits on {\CII} column densities.
 The `y $=$ x' equality is shown in black solid line. The black dotted lines show error range of
  $\pm$ 0.13 dex around this equality line due to the model uncertainties.}
\label{fig3577}
\end{figure}

In order to verify the results of our fiducial model \Mt\ we compare available observed column densities of several other
ions (such as {\CII}, {\SiIII} and {\SiIV}) associated with the components with our model predictions.
In Fig.~\ref{fig3577}, we show
observed (or limiting) column density ratio  of {\CII} to {{\CIII}}
with the predicted ratio from our model \Mt. Solid circles with error bars represent measurements.
Note the quoted errors take into account the uncertainties in the model predictions due to UVB (as discussed
earlier) and measurement uncertainties. The open circles with arrows are the upper limits. The black solid line represents
the equality in both observed and model predicted column density ratio in
logarithmic scale. The dotted lines give $\pm$0.13 dex uncertainty around this equality line. We show
data for both components from sub-sample {\sam} and {\san}.
Apart from four cases, {\CII} is not detected
\begin{figure}
 \centering
 \includegraphics[totalheight=0.287\textheight, trim=0cm 0.0cm 0cm .0cm, clip=true, angle=0]{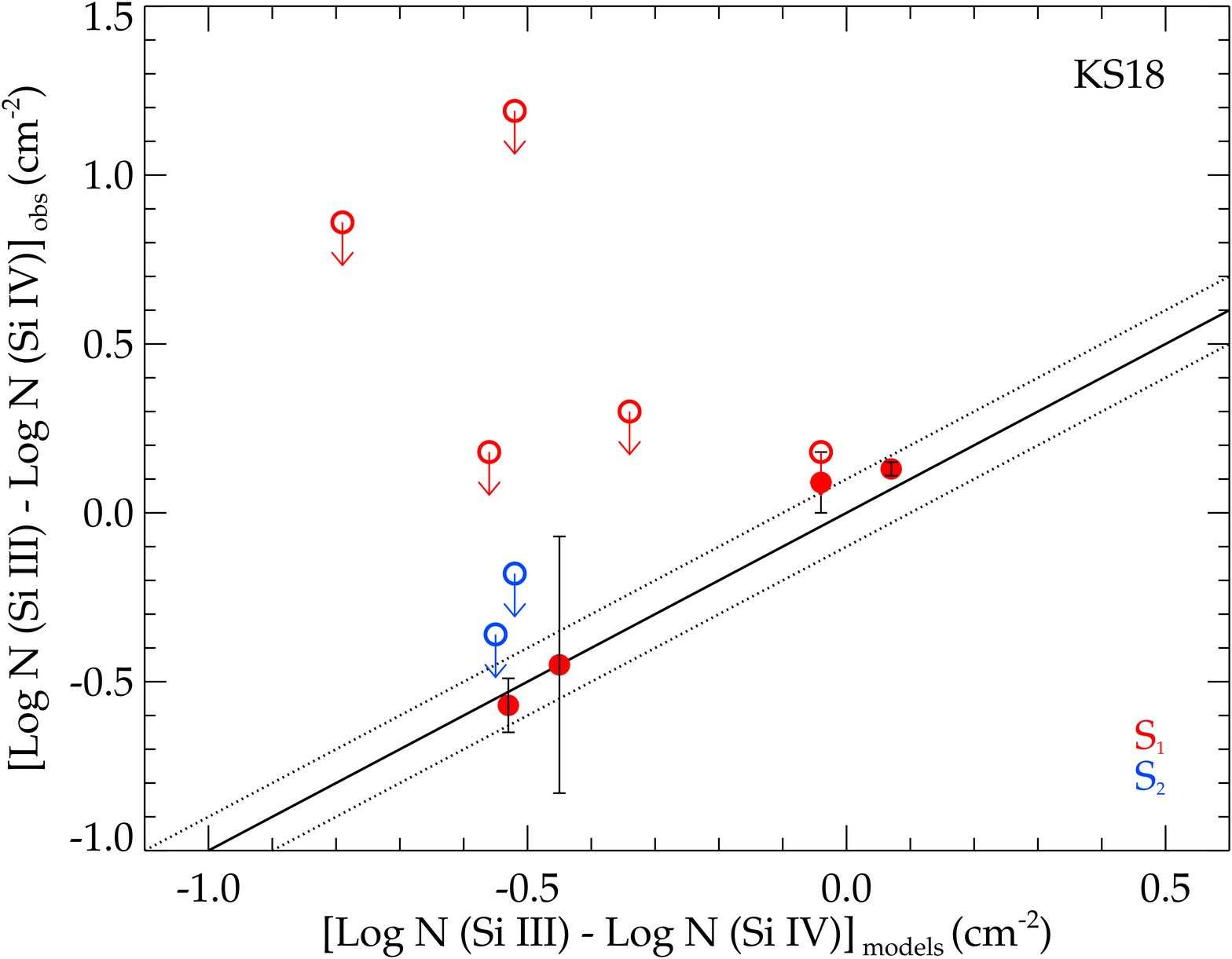}
 \caption{Comparison of observed column density ratio of {\SiIII} to {{\SiIV}} with the model
 predicted ratio for our fiducial \KS18\
 UVB. Black vertical bars show the errors associated with the observed column densities.
  Red and blue circles are used for {\sam} and {\san}, respectively. Filled circles are used
  to represent measurements and
  open symbols are used for upper limits on {\SiIII} column densities.
  The `y $=$ x' equality line is shown in
  black solid line. The black dotted lines show error range of
  $\pm$ 0.13 dex around this equality line due to the model uncertainties. }
 \label{fig101}
 \end{figure}
in the {\CIII} components. The derived upper limits on $N$(\CII)/$N$(\CIII) for these components are
consistent with our model predictions.  In the four cases where we have the {\CII} absorption line
detections our model predictions match with the observations within uncertainties. 

For nine components in the sample \sam\  (i.e component number 10, 13, 18, 19, 20, 23, 24, 29 and 30 in Table~\ref{tab1}), we
clearly detect {\SiIV}. Four of these components (i.e 10, 18, 19 and 23 in Table~\ref{tab1}) also show clear {\SiIII} absorption. 
We obtained column
densities of {\SiIV} and {\SiIII} by fitting Voigt profiles (with $b$ value consistent with the
fits from \citetalias{Kim2016})
using VPFIT \citep{vpfit}. For {\SiIII} non-detections we obtained upper limits assuming the $b$ value similar to {\CIII}.
There are two components in  \san\ for which we could get upper limits on the ratio of {\SiIII} to {\SiIV}
column densities. We show Voigt profile fit results for column densities and our fiducial model predicted
results in Table~{\color{blue}{C2}} (see online supplementary data).

In Fig.~\ref{fig101}, We show logarithm of observed column density ratio, ${\rm log}$~{\ratb} 
along with their predictions using model \Mt\ for fiducial \KS18\ UVB.
Black bars are used to show errors associated with the observed column density ratios. 
Red and blue color circles are used for samples {\sam} and {\san}, respectively. Filled circles are used
  to represent measurements and
  open symbols are used for components with upper limits on {\SiIII} column
densities and the black 
solid line shows equality in both the ratios and dotted lines give $\pm$0.13 dex uncertainty around this line. As can be seen from the figure
the model predictions agree with the observed ratios within 0.1 dex for the four components that have clear detections.
Also, all the upper limits are consistent with the model predictions. This confirms that the density constrains we
obtained for the components based on the column density ratio of {\CIII} to {\CIV} are consistent with the observed
{\SiIII} to {\SiIV} column density ratio. We next obtain the [Si/H] for each components by trying to reproduce the observed
column density of {\SiIV}. We find that [Si/C] $\sim 0.12\pm0.10$  for these nine components. This once again confirms the
consistency of metallicity we derived for individual components and [Si/C] in the {\CIII} absorbers being
close to solar value.

\noindent \section{Redshift evolution of parameters}
\begin{figure*}
  \centering
   \includegraphics[height=5.5cm, trim=2cm 0.0cm 2cm 2cm, clip=true, angle=0]{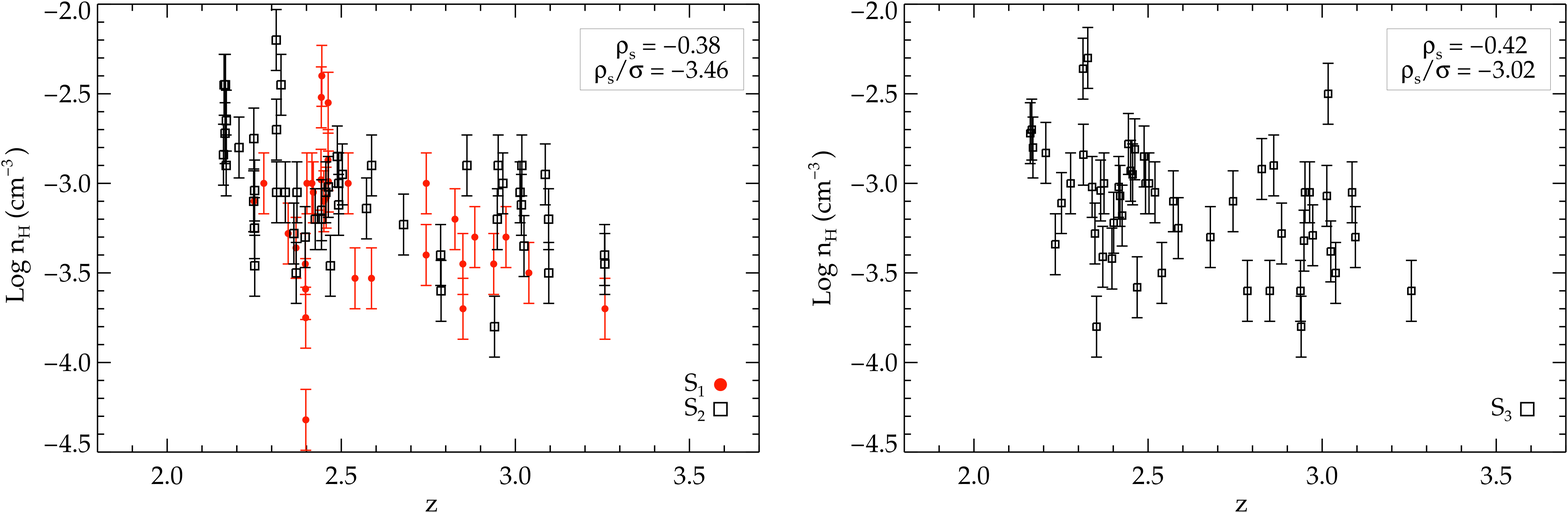}
   \includegraphics[height=5.5cm, trim=2cm 0.0cm 2cm 2cm, clip=true, angle=0]{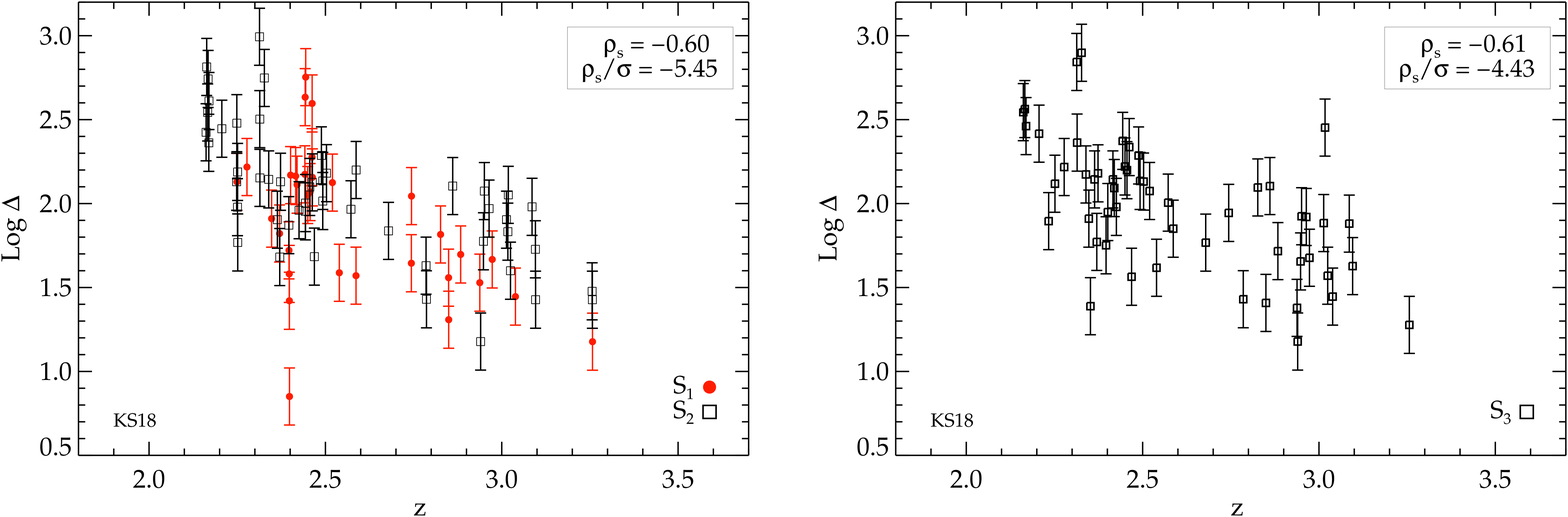}
   \includegraphics[height=5.5cm, trim=2cm 0.0cm 2cm 2cm, clip=true, angle=0]{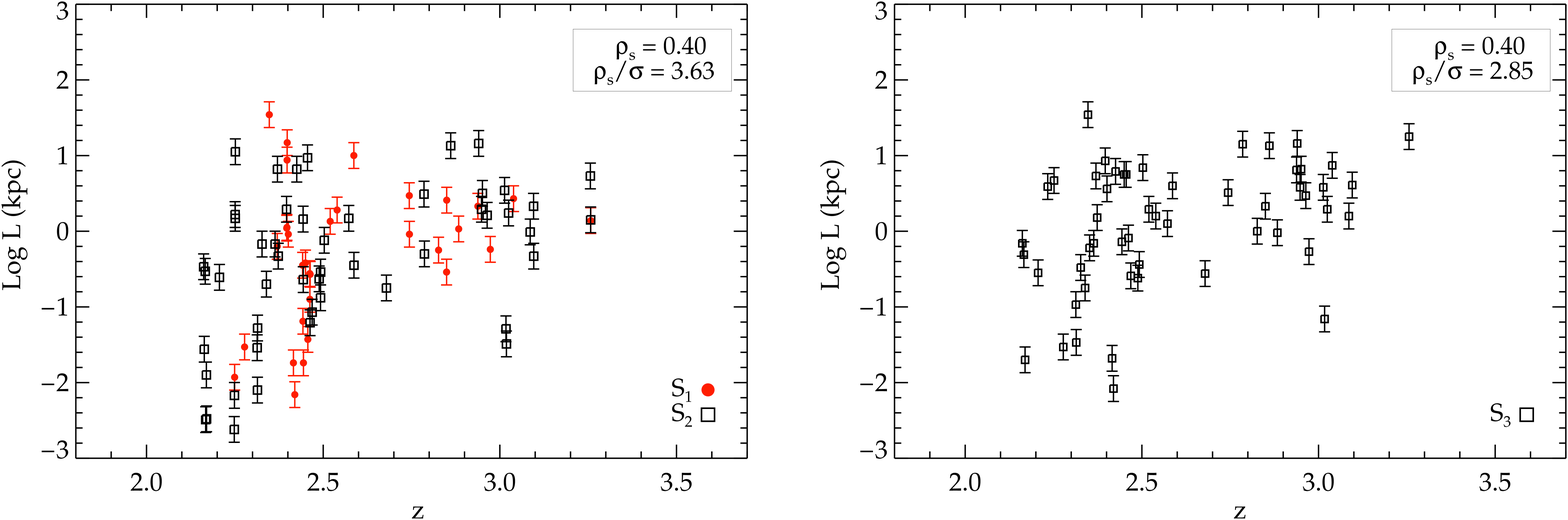}
   \includegraphics[height=5.5cm, trim=2cm 0.0cm 2cm 2cm, clip=true, angle=0]{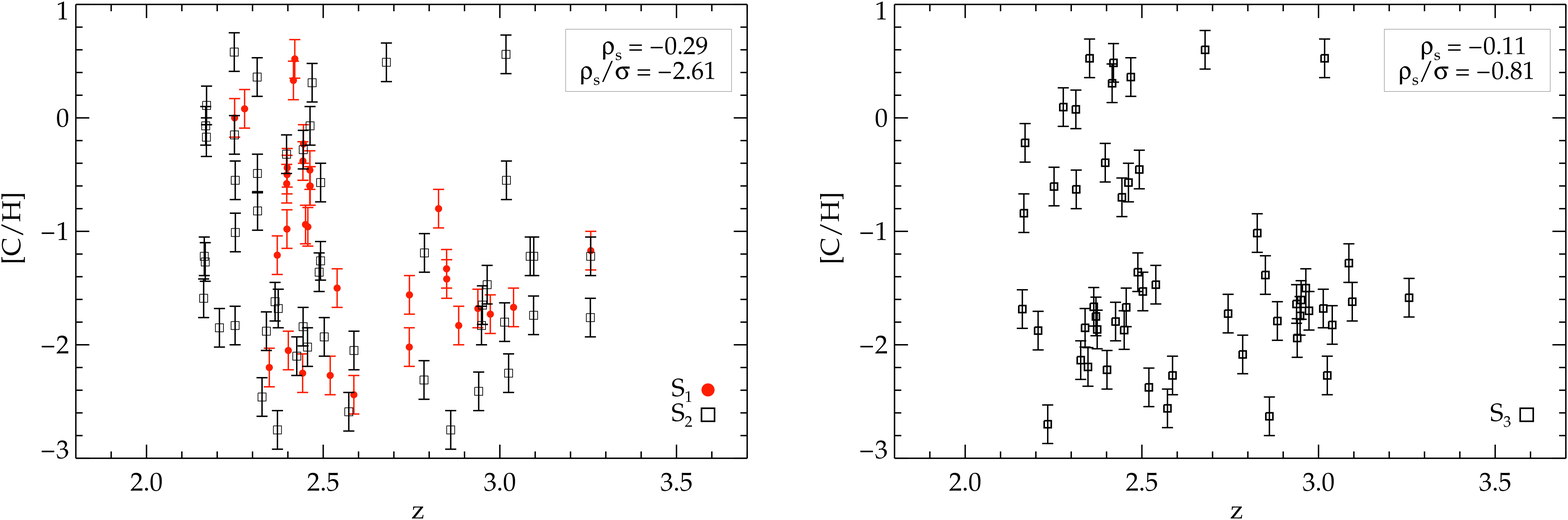}
   \caption{
     Redshift evolution of various parameter derived using our photoionization models (i.e model \Mt\ using our fiducial \KS18\ UVB).
     The evolution of
     hydrogen density (\nh), overdensity ($\Delta$), line-of-sight thickness (L) and metallicity (C/H)
     are plotted in different panels.
     In the left columns we plot the results for samples \sam\ (red solid circles)
     and \san\ (open squares). The right columns show
     the results for \sao\ where we have modeled the total column density for each system.
     In each panel we also show the rank correlation coefficient and its significance level.
}
\label{fig_zeve}
\end{figure*}

In this section, we investigate the possible redshift evolution of derived parameters.
In Fig.~\ref{fig_zeve}, we plot the redshift evolution
of hydrogen density (\nh), overdensity ($\Delta$), line-of-sight thickness ({\it L}) and metallicity ([{\it C/H}])
derived for our fiducial \KS18\ UVB and model \Mt.
It is clear from the top-left-hand panel of this figure that the average density of the {\CIII} components
decreases with increasing $z$.
Typical error bars shown take into account the uncertainty in our \nh\ measurements contributed by uncertainties
in the UVB and in the column density measurements. For the components in the combined sample of \sam\ and
\san\ we find the Spearman rank correlation coefficient, $\rho_s = -0.38$ with a
two-sided significance of its deviation from zero of $3\times10^{-4}$ (or the
anti-correlation is at 3.5$\sigma$ level). Combining \sam\ and \san\ are justified as in all cases \CIII\ and
\CIV\ are aligned. When we consider
the average density in each {\CIII} system (i.e sample \sao\ shown in right-hand panels)
the Spearman rank correlation coefficient is  $\rho_s = -0.42$ and the
two-sided significance of its deviation from zero is $2\times10^{-3}$ (or the
anti-correlation is at 3.0$\sigma$ level). {\it Thus we find a statistically significant
trend of increasing \nh\ associated with \CIII\ absorbers with decreasing redshift.} 

In the second row from the top in Fig.~\ref{fig_zeve}, we show
the gas overdensity $\Delta$ as a function of $z$. As expected this shows a much stronger anti-correlation
with redshift. For components in the combined sample of \sam\ and
\san\ we find the Spearman rank correlation coefficient, $\rho_s = -0.60$ with a
two-sided significance of its deviation from zero of $2\times10^{-9}$ (or the
anti-correlation is at 5.5$\sigma$ level). When we consider
the average overdensity of each {\CIII} system (i.e sample \sao\ shown in right-hand panels)
the Spearman rank correlation coefficient is  $\rho_s = -0.61$ and the
two-sided significance of its deviation from zero is $9\times10^{-7}$ (or the
anti-correlation is at 4.4$\sigma$ level).
As discussed before, for the measured $N$(\HI), the {\CIII} absorbers tend to
originate from gas having larger $\Delta$ compared to what one expects for a typical IGM gas.
However, the inferred \nh\ is less than what is seen in the ISM of galaxies. Therefore,
the optically thin {\CIII} components studied here are most probably originating from gas outside the
galactic discs (outflows, inflows or galactic halo gas).
{\it In summary, our results suggest that {\CIII} absorbers tend to probe regions of higher density (and hence higher
overdensity)  as we move towards lower redshifts.}

In the third row from the top in Fig.~\ref{fig_zeve}, we plot the line-of-sight thickness
({\it L}) 
as a function of $z$.  For components in the combined sample of \sam\ and
\san\ we find the Spearman rank correlation coefficient, $\rho_s = 0.40$ with a
two-sided significance of its deviation from zero of $2\times10^{-4}$ (or the
correlation is at 3.6$\sigma$ level). When we consider the full system
as a cloud (sample \sao), we have $\rho_s = 0.40$  with a
two-sided significance of its deviation from zero of $3\times10^{-3}$ (or the
correlation is at 2.9$\sigma$ level). {\it Thus there is a statistically significant evidence for
the low$-z$ (i.e $2.1\le $z$ \le 2.5$) {\CIII} absorbers being smaller in size compared
to those at high$-z$ (i.e $z\ge2.5$).}
However, unlike for \nh\ or $\Delta$ at low$-z$, in the case of {\it L}, we
notice that  
the spread in {\it L} is very large and a population of sub-kilo-parsec components are predominantly present.
Lack of such {\CIII} components at the high$-z$ could drive the observed $z$ evolution of $L$.
We discuss the origin of {\it L} in more details in the following section.

\begin{table*}
  \begin{center}
    \caption{Results of correlation analysis between different derived parameters and their $z$ dependence for sample \sam+\san\ }
  \begin{tabular}{ccccccc}
 \hline
 & \multicolumn{2}{c}{Full sample}& \multicolumn{2}{c}{For $z < 2.5$} &\multicolumn{2}{c}{For $z \ge 2.5$} \\
   Parameters &  $\rho_s$ &$\rho_s/\sigma$& $\rho_s$ &$\rho_s/\sigma$& $\rho_s$ &$\rho_s/\sigma$\\ \hline \hline
{${[C/H]}$} versus ${\rm log}$ $L$       & $-$0.68 & $-$6.09 &$-$0.66 & $-$4.60 &$-$0.62 & $-$3.50 \\ 
${\rm log}$ $\nc4$ versus ${\rm log}$ $L$    & +0.19 & +1.69 &+0.37 & +2.60 &+0.07 & +0.38 \\ 
${\rm log}$ $\nhi$ versus ${\rm log}$ $L$    & +0.59 & +5.26 &+0.60 & +4.30 &+0.57 & +3.21 \\ 
${\rm log}$ $\nc4$ versus ${[C/H]}$      & +0.42 & +3.76 &+0.29 & +2.00 &+0.52 & +2.94 \\ 
${\rm log}$ $\nhi$ versus ${[C/H]}$      & $-$0.65 & $-$5.80 &$-$0.64 & $-$4.5 &$-$0.59 & $-$3.35 \\ 
\hline
  \end{tabular}
  \label{tab_cor1}
  \end{center}
\end{table*}

\begin{table*}
  \caption{Correlation analysis for sub-samples from \sam+\san\ based on {\it L} and [{\it C/H}]}
\begin{center}
\begin{tabular}{ccccccccc}
\hline
& \multicolumn{2}{c}{For ${[C/H]}$ $<$ -1.22} & \multicolumn{2}{c}{For ${[C/H]}$ $\ge$ -1.22}&  \multicolumn{2}{c}{For ${\rm log}$ $L$ $<$ -1.0 kpc} & \multicolumn{2}{c}{For ${\rm log}$ $L$ $\ge$ -1.0 kpc} \\ 
Parameters & $\rho_s$ &$\rho_s/\sigma$& $\rho_s$ &$\rho_s/\sigma$&$\rho_s$  &$\rho_s/\sigma$& $\rho_s$ &$\rho_s/\sigma$\\ \hline \hline
${\rm log}$ $\nc4$ versus ${\rm log}$ $L$ & $+$0.50 & $+$3.16 &$+$0.54 &$+$3.44 &.... & .... & .... & .... \\      
${\rm log}$ $\nhi$ versus ${\rm log}$ $L$ & $+$0.41 & $+$2.59 &$+$0.51 &$+$3.22 &....&.... &....&.... \\      
${\rm log}$ $\nc4$ versus ${[C/H]}$ &   .... & .... &.... &.... &$+$0.59 &$+$2.58 &$+$0.65&$+$5.12 \\      
${\rm log}$ $\nhi$ versus ${[C/H]}$ &   .... & .... &....& ....&$-$0.47 &$-$2.03 &$-$0.50&$-$3.91 \\   
\hline
\end{tabular}
\label{tab_cor2}
\end{center}
\end{table*}

In the bottom row of Fig.~\ref{fig_zeve}, we plot the [{\it C/H}] as a function of $z$.
\citetalias{Kim2016} have noticed a possible $z$ evolution of [{\it C/H}] (see their Fig.~22).
When we consider individual
components (true even for systems as a whole), the derived [{\it C/H}] of $z<2.5$ components shows a large spread
compared to those at $z\ge2.5$. In particular, we find components with [{\it C/H}] $\ge -1$ are rare at $z\ge2.5$.
This trend is very much similar to the trend we notice for {\it L} above.
In the case of individual
components this lack of high metallicity components at high$-z$ causes an anti-correlation with
$\rho_s = -0.29$ having a  two-sided significance of its deviation from zero of 0.008 (or the
anti-correlation is at 2.6$\sigma$ level).
While the trend is apparent for the sample \textcolor{blue}{\sao}, we do not find any
significant anti-correlation (i.e $\rho_s = -0.11$ with a significance level of 0.8$\sigma$).
As suggested by \citetalias{Kim2016} the lack of high metallicity
points at the high$-z$ can not be attributed to the lack of sensitivity.
A linear regression fit to the left-hand panel gives [{\it C/H}] = ($-0.84\pm0.06$) $z$ + (1.01$\pm$0.16).
This is a steeper evolution compared to the redshift evolution measured in DLAs by \citet{Rafelski2012} ($<Z>$ = $(-0.22\pm0.03)$ $z - (0.65\pm0.09)$, 
see their {\it Fig.~11}) upto $z$ $\approx$ 5.

{\it In summary, we do find a strong evolution (i.e $>3\sigma$ level) in \nh, $\Delta$ and {\it L} as a function of $z$.
  [{\it C/H}] shows a moderate (i.e 2.5$\sigma$ level) increase with decreasing $z$.  It is quite possible that the the $z$ evolution
  of $L$ and [{\it C/H}] are driven by the appearance of compact clouds having high [{\it C/H}]
  only at $z<2.5$.} We explore this aspect
further in the following sections.
\noindent \section{Correlations between derived parameters}
\KIM16\ have found interesting correlations between different parameters derived for individual components.
As our predicted parameters are different from earlier results (in particular \nh\ and $L$, mainly
because we suspect \KIM16\ missed factor of 4$\pi$ in the UVB intensity, see Appendix~\ref{Ap6}.),
we investigate the correlations between different derived physical parameters in this section keeping
in mind the redshift evolution discussed above. In Table~\ref{tab_cor1}, we summarize the Spearman rank
correlation coefficient ($\rho_s$) and significance of $\rho_s$ in terms of $\sigma$ (i.e $\rho_s/\sigma$) for
different combinations of parameters and redshift ranges. This table also provides the correlation statistics for
low and high$-z$ sub-samples.

\begin{figure*}
  \centering
  \includegraphics[totalheight=0.22\textheight, trim=0cm 0cm 0cm 0cm, clip=true, angle=0]{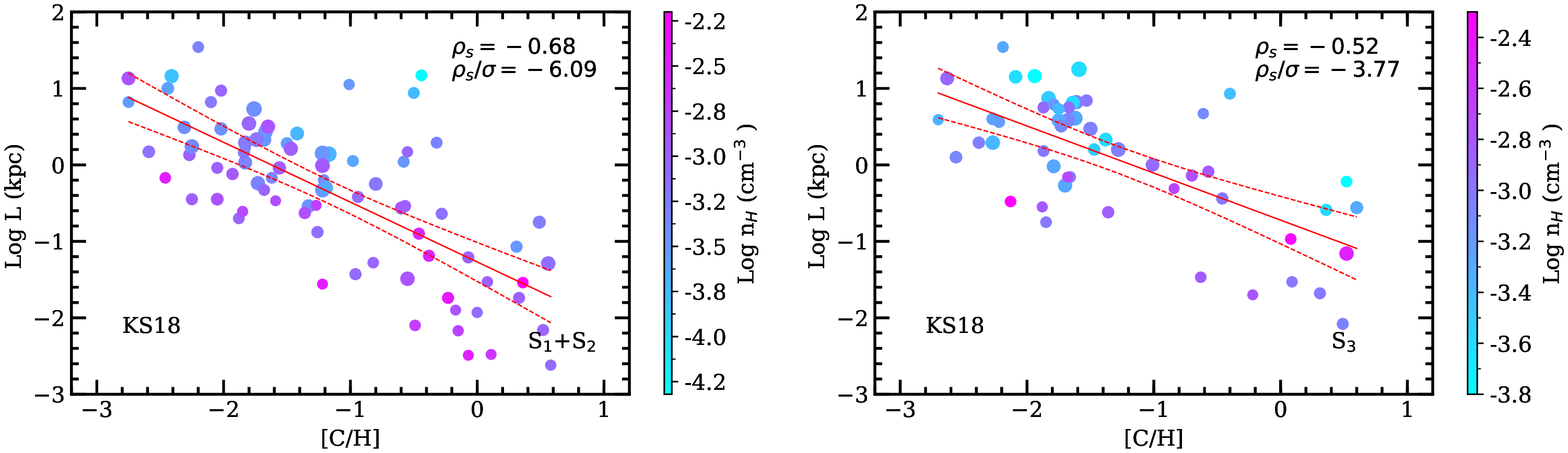}
  \includegraphics[totalheight=0.22\textheight, trim=0cm 0cm 0cm 0cm, clip=true, angle=0]{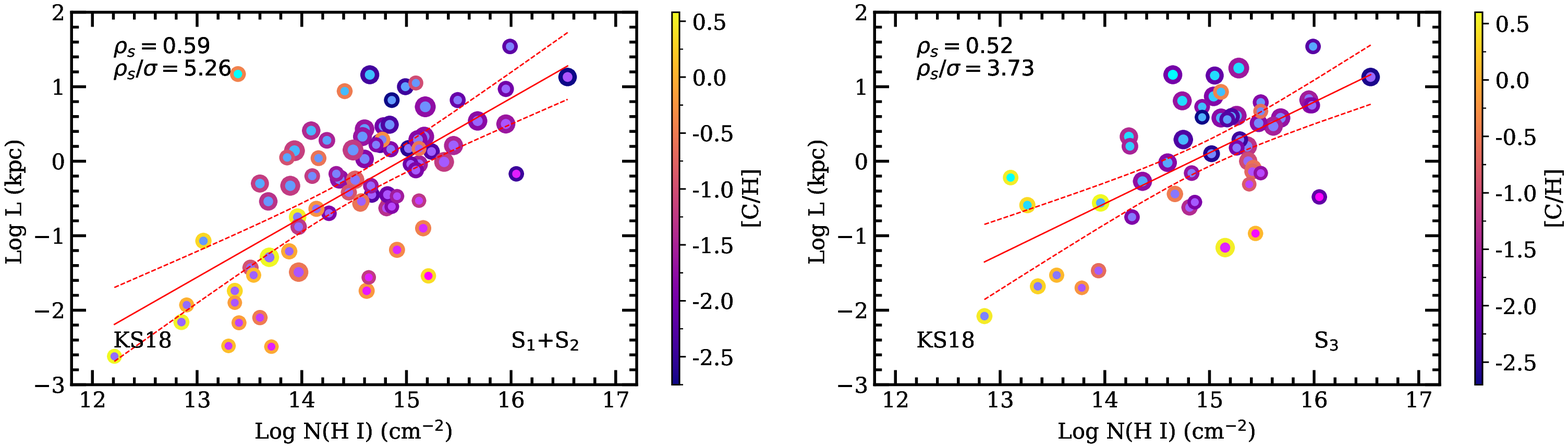}
  \includegraphics[totalheight=0.233\textheight, trim=0cm 0cm 0cm 0cm, clip=true, angle=0]{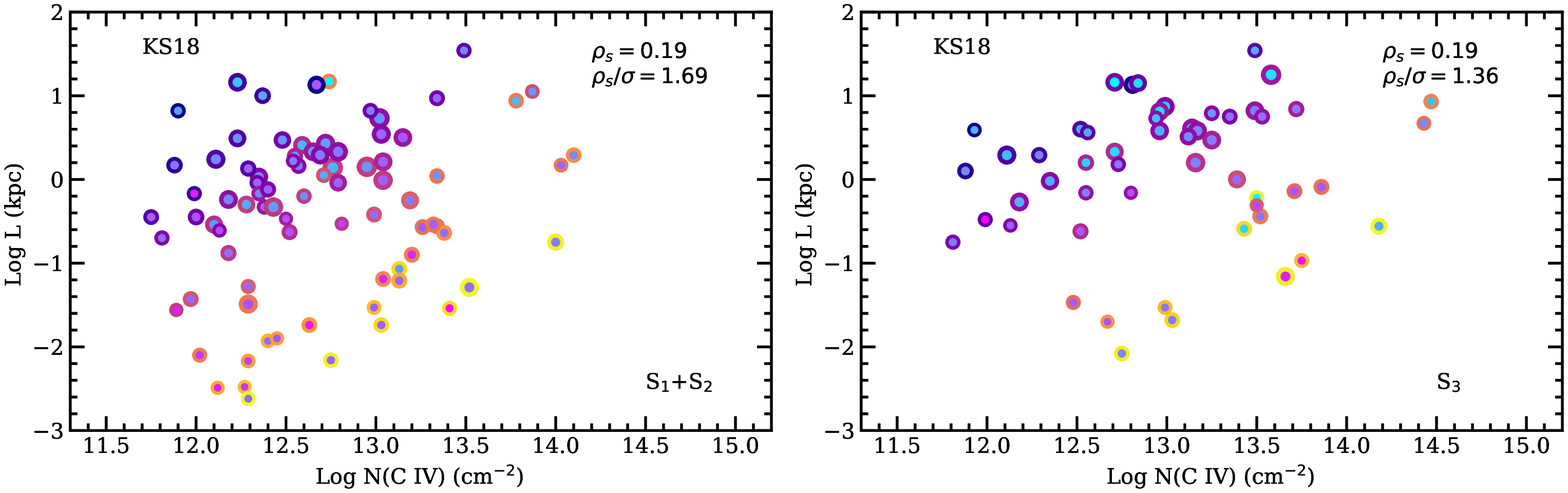}
  \includegraphics[totalheight=0.233\textheight, trim=0cm 0cm 0cm 0cm, clip=true, angle=0]{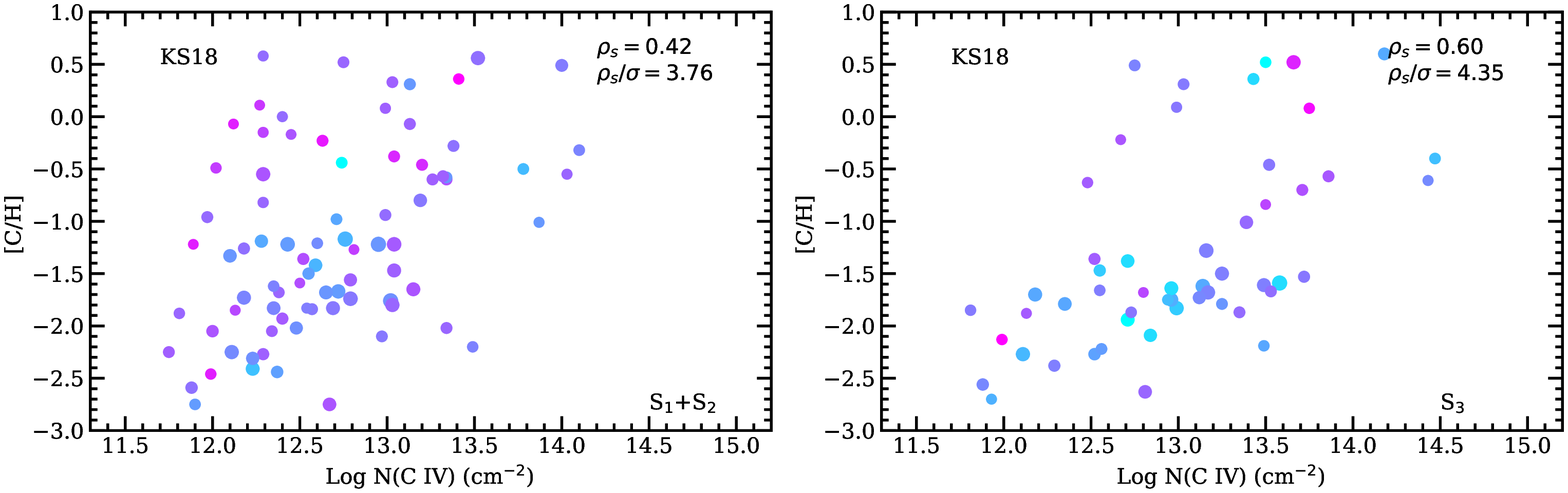}
 \caption{ Correlation study of different parameters predicted by our fiducial photoionization model \Mt.
 In the left columns we plot the results for samples \sam+\san\ (solid circles).
 The right columns show
     the results for \textcolor{blue}{\sao}.
 Top panels: the line-of-sight length {\it L} as a function of [{\it C/H}] , 
 second panels: {\it L} as a function of \nhi, third panels: {\it L} as a function of \nc4\ and bottom
 panels: [{\it C/H}] as a function of \nc4.
In the top panels the vertical color-bar represents \nh\ in
logarithmic unit. The solid circles sizes increase with increasing redshift.
In the second panel the
the vertical color-bar represents [{\it C/H}] values.
We show same color codings (for \nh\ and [{\it C/H}]) and corresponding symbol sizes (for $z$)
in the respective panels.
In each panel
we also show the rank correlation coefficient with significance level. The dashed solid line indicates
linear regression fit to the data in upper two left-hand panels where
we have found correlations significant at more than
5$\sigma$ level. The red dashed lines show the 1$\sigma$ range allowed by the regression analysis.}
\label{fig_corr}
\end{figure*}

In the top panel of Fig.~\ref{fig_corr}, we plot the $L$ vs. [{\it C/H}] for components in the combined sample
of \sam\ and \textcolor{blue}{\san}
in the left-hand panel and for \sao\ in the right-hand panel, respectively.
We indicate our derived $z$ scaling 
with symbol sizes (small size being smaller $z$ and vice-versa) and \nh\ in vertical color-bar as
shown in Fig.~\ref{fig_corr}.
A clear anti-correlation between [{\it C/H}] and {\it L} (as found by \citetalias{Kim2016}) is evident in this figure.
Spearman rank correlation analysis confirms an anti-correlation ($\rho_s = -0.68$) between these two quantities
at 6.09$\sigma$ level. Despite our individual {\it L} values being
smaller than that of \KIM16\ the anti-correlation found by them still
remains valid.
The anti-correlation still exists (albeit with slightly reduced significance level)
even when we divide the sample in to two redshift bins.
We find a strong anti-correlation with
$\rho_s = {-0.52}$ with a two-sided significance of its deviation from zero of  $4.0\times10^{-3}$
or the anti-correlation at 3.77$\sigma$ level is preset even for \sao. Note recently \citet[]{Muzahid2018}
showed that line-of-sight thickness of high metallicity \SiII\ and \CII\ absorbers at low $z$ tend to be smaller compared
to high $z$ ones. They use the integrated column density for those cases as in our sample \sao.
{\it The discussion presented here clearly confirms the existence of a correlation between metallicity and {\it L} among the \CIII\ components.
We discuss the possible origins for this correlation in the following section.}

In the second row from the top in Fig.~\ref{fig_corr},  we plot {\it L} versus $N$(\HI).
We use thick outer rim circles with
color coding to represent
individual [{\it C/H}] values with our above conventional symbol sizes for $z$ and same
color schemes for \nh\ as stated above.
We find a very strong correlation between these two quantities with $\rho_s = 0.59$ with 5.26$\sigma$ significant
level for the combined
sample \sam+\san. A strong correlation is also seen (albeit with reduced significance) when the sample
is divided in to two redshift bins
(see Table ~\ref{tab_cor1}) or based on [{\it C/H}] (see Table~\ref{tab_cor2}). We also see the same trend for sample \sao\ with
$\rho_s = 0.52$ at 3.73$\sigma$ significant level.
Given the two correlations discussed
till now we expect a strong anti-correlation between [{\it C/H}] and $N$(\HI). This is what we find for [{\it C/H}] and $N$(\HI)
with  $\rho_s = -0.65$
significant at 5.80$\sigma$ level for sample \sam+\san\ (see Table~\ref{tab_cor1}). However, the same is
slightly lower with $\rho_s = -0.35$
significant at 2.50$\sigma$ level for sample \sao\ due to the integrated column density.

In the third row from the top in Fig.~\ref{fig_corr}, we plot the observed \CIV\ column density vs.
the measured line-of-sight thickness ($L$)
(this is similar to Fig.~18 of \citetalias{Kim2016}).
For the full sample (\sam+\san), we do not find any statistically significant
correlation or anti-correlation between $N$(\CIV) and {\it L} (see Table~\ref{tab_cor1}).
We do not find any correlation (i.e more than 3$\sigma$ level) when we consider sub-samples
in two different redshift bins. \KIM16\ based on their {\it L} versus \nc4\ plot suggested the existence of two population of \CIII\
absorbers. In the case of large $L$ components, $L$ is found to be a weak function of \nc4\ whereas
for {\it L} $\leq$ 20 kpc, $L$ increase rapidly with increasing
\nc4. However, we find a $\sim 3\sigma$ correlation  when we
divide the sample based on metallicity (see Table~\ref{tab_cor2}). Note that the sub-sample based on metallicity is identical
to sub-sample based on {\it L} as there is a high anti-correlation between the two. Our study do not
firmly support the trend found by \KIM16 as contamination of
high $[C/H]$  components to larger $L$ is evident from the figure. 
We also notice the same for sample \sao. 
Note that \KIM16\ have found apparent lack of points in the {\it L} versus \nc4\ plane with
intermediate values of {\it L} when only \sam\ is considered. They interpreted this as the existence of
two different \CIII\  populations. However, in the combined sample the
distribution becomes more uniform. While the {\it L} measurements of \san\ are limits (in the absence 
of perfectly aligned \nhi\ measurements), the above formed demarcation could come from
L versus [{\it C/H}] seen in the full sample. It will be important to increase the numbers
of measurements to probe existence of this bimodal distribution at high statistical significance.
In the following section,
we address the lack or weak correlation between {\it L} and $N$(\CIV) of the full sample when there is more than 5$\sigma$
correlation between $N$(\HI) and {\it L} using simple photoionization considerations.

In the bottom panels of Fig.~\ref{fig_corr}, we plot $N$(\CIV) versus [{\it C/H}]. For any given $N$(\CIV), we notice a
large scatter in the measured [{\it C/H}]. However, there is a clear lack of  low metallicity and high $N$(\CIV) components.
This leads to an apparent correlation between the two quantities. We find a correlation with $\rho_s = 0.42$
significant at $\sim 3.76$ $\sigma$ level. It is evident from Table~\ref{tab_cor1} that the correlation is
slightly stronger for the high$-z$ sub-sample compared to the low$-z$ sub-sample.  It is also interesting to
note from Table~\ref{tab_cor2} that the $N$(\CIV) and [{\it C/H}] show a significant correlation (i.e 5.12$\sigma$ level)
when we consider only components having $L$ greater than median {\it L} of our sample.
The \nc4\ $-$ $[C/H]$ 
plane clearly does not show any distinct
population as mentioned above. Furthermore, we see mild evolution in \nc4\ as a function of $[C/H]$  for the
two metallicity ranges considered in the sub-samples
contrary to the strong correlation reported earlier. 

In summary, we find three combinations {\it L} versus [{\it C/H}], {\it L} versus $N$(\HI) and
{\it $[C/H]$} versus $N$(\HI) of parameters which show correlation or anti-correlation at
more than 5$\sigma$ level. In the
following section, using simple toy models, we try to understand these correlations and hence
the origin of {\CIII} absorbers.
\noindent \section{Simple phenomenological models}
\begin{figure*}
\centering
\includegraphics[totalheight=0.31\textheight, trim=.1cm .1cm .1cm .1cm, clip=true, angle=0]{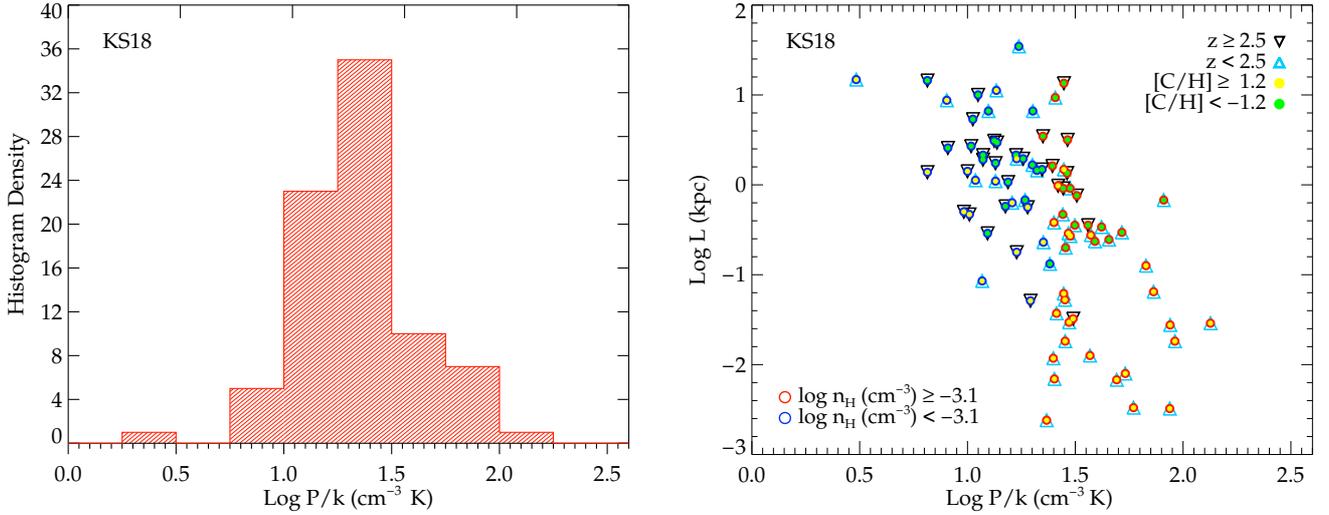}
\caption{{\it Left-hand panel}: Histogram of pressure (i.e P/k) in individual components in \sam\ and \textcolor{blue}{\san}.
  {\it Right-hand panel}:  ($P/k$) versus $L$. It is clear that there is a tendency for the high metallicity components
that also have slightly larger density to have higher pressure. 
Here we use
different symbols and colors to show the distributions of the combined sample (\sam\ + \san) on basis of the median values.
The red and blue circles represent high 
{\nh} (\nh\ $\geq$ $-3.1$), low {\nh} (\nh\ $<$ $-3.1$) components whereas the black and cyan open triangles show the high$-z$ (z $\geq$ 2.5)
and low$-z$ (z $<$ 2.5) absorbers,
respectively. Solid yellow and green circles
represent high and low [{\it C/H}] components with respect to the median value $-1.2$, respectively.}
\label{fig_p}
\end{figure*}

\begin{figure*}
\centering
\includegraphics[totalheight=0.24\textheight, trim=0cm 0cm 0cm 0cm, clip=true, angle=0]{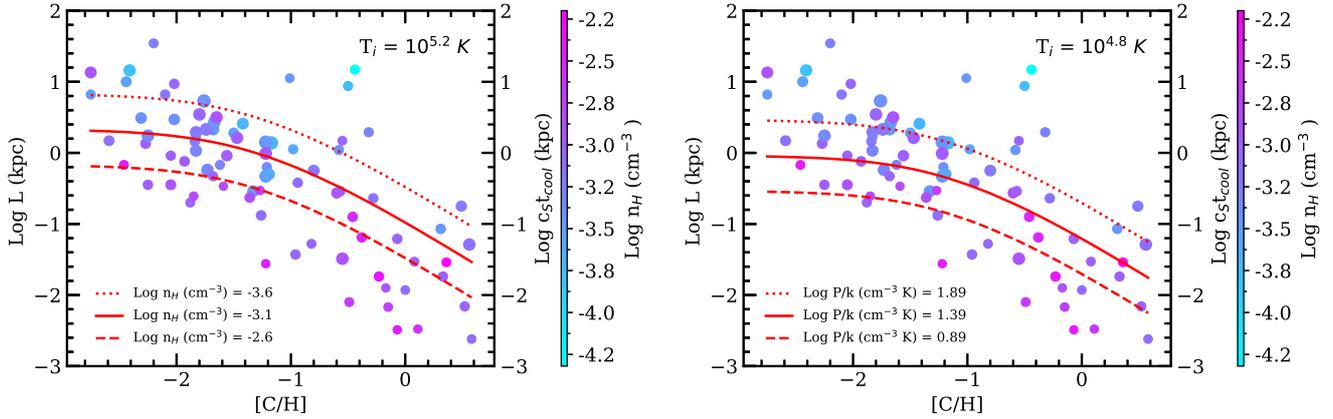}
\caption{{\it Left-hand panel}: Comparison of observed length (L) vs.  metallicity ([{\it C/H}]) 
relationship (color embedded circles for sample \sam\ and \textcolor{blue}{\san}) with the simple isochoric cooling case. 
The vertical color-bar represents \nh\ in
logarithmic unit. The solid circles sizes increase with increasing redshift.
The continuous line shows the cooling length scale ($c_st_{cool})$ with our median density $10^{-3.1}$ \cmcb, where $c_s$
is the sound speed at final temperature
($T_f$) = $10^{4.53}$ K  (we use $\gamma$ = 1.3 for this calculation) and $t_{cool}$ 
is the cooling time scale starting from an initial temperature of $10^{5.2} K$.
The dotted and dashed lines are used indicate the results for 1$\sigma$ range around this mean
density (i.e ${\rm log}$~\nh\ = $-3.6$ and $-2.6$).
{\it Right-hand panel}: Same as the left-hand panel but for isobaric case.
The continuous line shows the cooling length scale ($c_st_{cool})$ with our median pressure (${\rm log}$~$P/k$) = 1.39 \cmcb\ K, where $c_s$
is the sound speed at final temperature
($T_f$) = $10^{4.53}$ K  (we use $\gamma$ = 1.3 for this calculation) and $t_{cool}$ 
is the cooling time scale starting from an initial temperature of $10^{4.8} K$.
The dotted and dashed lines are used for two different range of high-pressure and low-pressure
(1$\sigma$ around median) which is, ${\rm log}$~$P/k$ (\cmcb\ K) = 1.89 and 0.89, respectively.}
\label{fig_lcool}
\end{figure*}

One of the main results from the correlation analysis discussed in the
previous section is the existence
of a strong anti-correlation between [{\it C/H}] and line-of-sight thickness, $L$.
All the discussions presented till now clearly suggest that the cloud sizes
are not driven by hydrostatic equilibrium considerations
(either with self-gravity or with the associated dark matter particles). Also the inferred over-densities of individual
clouds are much higher than what is expected for gas in the IGM but less than
what one expects in the interstellar medium of typical galaxies. Therefore,
it is reasonable to assume that the {\CIII} components are originating from the
halos (or CGM) of galaxies.

First, we show the distribution of thermal pressure, $P/k$ in $cm^{-3} K$
(where $k$ is the Boltzmann's constant) obtained from \CLOUDY\ for
individual components in left-hand panel of Fig.~\ref{fig_p}. We find the median ${\rm log}$~$P/k$ = $1.39 \pm 0.5$ (1$\sigma$ range). In the models
we consider below, we try to achieve  the final gas pressure to this median value. In the right
panel of Fig.~\ref{fig_p}, we plot ${\rm log}$~$P/k$ versus $L$. We also use different symbols to identify
high and low density (also metallicity) components depending on our median values. It is clear from this figure that high metallicity
components that are also smaller in size tend to have larger gas pressure
compared to their low metallicity counterparts. 

Metal enriched compact low temperature clouds in galactic halos can originate from winds that are ubiquitous in high$-z$
star forming galaxies \citep{Veilleux2005}. These winds could carry cold gas clouds from the
multi-phased ISM (i.e
the entrainment scenario), or clouds form in-situ in the CGM through thermal instabilities
\citep{Field1965,Meerson1989,Burkert2000} or condensation of cold gas clumps in-flows from hot halos
\citep{Hennebelle1999,Sharma2012}.  Recently, \citet[]{Mccourt2018}
have argued that the fragmentation is an easier way to reach equilibrium and they considered
this to be analogous to the Jeans instability in the gravitational collapse. They basically argued
that ${L} \propto c_{s} t_{cool}$, where $c_s$ and $t_{cool}$ are the local
sound speed and gas cooling time, respectively. They evaluated {\it L} at the temperature {\it T} where
the product ${c_{s}t_{cool}}$ is minimized.
To get the metallicity dependent on $L$,  we need to evaluate this at a temperature {\it T}  (or over the temperature range)
where cooling rate depends strongly on
the metallicity. In addition, in our case the gas of interest is also being heated and ionized continuously by the
meta-galactic UVB radiation. Thus the final gas temperature we observe will be the photoionization
equilibrium temperature of the gas.
Without going into the detailed modelling and just to capture
the basic picture, we consider a simple case where we calculated the ``isochoric'' and ``isobaric''
cooling time for a gas having initial temperature ($T_i$), [{\it C/H}] and \nh\ to reach the final
temperature and pressure close to the median temperature and pressure that we infer for the \CIII\ components. We use appropriate
cooling curves for different metallicities as given in \citet[]{Schure2009}.

In the ``isochoric'' case, we obtain the relationship between  ${c_{s}t_{cool}}$ and [{\it C/H}] for
different initial temperatures keeping the gas density to be constant at median density
(i.e ${\rm log}$~\nh\ = $-$3.1). In the left hand panels of
Fig.~\ref{fig_lcool}, we show the observed {\it L} versus [{\it C/H}] data overlayed with the isochoric model predictions
of ${c_{s}t_{cool}}$ for
three different \nh\ (i.e for the median and 1$\sigma$ range around it) obtained for Log~T$_i$ = 5.2.
These curves clearly predict an anti-correlation between {\it L} and [{\it C/H}] as observed in our data. Note that
the observed slope may be slightly steeper than the predictions of our constant density model but this
could be accommodated if we allow the metallicity dependent density as hinted by the data in Fig.~\ref{fig_p}.
For the assumed
temperature range and density the cooling time-scale is found to be 2.5$\times 10^6$ years and
5.0$\times 10^7$ years, respectively for [{\it C/H}] = 0 and $-2$. These time-scales are much smaller than
the typical free fall time-scale in the halos.
Just to check whether the above mentioned anti-correlation is generic prediction of cooling based
arguments we also consider the ``isobaric'' case (right-hand panel in Fig.~\ref{fig_lcool})

In the ``isobaric'' case,  for any assumed T$_i$ we fix the initial density of the gas in such a way
to maintain the pressure equal to the median observed $P/k$ $= 1.39 \pm 0.5$ ($1\sigma$ range). As the gas cools,
we readjust the density to keep the pressure constant. If we start with the same initial temperature
as we have for the ``isochoric'' case then the cooling time-scales are much longer as the gas
starts with much lower density (as we try to keep the pressure constant) that leads to a longer
cooling time-scale and larger cloud sizes. However, what is important to note here is that for a given choice of
initial temperature (and final pressure) anti-correlation between {\it L} and [{\it C/H}] is clearly evident.
As an illustration, we show models for three final $P/k$ and T$_i$ = $8\times10^4$ K. Here again
allowing for the final pressure to be higher for the high metallicity gas compare to the low metallicity
gas will make the curve more steep.

In summary, the toy models considered here provide the basic anti-correlation found between {\it L} and [{\it C/H}].
This lend supports to the idea that ${c_{s}t_{cool}}$ may be the main parameter deciding the
physical extent of the \CIII\ absorbers. We once again reiterate the fact that the calculations considered
here are not rigorous enough to draw more quantitative conclusions. The main inference to carry forward
is that for a given observed \nh\ the $L$  (or $N(H)$) depends on
metallicity,  perturbed density and temperature (i.e initial density and temperature) of the instability. Thus we expect
a lack of strong correlation between \nh\ and $N(H)$ unlike in the
case of hydrostatic equilibrium considerations. We shall keep this in mind while considering other correlations.
To proceed further,
we fit the observed correlation with a linear regression fit to obtain the following relationship,
\begin{equation}
  log~{L} = (-1.26\pm0.12) + (-0.78\pm0.08) {\rm {[C/H]}}.
  \label{eqn2}
\end{equation}
The linear regression fit with a 1$\sigma$ range is shown in Fig.~\ref{fig_corr}.
We will use this to understand correlations related to \CIV. 
In comparison, \KIM16\ have obtained line-of-sight thickness,
${\rm log}$~$L_{temp}$ = $-0.24$ $-$ 1.21 $\times$ ${[C/H]}$
with the $L_{temp}$ being {\it at least} one order of magnitude higher than our
values for any given value of [{\it C/H}].
We provide linear regression fit to three significant correlation in Table~\ref{tablek21},
\ref{tablek22} and \ref{tablek23}
which we found in the previous section for combined sample \sam+\san\ and individual sample \san.
\begin{figure}
\centering
\includegraphics[totalheight=0.287\textheight, trim=0cm 0.0cm 0cm .0cm, clip=true, angle=0]{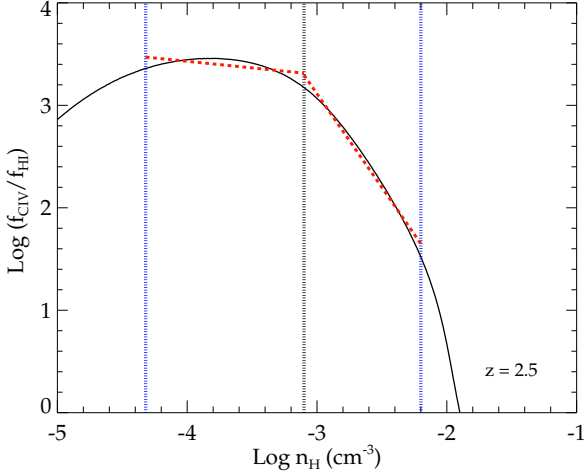}
\caption{The ratio of fraction of \CIV\ to fraction of \HI\ as function of \nh\ at our median redshift value z $=$ 2.5. 
The vertical black dotted line is used to show our median density value (${\rm log}$~\nh\ (\cmcb) $=$ $-3.1$) and 
the blue lines show the boundary of ${\rm log}$~\nh\ ($-4.32$ and $-2.2$, resp) for sample S1+S2. The red 
dotted lines show linear fits to the plot for two \nh\ ranges, ${\rm log}$~\nh\ (\cmcb) $\in$ [$-3.1$, $-2.2$] and [$-4.32$, $-3.1$] with slope $-1.82$ and $-0.12$, respectively.} 
\label{fig_frac}
\end{figure}

Next, we try to understand the implications of the strong correlation found between {\it L} and $N$(\HI).
For this we consider a constant density cloud in photoionization equilibrium.
To start with we get an expression for {\it L} in terms of $N$(\HI),
\begin{eqnarray}
  {L} & = &  {N_{\tiny H}\over n_{\tiny H}} ={N(\HI)\over f_{\tiny \HI} n_{\tiny H}} = {N(\HI)~\Gamma_{\tiny \HI} \over \alpha_0~n_{\tiny H}^2~T^{-0.78}} \nonumber \\
  & \propto & N(\HI)~n_{\tiny H}^{-2}~T^{0.78} \nonumber\\
  & \propto & N(\HI)~n_{\tiny H}^{0.78(\gamma - 1) -2}.
  \label{eqn3}
\end{eqnarray}
Here, $\Gamma_{\tiny \HI}$ is the H~{\sc i} photoionization rate, $f_{\tiny \HI}$ the neutral hydrogen
fraction and $\alpha_{\tiny 0}$ is the recombination coefficient
which depends on the kinetic temperature as $T^{-0.78}$. For the last step we have assumed that the gas follows
an equation of state (i.e $T\propto n^{\gamma - 1}$). In the case of isothermal equation of state (i.e $\gamma = 1$)
we expect L~$\propto N$(\HI)\nh$^{-2}$ and for $\gamma = 5/3$ we expect L~$\propto N$(\HI)\nh$^{-1.4}$. Therefore,
any deviation from a linear relationship between {\it L} and $N$(\HI) will be driven by the relationship between $N$(H~{\sc i}) and
\nh. Based on a linear regression fit we find L$\propto$ $N$(\HI)$^{0.8\pm0.1}$ (see Fig.~\ref{fig_corr}) which suggests that at best there is a very weak
correlation between \nh\ and $N$(\HI) (i.e \nh$\propto$ $N$(\HI)$^{0.10}$ or  $N$(\HI)$^{0.14}$
for the two equations of state discussed above). Note this lack of correlation is expected
in the framework of models considered above.
Our direct correlation analysis also confirms the lack of
correlation found between \nh\ and $N$(H~{\sc i}) in our data (with a correlation coefficient of 0.2
having a significant level of 1.9$\sigma$).

The third strongest correlation we see is between $N$(\HI) and [{\it C/H}]. Using Eq.~\ref{eqn2} and \ref{eqn3}
we can derive,
\begin{equation}
   Log~ N(\HI) \propto (-0.98\pm 0.18){[C/H]}.
   \label{eqn4}
  \end{equation}
Thus the observed strong anti-correlation between $N$(\HI) and [{\it C/H}] can also arise from simple
considerations. Indeed the linear regression fit between $N$(\HI) and [{\it C/H}] measurements is given
by, $ Log~N(\HI)= (13.77\pm0.11)-(0.62\pm0.07) {[C/H]}$ which is acceptable with the above expectations.

Lastly we try to understand the lack of correlation between $N$(\CIV) and L. From simple considerations of Eq.~\ref{eqn2} and \ref{eqn4}, we can write the column
density of \CIV\ in terms of $N$(\HI) as,
\begin{eqnarray}
   N(\CIV) & = & 10^{{[C/H]}}  N(\HI) (f_{\tiny \CIV}/f_{\tiny \HI}) \nonumber \\
   & \simeq & L^{-0.025 \pm 0.22} (f_{\tiny \CIV}/f_{\tiny \HI}).
   \label{eqn5}
  \end{eqnarray}
Here, $f_{\tiny \CIV}$ is the ion fraction of \CIV. 
Thus
if the ratio of ion fraction of \CIV\ to \HI\ remains constant we expect only
a weak correlation between $N$(\CIV) and L. 
Presence of a strong correlation/anti-correlations between $L$ and $f_{\tiny \CIV}/f_{\tiny \HI}$ alone
can set any possible correlation between \nc4 and $L$.
In Fig.~\ref{fig_frac}, we show the ion fraction ratio as a
function of density obtained from our fiducial
model at the mean redshift of the data used here. At low density 
(${\rm log}$~\nh\ $<$ $-3.1$) it is clearly evident that $L$ has 
no/weak dependent on \nc4\ as  $f_{\tiny \CIV}/f_{\tiny \HI}$ is nearly constant for the low \nh\ range. 
As expected we find $N(\CIV) \propto L^{-0.01 \pm 0.15}$ from the model predictions in these
regions for the optically thin \CIII\ components. However, at high \nh\ range (${\rm log}$~\nh\ $\ge$ $-3.1$) the ratio of ion fraction decreases
monotonically
with density i.e. $f_{\tiny \CIV}/f_{\tiny \HI} \propto$ \nh$^{-1.83\pm0.02 }$ as shown in the figure whereas we find
\nh\ $\propto L^{-0.07\pm0.03}$ with $\rho_s$ $=$ $-0.21$ at
$-1.4\sigma$ confidence level for the 
\sam+\san\ components. This results in a mild correlation with $N(\CIV) \propto L^{0.12 \pm 0.04}$ for these high \nh\ components. 
Despite the lack of a correlation for overall components,
we find 3$\sigma$ level 
correlation between these two with slope 0.96 and 0.73
for $log L= A + B$ ${\rm log}$~$N(\CIV) $ for high and low metallicity branch, respectively (see Table~\ref{tablek21}). 
Within errors the slope is consistent with the slope of {\it L} versus \nhi.
From Eq.~\ref{eqn5}, it is clear that when we restrict our samples
to smaller metallicity ranges, {\it L} dependence on [C/H]
is weak as compared to the higher metallicity sample.

\noindent \section {Summary and discussions}
We have presented detailed photoionization models for optically thin
\CIII\ absorption components in the redshift range $2.1 < z < 3.4$
along 19 quasar sightlines
analyzed by \citetalias{Kim2016}. The main motivations for this re-analysis
is to study the dependence of the assumed UVB on the derived parameters
and thereby quantify systematic uncertainties in these parameters and
understand various correlations between them.

We mainly focused on 53 absorption systems
where \CIII\ and \CIV\ column densities are measured using
Voigt profile fitting. Excluding the shifted or blended \CIII\ components from these
53 systems, we consider two sub-samples
\sam\ and \san\ where
Voigt profile fitting is performed using tied parameters for \HI\  
out of total 132 \CIII\ components. 
The sub-sample
\sam\ consists of 32 intervening {\CIII} components
that also have a co-aligned
{\HI} component whereas \san\ consists of 50 intervening
{\CIII} components with moderately-aligned \HI\ component or 
having upper limits on \CIII.
We also construct 
\textit{{\textcolor{blue}{\sao}}} which includes all
the 53 {\CIII} systems where we
consider the total column density of ions
(i.e sum of the column density in individual Voigt profile components)
to originate from single cloud in our models in order to account for
the possible complex velocity and density field of the gas.
We have analyzed these absorbers using
photoionization models with \CLOUDY\ for two UVBs, the \citetalias{Haardt2012} and \citetalias{Khaire2018}).
In case of \KS18, we use different
UVBs generated by varying spectral slope ($\alpha$) of extreme-UV quasar SEDs ($\alpha$ = $-1.6$, $-1.8$ and $-2.0$)
where $\alpha$ = $-1.8$ is our fiducial model.

  First, we considered that all the {\CIII} absorbers originating
  from metal enriched optically thin IGM. In this case,  we have used the cloud size to be the stopping
  criteria with its value being
  equal to Jeans length 
as suggested for the {\lya}
forest absorption by \citet{Schaye2001}.
Our fiducial \citetalias{Khaire2018} UVB models
reproduced the observed range in \nce/\nc4\ for
total hydrogen densities, {\nh} \si\ $6.3 \times 10^{-3} - 4.8 \times 10^{-5}$ {\cmcb}.
%
We see a mild evolution
in the median observed column density ratio {\nce}/{\nc4} with increasing
redshift. This is also captured by our models purely from the redshift
evolution of \citetalias{Khaire2018} UVB.
%
%
However, 
\nhi\ values predicted by these models are almost a factor \si\
40 higher than the observed \HI\ column density in sample \sam.
The same is also seen in sample \san\ despite the fact that we are probably considering upper limits for these cases.
This clearly suggests that the sizes of clouds are significantly smaller than what was given by hydrostatic equilibrium
arguments generally used for the IGM clouds.

Next, we considered models where the stopping criteria is the observed $N$(\HI). In this
case, the derived hydrogen density (\nh) for individual components
ranges from $10^{-4.3}$ $-$ $10^{-2.2}$ \cmcb\ with a median value of 
$10^{-3.1}$ \cmcb\ for combined sub-sample \sam+\san\ whereas
for sample \sao\ we have $10^{-3.9}$ $-$ $10^{-2.3}$ \cmcb\ with a median value of  $10^{-3.1}$ \cmcb.
[{\it C/H}] is found to be in the range
$-2.75$ to 0.58, with a median value of $-1.2$  for \sam+\san\ whereas the same range is
$-2.7$ to 0.6 with a median value of $-1.6$  for \sao.
The gas temperature obtained in our photoionization models do not match with that obtained
from the $b$ values for some components. For these components, we 
considered additional photoionization models where we fix the gas temperature to the one
obtained from the $b$. We find that the difference in temperature makes
negligible difference to the derived parameters with a median difference of $<$ $\Delta$ ${\rm log}$ ${n_{H}}>$ =  $0.1$ 
and $<\Delta$[{\it C/H}]$> = 0.05$ for our fiducial \KS18\ UVB. 

Using a range of UVB generated for the assumed range in the UV spectral energy
distribution of quasars we obtained a systematic uncertainty of $\pm0.17$ dex for the derived
\nh\ and $\pm0.2$ dex for [{\it C/H}]. Note that the UVB calculated by  \citetalias{Khaire2018}
is normalized by matching the \HI\ photoionization rate ($\Gamma_{\HI}$) measurements from
the Lyman-$\alpha$ forest observations \citep{Bolton2007, Becker2013}. Considering the uncertainties in the measurements
of $\Gamma_{\HI}$ will further increase the systematic uncertainties in the \nh\ measurements.
However this normalization uncertainties will have much less effect on the derived metallicities.

The \nh\ values measured in our study are typically one order of magnitude higher than
what has been derived by \KIM16 (see Appendix~\ref{Ap6} for possible explanation).
Because of this, \KIM16\ overestimated  the line of sight thickness of the \CIII\ absorbers.
We find the line-of-sight thickness of the clouds to be in the range of 2 pc to 35 kpc
with a median value of 0.63 kpc for the individual \CIII\ components in \sam\ and \san\ 
whereas the median value is slightly higher (1.6 kpc) for \sao\
as we consider total integrated column densities along the line-of-sight. For the measured \nhi\
the \CIII\ absorbers originate from gas having larger
$\Delta$ compared to what is expected in a typical IGM gas.
However, the inferred \nh\ is less than the observed
ISM of galaxies. Therefore, the optically thin
\CIII\ components studied here are most probably associated
with gas outside the galactic discs (outflows, inflows or
galactic halo gas). 
The derived sizes of the clouds are
consistent with the \CIII\ components originating from the CGM of high$-$z galaxies. However,
to confirm the association of \CIII\ clouds with the CGM, it is important to identify galaxies
at close impact parameters. \citet{Bibley2013} have presented Lyman break galaxies around two of these quasar sightlines (HE0940$-$1050 and PKS 2126$-$158)
of our sample. But there is no clear association found for the \CIII\ absorbers.
It will be important to have deep imaging and spectroscopic
observations in these quasar fields to identify galaxies associated with the \CIII\ absorbers
studied here.

We find that there are three combinations ({\it L} versus [{\it C/H}], {\it L} versus \nhi\
and [{\it C/H}] versus \nhi) of parameters that show correlation
or anti-correlation at more than 5$\sigma$ level. Using the linear regression analysis
we obtained relationship between L and [C/H] as well as between and L and \nhi. Based on simple photoionization
considerations this will mean a very weak correlation between \nhi\ and \nh~[i.e \nh $\propto$ \nhi$^{0.1}$].
We also show the expected relationship between \nhi\ and [{\it C/H}] based on the above two relationships which
is consistent with what is observed. While strong correlation is seen between L and \nhi, no such
correlation is seen between L and \CIV. This can also be easily understood in simple photoionization
models.

Using simple toy models we suggest   the basic idea that $c_st_{cool}$ may be
the main parameter deciding the physical size of the
\CIII\ absorbers. In particular the observed correlation between $L$ and [{\it C/H}] can be
obtained with a narrow range in \nh\ while considering ``isobaric'' or ``isochoric'' cooling. 
This supports fragmentation which is an easier way to reach
equilibrium and is considered analogous to
the Jeans instability in the gravitational collapse \citep{Mccourt2018}.
These metal enriched clouds
in galactic halos may originate from winds that are ubiquitous in high$-z$ star forming
galaxies. 
If cold clouds formed in-situ at large distances then their frequency of occurrence may have some links to the star formation
rate in the host galaxies. Therefore, it will be important to consider \CIII\ absorbers over large redshift ranges and associate
their evolution with the global star formation rate density found from high$-z$ galaxies. This we wish to pursue in the near
future.

Still there remain few uncertainties before drawing strong conclusions which are as follows: (i) 
velocity coincidence based on which we consider single cloud photoionization models to produce \rat\ ratio need not
corresponds to spatial coincidence, (ii) given the velocity resolution, single absorption
can come from a collection of multiple clouds and (iii) whether one can produce the
observed frequency of occurrence of \CIII\ absorbers consistently using the inferred cloud
sizes.  Another issue that needs to be addressed is the
survivability of these clouds. To address these, we need self-consistent CGM models.

\vspace{1cm}

\noindent  
{\it Acknowledgements:}
AM acknowledges the financial support by DST-INSPIRE fellowship program
of Govt. of India. AM and ACP are thankful to IUCAA
for providing free hospitality and travel grant during the visits. 
We thank Sowgat Muzahid for useful comments and suggestions.
The authors also wish to thank the anonymous
referee 
for providing valuable comments and
suggestions for improving
the manuscript.

\def\aj{AJ}%
\def\actaa{Acta Astron.}%
\def\araa{ARA\&A}%
\def\apj{ApJ}%
\def\apjl{ApJ}%
\def\apjs{ApJS}%
\def\ao{Appl.~Opt.}%
\def\apss{Ap\&SS}%
\def\aap{A\&A}%
\def\aapr{A\&A~Rev.}%
\def\aaps{A\&AS}%
\def\azh{AZh}%
\def\baas{BAAS}%
\def\bac{Bull. astr. Inst. Czechosl.}%
\def\caa{Chinese Astron. Astrophys.}%
\def\cjaa{Chinese J. Astron. Astrophys.}%
\def\icarus{Icarus}%
\def\jcap{J. Cosmology Astropart. Phys.}%
\def\jrasc{JRASC}%
\def\mnras{MNRAS}%
\def\memras{MmRAS}%
\def\na{New A}%
\def\nar{New A Rev.}%
\def\pasa{PASA}%
\def\pra{Phys.~Rev.~A}%
\def\prb{Phys.~Rev.~B}%
\def\prc{Phys.~Rev.~C}%
\def\prd{Phys.~Rev.~D}%
\def\pre{Phys.~Rev.~E}%
\def\prl{Phys.~Rev.~Lett.}%
\def\pasp{PASP}%
\def\pasj{PASJ}%
\def\qjras{QJRAS}%
\def\rmxaa{Rev. Mexicana Astron. Astrofis.}%
\def\skytel{S\&T}%
\def\solphys{Sol.~Phys.}%
\def\sovast{Soviet~Ast.}%
\def\ssr{Space~Sci.~Rev.}%
\def\zap{ZAp}%
\def\nat{Nature}%
\def\iaucirc{IAU~Circ.}%
\def\aplett{Astrophys.~Lett.}%
\def\apspr{Astrophys.~Space~Phys.~Res.}%
\def\bain{Bull.~Astron.~Inst.~Netherlands}%
\def\fcp{Fund.~Cosmic~Phys.}%
\def\gca{Geochim.~Cosmochim.~Acta}%
\def\grl{Geophys.~Res.~Lett.}%
\def\jcp{J.~Chem.~Phys.}%
\def\jgr{J.~Geophys.~Res.}%
\def\jqsrt{J.~Quant.~Spec.~Radiat.~Transf.}%
\def\memsai{Mem.~Soc.~Astron.~Italiana}%
\def\nphysa{Nucl.~Phys.~A}%
\def\physrep{Phys.~Rep.}%
\def\physscr{Phys.~Scr}%
\def\planss{Planet.~Space~Sci.}%
\def\procspie{Proc.~SPIE}%
\let\astap=\aap
\let\apjlett=\apjl
\let\apjsupp=\apjs
\let\applopt=\ao
\bibliographystyle{mnras}
\bibliography{paper}

\begin{thebibliography}{}
\makeatletter
\relax
\def\mn@urlcharsother{\let\do\@makeother \do\$\do\&\do\#\do\^\do\_\do\%\do\~}
\def\mn@doi{\begingroup\mn@urlcharsother \@ifnextchar [ {\mn@doi@}
  {\mn@doi@[]}}
\def\mn@doi@[#1]#2{\def\@tempa{#1}\ifx\@tempa\@empty \href
  {http://dx.doi.org/#2} {doi:#2}\else \href {http://dx.doi.org/#2} {#1}\fi
  \endgroup}
\def\mn@eprint#1#2{\mn@eprint@#1:#2::\@nil}
\def\mn@eprint@arXiv#1{\href {http://arxiv.org/abs/#1} {{\tt arXiv:#1}}}
\def\mn@eprint@dblp#1{\href {http://dblp.uni-trier.de/rec/bibtex/#1.xml}
  {dblp:#1}}
\def\mn@eprint@#1:#2:#3:#4\@nil{\def\@tempa {#1}\def\@tempb {#2}\def\@tempc
  {#3}\ifx \@tempc \@empty \let \@tempc \@tempb \let \@tempb \@tempa \fi \ifx
  \@tempb \@empty \def\@tempb {arXiv}\fi \@ifundefined
  {mn@eprint@\@tempb}{\@tempb:\@tempc}{\expandafter \expandafter \csname
  mn@eprint@\@tempb\endcsname \expandafter{\@tempc}}}

\bibitem[\protect\citeauthoryear{{Bechtold}, {Crotts}, {Duncan}  \&
  {Fang}}{{Bechtold} et~al.}{1994}]{Bechtold1994}
{Bechtold} J.,  {Crotts} A.~P.~S.,  {Duncan} R.~C.,   {Fang} Y.,  1994, \mn@doi
  [\apjl] {10.1086/187688}, \href
  {http://adsabs.harvard.edu/abs/1994ApJ...437L..83B} {437, L83}

\bibitem[\protect\citeauthoryear{{Becker} \& {Bolton}}{{Becker} \&
  {Bolton}}{2013}]{Becker2013}
{Becker} G.~D.,  {Bolton} J.~S.,  2013, \mn@doi [\mnras]
  {10.1093/mnras/stt1610}, \href
  {http://adsabs.harvard.edu/abs/2013MNRAS.436.1023B} {436, 1023}

\bibitem[\protect\citeauthoryear{{Bielby} et~al.,}{{Bielby}
  et~al.}{2013}]{Bibley2013}
{Bielby} R.,  et~al., 2013, \mn@doi [\mnras] {10.1093/mnras/sts639}, \href
  {http://adsabs.harvard.edu/abs/2013MNRAS.430..425B} {430, 425}

\bibitem[\protect\citeauthoryear{{Bolton} \& {Haehnelt}}{{Bolton} \&
  {Haehnelt}}{2007}]{Bolton2007}
{Bolton} J.~S.,  {Haehnelt} M.~G.,  2007, \mn@doi [\mnras]
  {10.1111/j.1365-2966.2007.12372.x}, \href
  {http://adsabs.harvard.edu/abs/2007MNRAS.382..325B} {382, 325}

\bibitem[\protect\citeauthoryear{{Burkert} \& {Lin}}{{Burkert} \&
  {Lin}}{2000}]{Burkert2000}
{Burkert} A.,  {Lin} D.~N.~C.,  2000, \mn@doi [\apj] {10.1086/308989}, \href
  {http://adsabs.harvard.edu/abs/2000ApJ...537..270B} {537, 270}

\bibitem[\protect\citeauthoryear{{Carswell} \& {Webb}}{{Carswell} \&
  {Webb}}{2014}]{vpfit}
{Carswell} R.~F.,  {Webb} J.~K.,  2014, {VPFIT: Voigt profile fitting program},
  Astrophysics Source Code Library (\mn@eprint {ascl} {1408.015})

\bibitem[\protect\citeauthoryear{{Dedikov} \& {Shchekinov}}{{Dedikov} \&
  {Shchekinov}}{2004}]{Dedikov2004}
{Dedikov} S.~Y.,  {Shchekinov} Y.~A.,  2004, \mn@doi [Astronomy Reports]
  {10.1134/1.1641117}, \href
  {http://adsabs.harvard.edu/abs/2004ARep...48....9D} {48, 9}

\bibitem[\protect\citeauthoryear{{Dinshaw}, {Weymann}, {Impey}, {Foltz},
  {Morris}  \& {Ake}}{{Dinshaw} et~al.}{1997}]{Dinshaw1997}
{Dinshaw} N.,  {Weymann} R.~J.,  {Impey} C.~D.,  {Foltz} C.~B.,  {Morris}
  S.~L.,   {Ake} T.,  1997, \mn@doi [\apj] {10.1086/304926}, \href
  {http://adsabs.harvard.edu/abs/1997ApJ...491...45D} {491, 45}

\bibitem[\protect\citeauthoryear{{Faucher-Gigu{\`e}re}, {Lidz}, {Zaldarriaga}
  \& {Hernquist}}{{Faucher-Gigu{\`e}re} et~al.}{2009}]{Faucher2009}
{Faucher-Gigu{\`e}re} C.-A.,  {Lidz} A.,  {Zaldarriaga} M.,   {Hernquist} L.,
  2009, \mn@doi [\apj] {10.1088/0004-637X/703/2/1416}, \href
  {http://adsabs.harvard.edu/abs/2009ApJ...703.1416F} {703, 1416}

\bibitem[\protect\citeauthoryear{{Fechner}}{{Fechner}}{2011}]{Fechner2011}
{Fechner} C.,  2011, \mn@doi [\aap] {10.1051/0004-6361/201117080}, \href
  {http://adsabs.harvard.edu/abs/2011A%26A...532A..62F} {532, A62}

\bibitem[\protect\citeauthoryear{{Ferland} et~al.,}{{Ferland}
  et~al.}{2013}]{Ferland2013}
{Ferland} G.~J.,  et~al., 2013, \rmxaa, \href
  {http://adsabs.harvard.edu/abs/2013RMxAA..49..137F} {49, 137}

\bibitem[\protect\citeauthoryear{{Field}}{{Field}}{1965}]{Field1965}
{Field} G.~B.,  1965, \mn@doi [\apj] {10.1086/148317}, \href
  {http://adsabs.harvard.edu/abs/1965ApJ...142..531F} {142, 531}

\bibitem[\protect\citeauthoryear{{Francis}, {Hewett}, {Foltz}, {Chaffee},
  {Weymann}  \& {Morris}}{{Francis} et~al.}{1991}]{Francis1991}
{Francis} P.~J.,  {Hewett} P.~C.,  {Foltz} C.~B.,  {Chaffee} F.~H.,  {Weymann}
  R.~J.,   {Morris} S.~L.,  1991, \mn@doi [\apj] {10.1086/170066}, \href
  {http://adsabs.harvard.edu/abs/1991ApJ...373..465F} {373, 465}

\bibitem[\protect\citeauthoryear{{Gaikwad}, {Choudhury}, {Srianand}  \&
  {Khaire}}{{Gaikwad} et~al.}{2017a}]{Gaikwad2017}
{Gaikwad} P.,  {Choudhury} T.~R.,  {Srianand} R.,   {Khaire} V.,  2017a,
  preprint, \href {http://adsabs.harvard.edu/abs/2017arXiv170505374G} {}
  (\mn@eprint {arXiv} {1705.05374})

\bibitem[\protect\citeauthoryear{{Gaikwad}, {Srianand}, {Choudhury}  \&
  {Khaire}}{{Gaikwad} et~al.}{2017b}]{UVGaikwad2017}
{Gaikwad} P.,  {Srianand} R.,  {Choudhury} T.~R.,   {Khaire} V.,  2017b,
  \mn@doi [\mnras] {10.1093/mnras/stx248}, \href
  {http://adsabs.harvard.edu/abs/2017MNRAS.467.3172G} {467, 3172}

\bibitem[\protect\citeauthoryear{{Gaikwad}, {Srianand}, {Khaire}  \&
  {Choudhury}}{{Gaikwad} et~al.}{2018a}]{Gaikwad2018b}
{Gaikwad} P.,  {Srianand} R.,  {Khaire} V.,   {Choudhury} T.~R.,  2018a, arXiv
  e-prints, \href {http://adsabs.harvard.edu/abs/2018arXiv181201016G} {}

\bibitem[\protect\citeauthoryear{{Gaikwad}, {Choudhury}, {Srianand}  \&
  {Khaire}}{{Gaikwad} et~al.}{2018b}]{Gaikwad2018a}
{Gaikwad} P.,  {Choudhury} T.~R.,  {Srianand} R.,   {Khaire} V.,  2018b,
  \mn@doi [\mnras] {10.1093/mnras/stx2859}, \href
  {http://adsabs.harvard.edu/abs/2018MNRAS.474.2233G} {474, 2233}

\bibitem[\protect\citeauthoryear{Grevesse, Asplund, Sauval  \& Scott}{Grevesse
  et~al.}{2010}]{Grevesse2010}
Grevesse N.,  Asplund M.,  Sauval A.~J.,   Scott P.,  2010, \mn@doi
  [Astrophysics and Space Science] {10.1007/s10509-010-0288-z}, 328, 179

\bibitem[\protect\citeauthoryear{{Gronke} \& {Oh}}{{Gronke} \&
  {Oh}}{2018}]{Gronke2018}
{Gronke} M.,  {Oh} S.~P.,  2018, \mn@doi [\mnras] {10.1093/mnrasl/sly131},
  \href {http://adsabs.harvard.edu/abs/2018MNRAS.480L.111G} {480, L111}

\bibitem[\protect\citeauthoryear{{Haardt} \& {Madau}}{{Haardt} \&
  {Madau}}{1996}]{Haardt1996}
{Haardt} F.,  {Madau} P.,  1996, \mn@doi [\apj] {10.1086/177035}, \href
  {http://adsabs.harvard.edu/abs/1996ApJ...461...20H} {461, 20}

\bibitem[\protect\citeauthoryear{{Haardt} \& {Madau}}{{Haardt} \&
  {Madau}}{2012}]{Haardt2012}
{Haardt} F.,  {Madau} P.,  2012, \mn@doi [\apj] {10.1088/0004-637X/746/2/125},
  \href {http://adsabs.harvard.edu/abs/2012ApJ...746..125H} {746, 125}

\bibitem[\protect\citeauthoryear{{Heckman}, {Borthakur}, {Wild}, {Schiminovich}
   \& {Bordoloi}}{{Heckman} et~al.}{2017}]{Heckman2017}
{Heckman} T.,  {Borthakur} S.,  {Wild} V.,  {Schiminovich} D.,   {Bordoloi} R.,
   2017, \mn@doi [\apj] {10.3847/1538-4357/aa80dc}, \href
  {http://adsabs.harvard.edu/abs/2017ApJ...846..151H} {846, 151}

\bibitem[\protect\citeauthoryear{{Hennebelle} \& {P{\'e}rault}}{{Hennebelle} \&
  {P{\'e}rault}}{1999}]{Hennebelle1999}
{Hennebelle} P.,  {P{\'e}rault} M.,  1999, \aap, \href
  {http://adsabs.harvard.edu/abs/1999A%26A...351..309H} {351, 309}

\bibitem[\protect\citeauthoryear{{Hui} \& {Gnedin}}{{Hui} \&
  {Gnedin}}{1997}]{Hui1997}
{Hui} L.,  {Gnedin} N.~Y.,  1997, \mn@doi [\mnras] {10.1093/mnras/292.1.27},
  \href {http://adsabs.harvard.edu/abs/1997MNRAS.292...27H} {292, 27}

\bibitem[\protect\citeauthoryear{{Hussain}, {Khaire}, {Srianand}, {Muzahid}  \&
  {Pathak}}{{Hussain} et~al.}{2017}]{Hussain2017}
{Hussain} T.,  {Khaire} V.,  {Srianand} R.,  {Muzahid} S.,   {Pathak} A.,
  2017, \mn@doi [\mnras] {10.1093/mnras/stw3265}, \href
  {http://adsabs.harvard.edu/abs/2017MNRAS.466.3133H} {466, 3133}

\bibitem[\protect\citeauthoryear{{Ikeuchi}}{{Ikeuchi}}{1986}]{Ikeuchi1986}
{Ikeuchi} S.,  1986, \mn@doi [\apss] {10.1007/BF00651178}, \href
  {http://adsabs.harvard.edu/abs/1986Ap%26SS.118..509I} {118, 509}

\bibitem[\protect\citeauthoryear{{Inoue}, {Shimizu}, {Iwata}  \&
  {Tanaka}}{{Inoue} et~al.}{2014}]{Inoue2014}
{Inoue} A.~K.,  {Shimizu} I.,  {Iwata} I.,   {Tanaka} M.,  2014, \mn@doi
  [\mnras] {10.1093/mnras/stu936}, \href
  {http://adsabs.harvard.edu/abs/2014MNRAS.442.1805I} {442, 1805}

\bibitem[\protect\citeauthoryear{{Khaire}}{{Khaire}}{2017}]{Khaire2017}
{Khaire} V.,  2017, preprint, \href
  {http://adsabs.harvard.edu/abs/2017arXiv170203937K} {} (\mn@eprint {arXiv}
  {1702.03937})

\bibitem[\protect\citeauthoryear{{Khaire} \& {Srianand}}{{Khaire} \&
  {Srianand}}{2013}]{Khaire2013}
{Khaire} V.,  {Srianand} R.,  2013, \mn@doi [\mnras] {10.1093/mnrasl/slt007},
  \href {http://adsabs.harvard.edu/abs/2013MNRAS.431L..53K} {431, L53}

\bibitem[\protect\citeauthoryear{{Khaire} \& {Srianand}}{{Khaire} \&
  {Srianand}}{2015a}]{UVKhaire2015}
{Khaire} V.,  {Srianand} R.,  2015a, \mn@doi [\mnras] {10.1093/mnrasl/slv060},
  \href {http://adsabs.harvard.edu/abs/2015MNRAS.451L..30K} {451, L30}

\bibitem[\protect\citeauthoryear{{Khaire} \& {Srianand}}{{Khaire} \&
  {Srianand}}{2015b}]{Khaire2015}
{Khaire} V.,  {Srianand} R.,  2015b, \mn@doi [\apj]
  {10.1088/0004-637X/805/1/33}, \href
  {http://adsabs.harvard.edu/abs/2015ApJ...805...33K} {805, 33}

\bibitem[\protect\citeauthoryear{{Khaire} \& {Srianand}}{{Khaire} \&
  {Srianand}}{2018}]{Khaire2018}
{Khaire} V.,  {Srianand} R.,  2018, preprint, \href
  {http://adsabs.harvard.edu/abs/2018arXiv180109693K} {} (\mn@eprint {arXiv}
  {1801.09693})

\bibitem[\protect\citeauthoryear{{Khaire}, {Srianand}, {Choudhury}  \&
  {Gaikwad}}{{Khaire} et~al.}{2016}]{Khaire2016}
{Khaire} V.,  {Srianand} R.,  {Choudhury} T.~R.,   {Gaikwad} P.,  2016, \mn@doi
  [\mnras] {10.1093/mnras/stw192}, \href
  {http://adsabs.harvard.edu/abs/2016MNRAS.457.4051K} {457, 4051}

\bibitem[\protect\citeauthoryear{{Kim}, {Carswell}  \& {Ranquist}}{{Kim}
  et~al.}{2016}]{Kim2016}
{Kim} T.-S.,  {Carswell} R.~F.,   {Ranquist} D.,  2016, \mn@doi [\mnras]
  {10.1093/mnras/stv2847}, \href
  {http://adsabs.harvard.edu/abs/2016MNRAS.456.3509K} {456, 3509}

\bibitem[\protect\citeauthoryear{{Liang} \& {Remming}}{{Liang} \&
  {Remming}}{2018}]{Liang2018}
{Liang} C.~J.,  {Remming} I.~S.,  2018, preprint, \href
  {http://adsabs.harvard.edu/abs/2018arXiv180610688L} {} (\mn@eprint {arXiv}
  {1806.10688})

\bibitem[\protect\citeauthoryear{{McCourt}, {O'Leary}, {Madigan}  \&
  {Quataert}}{{McCourt} et~al.}{2015}]{Mccourt2015}
{McCourt} M.,  {O'Leary} R.~M.,  {Madigan} A.-M.,   {Quataert} E.,  2015,
  \mn@doi [\mnras] {10.1093/mnras/stv355}, \href
  {http://adsabs.harvard.edu/abs/2015MNRAS.449....2M} {449, 2}

\bibitem[\protect\citeauthoryear{{McCourt}, {Oh}, {O'Leary}  \&
  {Madigan}}{{McCourt} et~al.}{2018}]{Mccourt2018}
{McCourt} M.,  {Oh} S.~P.,  {O'Leary} R.,   {Madigan} A.-M.,  2018, \mn@doi
  [\mnras] {10.1093/mnras/stx2687}, \href
  {http://adsabs.harvard.edu/abs/2018MNRAS.473.5407M} {473, 5407}

\bibitem[\protect\citeauthoryear{{Meerson}}{{Meerson}}{1989}]{Meerson1989}
{Meerson} B.,  1989, \mn@doi [\apj] {10.1086/168191}, \href
  {http://adsabs.harvard.edu/abs/1989ApJ...347.1012M} {347, 1012}

\bibitem[\protect\citeauthoryear{{Miralda-Escude} \&
  {Ostriker}}{{Miralda-Escude} \& {Ostriker}}{1990}]{Miralda1990}
{Miralda-Escude} J.,  {Ostriker} J.~P.,  1990, \mn@doi [\apj] {10.1086/168358},
  \href {http://adsabs.harvard.edu/abs/1990ApJ...350....1M} {350, 1}

\bibitem[\protect\citeauthoryear{{Muzahid}, {Fonseca}, {Roberts},
  {Rosenwasser}, {Richter}, {Narayanan}, {Churchill}  \& {Charlton}}{{Muzahid}
  et~al.}{2018}]{Muzahid2018}
{Muzahid} S.,  {Fonseca} G.,  {Roberts} A.,  {Rosenwasser} B.,  {Richter} P.,
  {Narayanan} A.,  {Churchill} C.,   {Charlton} J.,  2018, \mn@doi [\mnras]
  {10.1093/mnras/sty529}, \href
  {http://adsabs.harvard.edu/abs/2018MNRAS.476.4965M} {476, 4965}

\bibitem[\protect\citeauthoryear{{Pachat}, {Narayanan}, {Muzahid}, {Khaire},
  {Srianand}, {Wakker}  \& {Savage}}{{Pachat} et~al.}{2016}]{Pachat2016}
{Pachat} S.,  {Narayanan} A.,  {Muzahid} S.,  {Khaire} V.,  {Srianand} R.,
  {Wakker} B.~P.,   {Savage} B.~D.,  2016, \mn@doi [\mnras]
  {10.1093/mnras/stw194}, \href
  {http://adsabs.harvard.edu/abs/2016MNRAS.458..733P} {458, 733}

\bibitem[\protect\citeauthoryear{{Pachat}, {Narayanan}, {Khaire}, {Savage},
  {Muzahid}  \& {Wakker}}{{Pachat} et~al.}{2017}]{Pachat2017a}
{Pachat} S.,  {Narayanan} A.,  {Khaire} V.,  {Savage} B.~D.,  {Muzahid} S.,
  {Wakker} B.~P.,  2017, \mn@doi [\mnras] {10.1093/mnras/stx1435}, \href
  {http://adsabs.harvard.edu/abs/2017MNRAS.471..792P} {471, 792}

\bibitem[\protect\citeauthoryear{{Planck Collaboration} et~al.,}{{Planck
  Collaboration} et~al.}{2016}]{Plank2016}
{Planck Collaboration} et~al., 2016, \mn@doi [\aap]
  {10.1051/0004-6361/201525830}, \href
  {http://adsabs.harvard.edu/abs/2016A%26A...594A..13P} {594, A13}

\bibitem[\protect\citeauthoryear{{Rafelski}, {Wolfe}, {Prochaska}, {Neeleman}
  \& {Mendez}}{{Rafelski} et~al.}{2012}]{Rafelski2012}
{Rafelski} M.,  {Wolfe} A.~M.,  {Prochaska} J.~X.,  {Neeleman} M.,   {Mendez}
  A.~J.,  2012, \mn@doi [\apj] {10.1088/0004-637X/755/2/89}, \href
  {http://adsabs.harvard.edu/abs/2012ApJ...755...89R} {755, 89}

\bibitem[\protect\citeauthoryear{{Rees}}{{Rees}}{1986}]{Rees1986}
{Rees} M.~J.,  1986, \mn@doi [\mnras] {10.1093/mnras/218.1.25P}, \href
  {http://adsabs.harvard.edu/abs/1986MNRAS.218P..25R} {218, 25P}

\bibitem[\protect\citeauthoryear{{Sargent}, {Young}, {Boksenberg}  \&
  {Tytler}}{{Sargent} et~al.}{1980}]{Sargent1980}
{Sargent} W.~L.~W.,  {Young} P.~J.,  {Boksenberg} A.,   {Tytler} D.,  1980,
  \mn@doi [\apjs] {10.1086/190644}, \href
  {http://adsabs.harvard.edu/abs/1980ApJS...42...41S} {42, 41}

\bibitem[\protect\citeauthoryear{{Schaye}}{{Schaye}}{2001}]{Schaye2001}
{Schaye} J.,  2001, \mn@doi [\apj] {10.1086/322421}, \href
  {http://adsabs.harvard.edu/abs/2001ApJ...559..507S} {559, 507}

\bibitem[\protect\citeauthoryear{{Schaye}, {Carswell}  \& {Kim}}{{Schaye}
  et~al.}{2007}]{Schaye2007}
{Schaye} J.,  {Carswell} R.~F.,   {Kim} T.-S.,  2007, \mn@doi [\mnras]
  {10.1111/j.1365-2966.2007.12005.x}, \href
  {http://adsabs.harvard.edu/abs/2007MNRAS.379.1169S} {379, 1169}

\bibitem[\protect\citeauthoryear{{Schure}, {Kosenko}, {Kaastra}, {Keppens}  \&
  {Vink}}{{Schure} et~al.}{2009}]{Schure2009}
{Schure} K.~M.,  {Kosenko} D.,  {Kaastra} J.~S.,  {Keppens} R.,   {Vink} J.,
  2009, \mn@doi [\aap] {10.1051/0004-6361/200912495}, \href
  {http://adsabs.harvard.edu/abs/2009A%26A...508..751S} {508, 751}

\bibitem[\protect\citeauthoryear{{Scott}, {Kriss}, {Brotherton}, {Green},
  {Hutchings}, {Shull}  \& {Zheng}}{{Scott} et~al.}{2004}]{Scott2004}
{Scott} J.~E.,  {Kriss} G.~A.,  {Brotherton} M.,  {Green} R.~F.,  {Hutchings}
  J.,  {Shull} J.~M.,   {Zheng} W.,  2004, \mn@doi [\apj] {10.1086/422336},
  \href {http://adsabs.harvard.edu/abs/2004ApJ...615..135S} {615, 135}

\bibitem[\protect\citeauthoryear{{Shapiro}, {Giroux}  \& {Babul}}{{Shapiro}
  et~al.}{1994}]{Shapiro1994}
{Shapiro} P.~R.,  {Giroux} M.~L.,   {Babul} A.,  1994, \mn@doi [\apj]
  {10.1086/174120}, \href {http://adsabs.harvard.edu/abs/1994ApJ...427...25S}
  {427, 25}

\bibitem[\protect\citeauthoryear{{Sharma}, {McCourt}, {Quataert}  \&
  {Parrish}}{{Sharma} et~al.}{2012}]{Sharma2012}
{Sharma} P.,  {McCourt} M.,  {Quataert} E.,   {Parrish} I.~J.,  2012, \mn@doi
  [\mnras] {10.1111/j.1365-2966.2011.20246.x}, \href
  {http://adsabs.harvard.edu/abs/2012MNRAS.420.3174S} {420, 3174}

\bibitem[\protect\citeauthoryear{{Shull}, {Roberts}, {Giroux}, {Penton}  \&
  {Fardal}}{{Shull} et~al.}{1999}]{Shull1999}
{Shull} J.~M.,  {Roberts} D.,  {Giroux} M.~L.,  {Penton} S.~V.,   {Fardal}
  M.~A.,  1999, \mn@doi [\aj] {10.1086/301053}, \href
  {http://adsabs.harvard.edu/abs/1999AJ....118.1450S} {118, 1450}

\bibitem[\protect\citeauthoryear{{Shull}, {Stevans}  \& {Danforth}}{{Shull}
  et~al.}{2012}]{Shull2012}
{Shull} J.~M.,  {Stevans} M.,   {Danforth} C.~W.,  2012, \mn@doi [\apj]
  {10.1088/0004-637X/752/2/162}, \href
  {http://adsabs.harvard.edu/abs/2012ApJ...752..162S} {752, 162}

\bibitem[\protect\citeauthoryear{{Smette}, {Surdej}, {Shaver}, {Foltz},
  {Chaffee}, {Weymann}, {Williams}  \& {Magain}}{{Smette}
  et~al.}{1992}]{Smette1992}
{Smette} A.,  {Surdej} J.,  {Shaver} P.~A.,  {Foltz} C.~B.,  {Chaffee} F.~H.,
  {Weymann} R.~J.,  {Williams} R.~E.,   {Magain} P.,  1992, \mn@doi [\apj]
  {10.1086/171187}, \href {http://adsabs.harvard.edu/abs/1992ApJ...389...39S}
  {389, 39}

\bibitem[\protect\citeauthoryear{{Sparre}, {Pfrommer}  \&
  {Vogelsberger}}{{Sparre} et~al.}{2018}]{Sparre2018}
{Sparre} M.,  {Pfrommer} C.,   {Vogelsberger} M.,  2018, preprint, \href
  {http://adsabs.harvard.edu/abs/2018arXiv180707971S} {} (\mn@eprint {arXiv}
  {1807.07971})

\bibitem[\protect\citeauthoryear{{Stevans}, {Shull}, {Danforth}  \&
  {Tilton}}{{Stevans} et~al.}{2014}]{Stevans2014}
{Stevans} M.~L.,  {Shull} J.~M.,  {Danforth} C.~W.,   {Tilton} E.~M.,  2014,
  \mn@doi [\apj] {10.1088/0004-637X/794/1/75}, \href
  {http://adsabs.harvard.edu/abs/2014ApJ...794...75S} {794, 75}

\bibitem[\protect\citeauthoryear{{Telfer}, {Zheng}, {Kriss}  \&
  {Davidsen}}{{Telfer} et~al.}{2002}]{Telfer2002}
{Telfer} R.~C.,  {Zheng} W.,  {Kriss} G.~A.,   {Davidsen} A.~F.,  2002, \mn@doi
  [\apj] {10.1086/324689}, \href
  {http://adsabs.harvard.edu/abs/2002ApJ...565..773T} {565, 773}

\bibitem[\protect\citeauthoryear{{Thompson}, {Quataert}, {Zhang}  \&
  {Weinberg}}{{Thompson} et~al.}{2016}]{Thompson2016}
{Thompson} T.~A.,  {Quataert} E.,  {Zhang} D.,   {Weinberg} D.~H.,  2016,
  \mn@doi [\mnras] {10.1093/mnras/stv2428}, \href
  {http://adsabs.harvard.edu/abs/2016MNRAS.455.1830T} {455, 1830}

\bibitem[\protect\citeauthoryear{{Vanden Berk} et~al.,}{{Vanden Berk}
  et~al.}{2001}]{Vanden2001}
{Vanden Berk} D.~E.,  et~al., 2001, \mn@doi [\aj] {10.1086/321167}, \href
  {http://adsabs.harvard.edu/abs/2001AJ....122..549V} {122, 549}

\bibitem[\protect\citeauthoryear{{Veilleux}, {Cecil}  \&
  {Bland-Hawthorn}}{{Veilleux} et~al.}{2005}]{Veilleux2005}
{Veilleux} S.,  {Cecil} G.,   {Bland-Hawthorn} J.,  2005, \mn@doi [\araa]
  {10.1146/annurev.astro.43.072103.150610}, \href
  {http://adsabs.harvard.edu/abs/2005ARA%26A..43..769V} {43, 769}

\bibitem[\protect\citeauthoryear{{Williger} \& {Babul}}{{Williger} \&
  {Babul}}{1992}]{Williger1992}
{Williger} G.~M.,  {Babul} A.,  1992, \mn@doi [\apj] {10.1086/171935}, \href
  {http://adsabs.harvard.edu/abs/1992ApJ...399..385W} {399, 385}

\bibitem[\protect\citeauthoryear{{van de Voort}, {Springel}, {Mandelker}, {van
  den Bosch}  \& {Pakmor}}{{van de Voort} et~al.}{2018}]{Voort2018}
{van de Voort} F.,  {Springel} V.,  {Mandelker} N.,  {van den Bosch} F.~C.,
  {Pakmor} R.,  2018, preprint, \href
  {http://adsabs.harvard.edu/abs/2018arXiv180804369V} {} (\mn@eprint {arXiv}
  {1808.04369})

\makeatother
\end{thebibliography}
\appendix 

\section{Understanding the possible cause of difference in \nh\ between our calculations and that of \KIM16\ }\label{Ap6}
\begin{figure}
  \centering
  \includegraphics[totalheight=0.26\textheight, trim=.1cm 2.1cm 0cm 0.6cm, clip=true, angle=0]{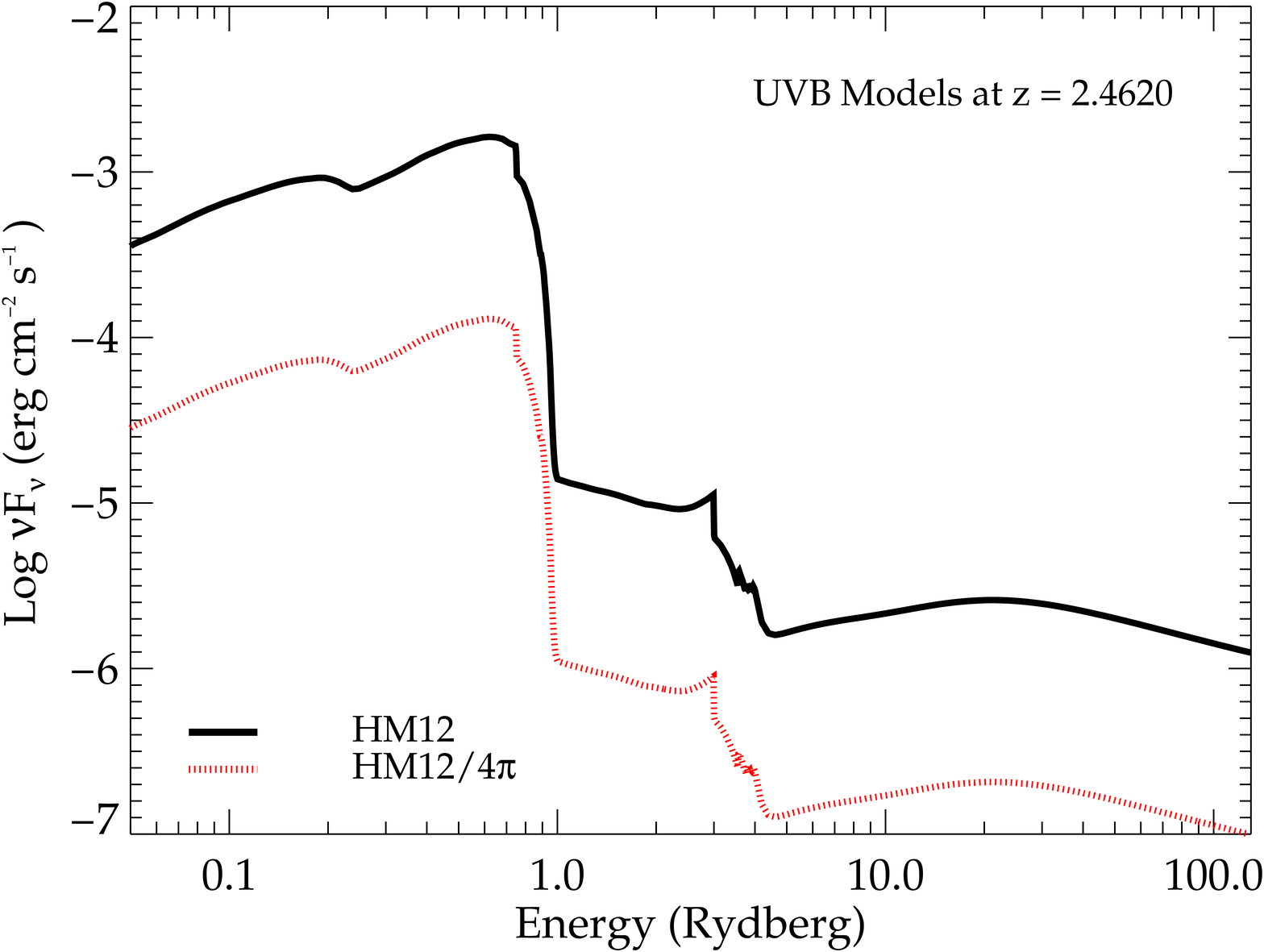}
\caption{Comparison of \HM12\ UVB (black solid curve) with the one rescaled lower by a factor 4$\pi$ (red dotted curve)  at
  $z$ $\approx$  2.4620. Red dotted curve resembles
  the UVB shown by \KIM16\ in their figure~12.}
\label{figa6}
\end{figure}

\begin{figure}
  \centering
  \includegraphics[totalheight=0.26\textheight, trim=.1cm 0.9cm 0cm 0.6cm, clip=true, angle=0]{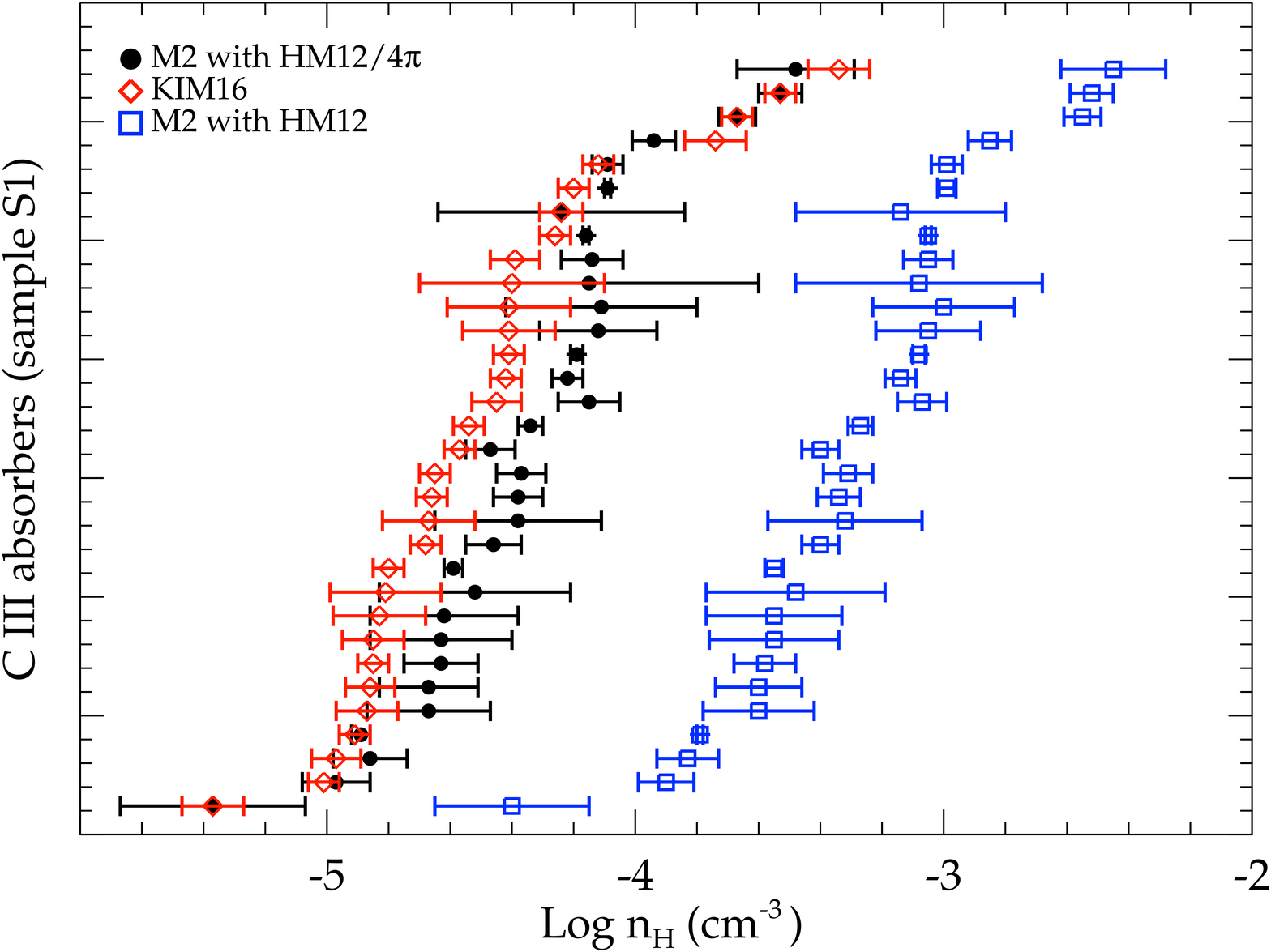}
\caption{Comparison of \nh\ from different calculations. Black circles show our model \Mt\ predicted \nh\ using rescaled \HM12\ UVB with a factor of 4$\pi$ lower intensity.
Red diamonds show the calculations from \KIM16\ and blue squares show our model \Mt\ derived \nh\ for \HM12\ UVB. The horizontal
bars are errors associated with \nh\ in the respective calculations considering systematic
uncertainties.
Note that our derived \nh\ using \HM12\ UVB are consistently
$\sim$1.1 dex higher than that inferred from rescaled \HM12\ UVB.}
\label{a61}
\end{figure}

  \KIM16\ mentioned that they have used the \HM12\ UVB in their \CLOUDY\ models. However,
  when we use the \HM12\ UVB in our models, we find our \nh\ values are systematically higher by at least one order of magnitude
  than their reported values.
  In this work, we obtain the best fit \nh\ and [{\it C/H}] by constructing grids of \nh\ and [{\it C/H}] in \CLOUDY\ models
  for each system while \KIM16\ used grids of ionization parameter and [{\it C/H}]. 
Thus, it is possible
that some mistake may have occurred in \KIM16\ while \nh\ was obtained
from the best fitted ionization parameter. Alternatively, the difference in the derived \nh\ for each component
  between this paper and that of \KIM16\ could also come from the \HM12\ UVB used by them in their
  \CLOUDY\ models which is less by a normalization factor (approximately $4\pi$).
  In Fig.~\ref{figa6}, we show the comparison of \HM12\ UVB{\footnote{taken from \url{http://www.ucolick.org/~pmadau/CUBA}}} and the same UVB with a factor
  of 4$\pi$ lower intensity 
at $z$ $\approx$  2.4620. A cursory look at the Fig.~\ref{figa6} and the figure~12 of
  \KIM16\ reveals that \HM12\ UVB intensity lower by a factor 4$\pi$ gives a good representation of
  the UVB used by \KIM16.
 To just check our conjuncture, we run photoionization models with the rescaled \HM12\ UVB by a factor of 4$\pi$ lower intensity and
  calculate the \nh\ for all the individual \CIII\ components. We produce the results in Table.~\ref{tabc2} and show a comparison plot
  in Fig.~\ref{a61} along with our derived \nh\ from \HM12\ UVB (see Table~\ref{tab1}). 
  It can be seen from the Table.~\ref{tabc2} and Fig.~\ref{a61} that for 60\% of the components, our model predicted \nh\
  values using rescaled \HM12\ UVB match with that of \KIM16\
  within $1\sigma$ errors.
  Also, the results are consistent with each other within $2\sigma$ errors in 80\% of the cases. Even in the remaining cases,  
  the values we derive for \nh\ are typically 0.2 dex higher than the values derived by \KIM16. We also
  find our [{\it C/H}] estimates are within 0.1 dex to the values derived by \KIM16.
  The small differences can come from various reasons, such as the
  different modelling procedure or providing UVB at nearest redshift where the original \HM12\
  tables exist instead of interpolating to the exact redshift of absorbers. 
  Nevertheless, this exercise suggests that \KIM16\ might have missed 4$\pi$ factor in the \HM12\ UVB somewhere in their
  calculations which gives rise to one order of magnitude difference in the \nh\ values. 

\begin{table*}
\caption{Comparison of \nh derived by \KIM16\ and by us using rescaled \HM12\ UVB. }
\begin{tabular}{c c c c c c }
\hline \hline

{\#} & {quasar name} &  {$z_{abs}$} & \multicolumn{2}{c} {log~\nh} & \\
 &  &  & \KIM16\ &This work \\ \hline \hline
1 & Q0055-269 & 3.257359 & -5.01$\pm$0.05 & -4.97$\pm$0.11 \\ 
2 & Q0055-269 & 3.038795 & -4.87$\pm$0.10 & -4.67$\pm$0.20 \\ 
3 & Q0055-269 & 2.744100 & -4.39$\pm$0.08 & -4.14$\pm$0.10 \\ 
4 & Q0055-269 & 2.743720 & -4.81$\pm$0.18 & -4.52$\pm$0.31 \\ 
5 & PKS2126-158 & 2.973015 & -4.66$\pm$0.05 & -4.38$\pm$0.08 \\ 
6 & Q0420-388 & 2.849598 & -4.97$\pm$0.08 & -4.86$\pm$0.12 \\ 
7 & Q0420-388 & 2.849229 & -4.83$\pm$0.15 & -4.62$\pm$0.24 \\ 
8 & HE0940-1050 & 2.937755 & -4.8$\pm$0.05 & -4.59$\pm$0.03 \\ 
9 & HE0940-1050 & 2.883509 & -4.65$\pm$0.05 & -4.37$\pm$0.08 \\ 
10 & HE0940-1050 & 2.826555 & -4.54$\pm$0.05 & -4.34$\pm$0.04 \\ 
11 & HE2347-4342 & 2.347467 & -4.67$\pm$0.15 & -4.38$\pm$0.27 \\ 
12 & HE0151-4326 & 2.519825 & -4.45$\pm$0.08 & -4.15$\pm$0.1 \\ 
13 & HE0151-4326 & 2.449902 & -4.42$\pm$0.05 & -4.22$\pm$0.05 \\ 
14 & HE0151-4326 & 2.419676 & -4.41$\pm$0.05 & -4.19$\pm$0.02 \\ 
15 & HE0151-4326 & 2.415718 & -4.26$\pm$0.05 & -4.16$\pm$0.01 \\ 
16 & HE0151-4326 & 2.401315 & -4.41$\pm$0.15 & -4.12$\pm$0.19 \\ 
17 & Q0002-422 & 2.539455 & -4.85$\pm$0.05 & -4.63$\pm$0.12 \\ 
18 & Q0002-422 & 2.463222 & -4.2$\pm$0.05 & -4.09$\pm$0.01 \\ 
19 & Q0002-422 & 2.462358 & -3.53$\pm$0.05 & -3.53$\pm$0.07 \\ 
20 & Q0002-422 & 2.462044 & -3.74$\pm$0.10 & -3.94$\pm$0.07 \\ 
21 & PKS0329-255 & 2.586757 & -4.85$\pm$0.10 & -4.63$\pm$0.23 \\ 
22 & PKS0329-255 & 2.456581 & -4.4$\pm$0.30 & -4.15$\pm$0.55 \\ 
23 & Q0453-423 & 2.444109 & -3.34$\pm$0.10 & -3.48$\pm$0.19 \\ 
24 & Q0453-423 & 2.442644 & -3.67$\pm$0.05 & -3.67$\pm$0.06 \\ 
25 & Q0453-423 & 2.441813 & -4.41$\pm$0.20 & -4.11$\pm$0.31 \\ 
26 & Q0453-423 & 2.398159 & -5.37$\pm$0.10 & -5.37$\pm$0.30 \\ 
27 & Q0453-423 & 2.397801 & -4.91$\pm$0.05 & -4.89$\pm$0.03 \\ 
28 & Q0453-423 & 2.397447 & -4.86$\pm$0.08 & -4.67$\pm$0.16 \\ 
29 & Q0453-423 & 2.396755 & -4.57$\pm$0.05 & -4.47$\pm$0.08 \\ 
30 & Q0453-423 & 2.277569 & -4.12$\pm$0.05 & -4.09$\pm$0.05 \\ 
31 & HE1347-2457 & 2.370003 & -4.68$\pm$0.05 & -4.46$\pm$0.07 \\ 
32 & Q0329-385 & 2.249389 & -4.24$\pm$0.07 & -4.24$\pm$0.40 \\ \hline \hline
\end{tabular}
\label{tabc2}
\end{table*}

\section{Redshift Evolution using \KS18\ UVB} \label{Ap1}
\begin{figure*}
  \centering 
 \includegraphics[totalheight=0.45\textheight, trim=0.5cm 9cm 0cm 0.559cm, clip=true, angle=0]{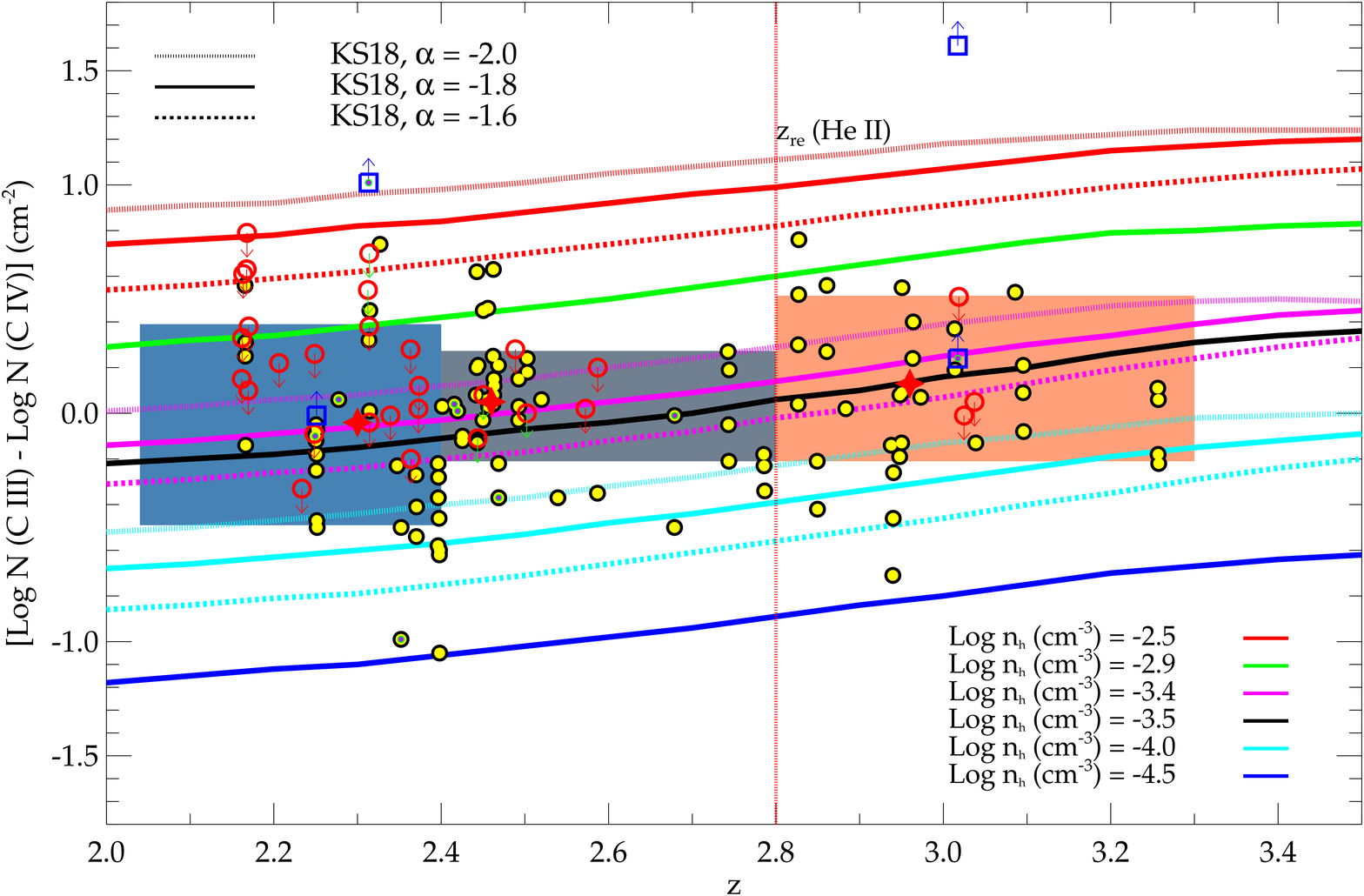}
\caption{\small{The observed ratio of $N$({\CIII}) to $N$({\CIV}) for 132 {\CIII} components as a function
of redshift [104 clean detections (scattered yellow filled black circles), 24 components with
upper limit on {\CIII} (red open circles) and 4 components with
lower limit on {\CIII} (blue open squares)]. 
\Mo\ generated column density ratio of {\CIII} to {\CIV} for two different UVBs is over plotted on top of the observed data.
The solid lines are used to show the model predictions using our fiducial \citetalias{Khaire2018} with $\alpha$=-1.8.
We also use \citetalias{Khaire2018} UVB with $\alpha$ values -2.0 and -1.8 a shown by dotted and dash-dotted lines, respectability
to compare the uncertainties in 
densities. From top to bottom  
\nh\ is increasing as shown in the legends.
The shaded blue, gray and orange
regions show the range for redshift 
2.1 < $z$ $\le$ 2.4, 2.4 < $z$ $\le$ 2.8 and 2.8 < $z$ $\le$ 3.4
where 68\% of observed data lies as obtained from the cumulative probability distribution of the sample data.
The red filled stars are used to mark the median values of the column
density ratios at each median redshift of the above three bins. The vertical red dotted line is used to
mark the reionization redshift of \HeII\ ($z_{re}$(\HeII))
for our fiducial \KS18\ UVB model.
}}
\label{figA111}
\end{figure*}
\section{Predictions by \CLOUDY\ photoionization models and correlation analysis tables of derived parameters}\label{Prdection}
\begin{table*} 
\caption{\Mt\ Cloudy-prediction column densities for the tied {\HI}+{\CIV}+{\CIII} components of \sam\ for
the \citetalias{Haardt2012} and \citetalias{Khaire2018} UVBs.}
\noindent
\scriptsize{
\begin{tabular}{ccccccccccccccc}
 \hline \hline
\textbf{Sl no.} & \textbf{quasar name} & \textbf{$z_{abs}$} & \multicolumn{ 4}{c}{\textbf{Observed}} & \multicolumn{ 8}{c}{\textbf{Photoionization Model}} \\ \hline \hline
\textbf{} & \textbf{} & \textbf{} & \textbf{log \nhi} & \textbf{log $N (\CIV)$} & \textbf{log $N (\CIII)$} & \textbf{log $N (\CII)$} & \multicolumn{ 2}{c}{\textbf{log $N (\CIV)$}} & \multicolumn{ 2}{c}{\textbf{log $N (\CIII)$}} & \multicolumn{ 2}{c}{\textbf{log $N (\CII)$}} & \multicolumn{ 2}{c}{\textbf{log $N (H)$}} \\ \hline
\textbf{} & \textbf{} & \textbf{} & \multicolumn{ 10}{c}{\textbf{in \cmsq}} &  &  \\ \hline
\textbf{} & \textbf{} & \textbf{} &  &  &  &  & \textbf{HM12} & \textbf{KS18} & \textbf{HM12} & \textbf{KS18} & \textbf{HM12} & \textbf{KS18} & \textbf{HM12} & \textbf{KS18} \\  \hline \hline
1 & Q0055-269 & 3.25736 & 13.93$\pm$0.02 & 12.76$\pm$0.03 & 12.54$\pm$0.06 & $\leq$12.3 & 12.76 & 12.76 & 12.54 & 12.56 & 9.98 & 10.17 & 18.24 & 17.93 \\ 
2 & Q0055-269 & 3.03880 & 14.60$\pm$0.04 & 12.72$\pm$0.04 & 12.59$\pm$0.15 & $\leq$12.7 & 12.72 & 12.72 & 12.59 & 12.57 & 10.25 & 10.35 & 18.61 & 18.42 \\ 
3 & Q0055-269 & 2.74410 & 15.12$\pm$0.13 & 12.79$\pm$0.05 & 12.98$\pm$0.06 & ... & 12.79 & 12.79 & 12.97 & 13 & 11.13 & 11.24 & 18.54 & 18.45 \\ 
4 & Q0055-269 & 2.74372 & 14.78$\pm$0.39 & 12.48$\pm$0.09 & 12.27$\pm$0.19 & ... & 12.48 & 12.48 & 12.28 & 12.28 & 10.01 & 10.1 & 18.71 & 18.56 \\ 
5 & PKS2126-158 & 2.97302 & 14.36$\pm$0.01 & 12.18$\pm$0.03 & 12.25$\pm$0.03 & $\leq$11.45 & 12.18 & 12.18 & 12.23 & 12.23 & 10.12 & 10.22 & 18.08 & 17.95 \\ 
6 & Q0420-388 & 2.84960 & 14.09$\pm$0.02 & 12.59$\pm$0.04 & 12.17$\pm$0.06 & ... & 12.59 & 12.59 & 12.17 & 12.17 & 9.64 & 9.73 & 18.41 & 18.2 \\ 
7 & Q0420-388 & 2.84923 & 13.68$\pm$0.06 & 12.10$\pm$0.10 & 11.89$\pm$0.14 & ... & 12.10 & 12.10 & 11.9 & 11.9 & 9.59 & 9.7 & 17.67 & 17.5 \\ 
8 & HE0940-1050 & 2.93776 & 14.58$\pm$0.02 & 12.65$\pm$0.02 & 12.51$\pm$0.01 & $\leq$11.95 & 12.65 & 12.65 & 12.49 & 12.49 & 10.19 & 10.31 & 18.55 & 18.37 \\ 
9 & HE0940-1050 & 2.88351 & 14.60$\pm$0.01 & 12.35$\pm$0.03 & 12.37$\pm$0.02 & $\leq$12.25 & 12.35 & 12.35 & 12.36 & 12.34 & 10.28 & 10.32 & 18.3 & 18.22 \\ 
10 & HE0940-1050 & 2.82656 & 14.51$\pm$0.21 & 13.19$\pm$0.02 & 13.23$\pm$0.03 & $\leq$12 & 13.19 & 13.19 & 13.2 & 13.23 & 11.15 & 11.29 & 18.17 & 18.04 \\ 
11 & HE2347-4342 & 2.34747 & 15.99$\pm$0.20 & 13.49$\pm$0.00 & 13.26$\pm$0.22 & $\leq$11.75 & 13.49 & 13.49 & 13.25 & 13.24 & 11.08 & 11.09 & 19.79 & 19.75 \\ 
12 & HE0151-4326 & 2.51983 & 15.24$\pm$0.02 & 12.29$\pm$0.04 & 12.35$\pm$0.06 & $\leq$11.7 & 12.29 & 12.29 & 12.34 & 12.4 & 10.45 & 10.59 & 18.72 & 18.62 \\ 
13 & HE0151-4326 & 2.44990 & 14.45$\pm$0.02 & 12.99$\pm$0.01 & 12.96$\pm$0.04 & $\leq$12.2 & 12.99 & 12.98 & 12.95 & 12.95 & 10.98 & 11.01 & 18.02 & 17.97 \\ 
14 & HE0151-4326 & 2.41968 & 12.85$\pm$0.04 & 12.75$\pm$0.01 & 12.76$\pm$0.02 & ... & 12.75 & 12.75 & 12.75 & 12.78 & 10.87 & 10.92 & 16.31 & 16.28 \\ 
15 & HE0151-4326 & 2.41572 & 13.36$\pm$0.01 & 13.03$\pm$0.01 & 13.07$\pm$0.02 & ... & 13.02 & 13.01 & 13.06 & 13.09 & 11.19 & 11.28 & 16.79 & 16.75 \\ 
16 & HE0151-4326 & 2.40132 & 15.04$\pm$0.02 & 12.34$\pm$0.11 & 12.37$\pm$0.06 & ... & 12.34 & 12.34 & 12.37 & 12.41 & 10.49 & 10.58 & 18.51 & 18.45 \\ 
17 & Q0002-422 & 2.53946 & 14.24$\pm$0.03 & 12.55$\pm$0.02 & 12.18$\pm$0.06 & $\leq$11.95 & 12.55 & 12.55 & 12.18 & 12.16 & 9.83 & 9.81 & 18.32 & 18.24 \\ 
18 & Q0002-422 & 2.46322 & 14.56$\pm$0.06 & 13.26$\pm$0.01 & 13.38$\pm$0.02 & $\leq$11.75 & 13.26 & 13.26 & 13.36 & 13.38 & 11.55 & 11.59 & 17.95 & 17.93 \\ 
19 & Q0002-422 & 2.46236 & 15.16$\pm$0.02 & 13.20$\pm$0.01 & 13.83$\pm$0.03 & 12.60$\pm$0.03 & 13.20 & 13.20 & 13.85 & 13.85 & 12.54 & 12.53 & 18.02 & 18.04 \\ 
20 & Q0002-422 & 2.46204 & 14.83$\pm$0.02 & 13.34$\pm$0.02 & 13.59$\pm$0.02 & 12.19$\pm$0.08 & 13.34 & 13.34 & 13.59 & 13.6 & 11.93 & 11.94 & 18.06 & 18.06 \\ 
21 & PKS0329-255 & 2.58676 & 14.99$\pm$0.03 & 12.37$\pm$0.06 & 12.02$\pm$0.11 & $\leq$11.9 & 12.37 & 12.37 & 12.03 & 12.01 & 9.7 & 9.68 & 19.03 & 18.96 \\ 
22 & PKS0329-255 & 2.45658 & 13.51$\pm$0.24 & 11.97$\pm$0.25 & 12.00$\pm$0.29 & $\leq$12.3 & 11.97 & 11.97 & 12 & 12 & 10.09 & 10.11 & 17.01 & 16.98 \\ 
23 & Q0453-423 & 2.44411 & 14.62$\pm$0.01 & 12.63$\pm$0.06 & 13.41$\pm$0.08 & 12.66$\pm$0.11 & 12.63 & 12.64 & 13.36 & 13.42 & 12.12 & 12.23 & 17.4 & 17.35 \\ 
24 & Q0453-423 & 2.44264 & 14.91$\pm$0.01 & 13.04$\pm$0.03 & 13.66$\pm$0.07 & 12.18$\pm$0.07 & 13.04 & 13.04 & 13.65 & 13.66 & 12.3 & 12.35 & 17.8 & 17.78 \\ 
25 & Q0453-423 & 2.44181 & 14.67$\pm$0.01 & 11.75$\pm$0.13 & 11.83$\pm$0.14 & $\leq$11.65 & 11.75 & 11.75 & 11.84 & 11.83 & 10.01 & 10.03 & 18.08 & 18.06 \\ 
26 & Q0453-423 & 2.39816 & 13.39$\pm$0.07 & 12.74$\pm$0.04 & 11.69$\pm$0.25 & $\leq$12.05 & 12.74 & 12.74 & 11.71 & 11.71 & 8.87 & 8.79 & 18.45 & 18.34 \\ 
27 & Q0453-423 & 2.39780 & 14.41$\pm$0.01 & 13.78$\pm$0.01 & 13.16$\pm$0.02 & $\leq$11.7 & 13.78 & 13.78 & 13.18 & 13.16 & 10.65 & 10.6 & 18.73 & 18.68 \\ 
28 & Q0453-423 & 2.39745 & 13.86$\pm$0.05 & 12.71$\pm$0.03 & 12.25$\pm$0.09 & $\leq$11.8 & 12.71 & 12.71 & 12.27 & 12.23 & 9.9 & 9.81 & 17.97 & 17.95 \\ 
29 & Q0453-423 & 2.39676 & 14.16$\pm$0.03 & 13.34$\pm$0.01 & 13.06$\pm$0.04 & $\leq$11.6 & 13.34 & 13.34 & 13.05 & 12.97 & 10.83 & 10.69 & 18.03 & 18.08 \\ 
30 & Q0453-423 & 2.27757 & 13.54$\pm$0.03 & 12.99$\pm$0.01 & 13.05$\pm$0.05 & ... & 13.00 & 12.98 & 13.05 & 13.02 & 11.24 & 11.18 & 16.92 & 16.96 \\ 
31 & HE1347-2457 & 2.37000 & 14.10$\pm$0.09 & 12.60$\pm$0.03 & 12.33$\pm$0.04 & $\leq$11.55 & 12.60 & 12.60 & 12.31 & 12.31 & 10.1 & 10.12 & 17.98 & 17.93 \\ 
32 & Q0329-385 & 2.24939 & 12.90$\pm$0.10 & 12.40$\pm$0.06 & 12.30$\pm$0.34 & $\leq$12.55 & 12.41 & 12.40 & 12.3 & 12.3 & 10.33 & 10.33 & 16.45 & 16.46 \\ \hline \hline
\end{tabular}
\label{tab1}
}
\end{table*}

\begin{table*}
{\scriptsize {
\begin{tabular}{ccccccccccccc}
{\small { {\bf Table~\ref{tab1}}}}&{\small{{\bf Continue...}}}&   \\
& & \\ 
\\
\hline \hline
\textbf{Sl. no.} & \textbf{quasar name} & \textbf{$z_{abs}$} & \multicolumn{ 9}{c}{\textbf{Photoionization Model}} \\ \hline \hline
\textbf{} & \textbf{} & \textbf{} & \multicolumn{ 2}{c}{\textbf{log \nh}} & \multicolumn{ 2}{c}{\textbf{$[C/H]$}} & \multicolumn{ 2}{c}{\textbf{log $\Delta$}} & \multicolumn{ 2}{c}{\textbf{log $L$}} & \multicolumn{ 2}{c}{\textbf{log $P/k$}} \\  \hline
\textbf{} & \textbf{} & \textbf{} & \multicolumn{ 2}{c}{\textbf{in {\cmcb}}} & \multicolumn{ 2}{c}{\textbf{}} & \multicolumn{ 2}{c}{\textbf{}} & \multicolumn{ 2}{c}{\textbf{in kpc}} & \multicolumn{ 2}{c}{\textbf{in \cmcb\ K}} \\ \hline
\textbf{} & \textbf{} & \textbf{} & \textbf{HM12} & \textbf{KS18} & \textbf{HM12} & \textbf{KS18} & \textbf{HM12} & \textbf{KS18} & \textbf{HM12} & \textbf{KS18} & \textbf{HM12} & \textbf{KS18} \\ \hline \hline
1 & Q0055-269 & 3.25736 & -3.90 & -3.70 & -1.18 & -1.17 & 0.98 & 1.18 & 0.65 & 0.14 & 0.80 & 0.81 \\ 
2 & Q0055-269 & 3.03880 & -3.60 & -3.50 & -1.7 & -1.67 & 1.35 & 1.45 & 0.72 & 0.43 & 1.05 & 1.02 \\ 
3 & Q0055-269 & 2.74410 & -3.05 & -3.00 & -1.63 & -1.56 & 1.99 & 2.04 & 0.10 & -0.04 & 1.49 & 1.44 \\ 
4 & Q0055-269 & 2.74372 & -3.48 & -3.40 & -2.02 & -2.02 & 1.56 & 1.64 & 0.70 & 0.47 & 1.17 & 1.14 \\ 
5 & PKS2126-158 & 2.97302 & -3.34 & -3.30 & -1.77 & -1.73 & 1.63 & 1.67 & -0.07 & -0.24 & 1.25 & 1.18 \\ 
6 & Q0420-388 & 2.84960 & -3.83 & -3.70 & -1.37 & -1.42 & 1.18 & 1.31 & 0.75 & 0.41 & 0.90 & 0.91 \\ 
7 & Q0420-388 & 2.84923 & -3.55 & -3.45 & -1.35 & -1.33 & 1.46 & 1.56 & -0.27 & -0.54 & 1.11 & 1.09 \\ 
8 & HE0940-1050 & 2.93776 & -3.55 & -3.45 & -1.71 & -1.68 & 1.43 & 1.53 & 0.61 & 0.33 & 1.10 & 1.07 \\ 
9 & HE0940-1050 & 2.88351 & -3.31 & -3.30 & -1.83 & -1.83 & 1.69 & 1.7 & 0.12 & 0.03 & 1.28 & 1.19 \\ 
10 & HE0940-1050 & 2.82656 & -3.27 & -3.20 & -0.87 & -0.8 & 1.75 & 1.82 & -0.05 & -0.25 & 1.31 & 1.28 \\ 
11 & HE2347-4342 & 2.34747 & -3.32 & -3.28 & -2.14 & -2.2 & 1.87 & 1.91 & 1.62 & 1.54 & 1.32 & 1.24 \\ 
12 & HE0151-4326 & 2.51983 & -3.07 & -3.00 & -2.33 & -2.27 & 2.05 & 2.12 & 0.30 & 0.13 & 1.50 & 1.46 \\ 
13 & HE0151-4326 & 2.44990 & -3.14 & -3.10 & -0.92 & -0.94 & 2.01 & 2.05 & -0.33 & -0.42 & 1.44 & 1.40 \\ 
14 & HE0151-4326 & 2.41968 & -3.08 & -3.05 & 0.53 & 0.52 & 2.08 & 2.11 & -2.10 & -2.16 & 1.43 & 1.40 \\ 
15 & HE0151-4326 & 2.41572 & -3.05 & -3.00 & 0.32 & 0.33 & 2.11 & 2.16 & -1.65 & -1.74 & 1.47 & 1.45 \\ 
16 & HE0151-4326 & 2.40132 & -3.05 & -3.00 & -2.07 & -2.05 & 2.12 & 2.17 & 0.07 & -0.04 & 1.52 & 1.48 \\ 
17 & Q0002-422 & 2.53946 & -3.58 & -3.53 & -1.42 & -1.5 & 1.54 & 1.59 & 0.41 & 0.28 & 1.11 & 1.07 \\ 
18 & Q0002-422 & 2.46322 & -2.99 & -2.99 & -0.59 & -0.6 & 2.16 & 2.16 & -0.55 & -0.57 & 1.55 & 1.48 \\ 
19 & Q0002-422 & 2.46236 & -2.52 & -2.55 & -0.43 & -0.46 & 2.63 & 2.6 & -0.95 & -0.90 & 1.91 & 1.83 \\ 
20 & Q0002-422 & 2.46204 & -2.85 & -2.87 & -0.58 & -0.6 & 2.3 & 2.28 & -0.58 & -0.56 & 1.66 & 1.57 \\ 
21 & PKS0329-255 & 2.58676 & -3.55 & -3.53 & -2.36 & -2.44 & 1.55 & 1.57 & 1.09 & 1.00 & 1.13 & 1.05 \\ 
22 & PKS0329-255 & 2.45658 & -3.08 & -3.08 & -0.93 & -0.96 & 2.07 & 2.07 & -1.40 & -1.43 & 1.49 & 1.41 \\ 
23 & Q0453-423 & 2.44411 & -2.45 & -2.40 & -0.32 & -0.23 & 2.75 & 2.75 & -1.64 & -1.74 & 1.96 & 1.96 \\ 
24 & Q0453-423 & 2.44264 & -2.55 & -2.52 & -0.41 & -0.38 & 2.6 & 2.63 & -1.14 & -1.19 & 1.89 & 1.86 \\ 
25 & Q0453-423 & 2.44181 & -3.00 & -2.98 & -2.23 & -2.25 & 2.15 & 2.17 & -0.41 & -0.45 & 1.55 & 1.50 \\ 
26 & Q0453-423 & 2.39816 & -4.40 & -4.32 & -0.13 & -0.44 & 0.77 & 0.85 & 1.36 & 1.17 & 0.50 & 0.48 \\ 
27 & Q0453-423 & 2.39780 & -3.79 & -3.75 & -0.36 & -0.5 & 1.38 & 1.42 & 1.03 & 0.94 & 0.94 & 0.90 \\ 
28 & Q0453-423 & 2.39745 & -3.60 & -3.59 & -0.86 & -0.98 & 1.57 & 1.58 & 0.08 & 0.05 & 1.10 & 1.04 \\ 
29 & Q0453-423 & 2.39676 & -3.40 & -3.45 & -0.46 & -0.58 & 1.77 & 1.72 & -0.06 & 0.04 & 1.24 & 1.13 \\ 
30 & Q0453-423 & 2.27757 & -2.99 & -3.00 & 0.17 & 0.08 & 2.23 & 2.22 & -1.58 & -1.53 & 1.53 & 1.47 \\ 
31 & HE1347-2457 & 2.37000 & -3.40 & -3.36 & -1.14 & -1.21 & 1.78 & 1.82 & -0.11 & -0.20 & 1.25 & 1.21 \\ 
32 & Q0329-385 & 2.24939 & -3.14 & -3.10 & 0.06 & 0 & 2.09 & 2.13 & -1.90 & -1.93 & 1.42 & 1.40 \\ \hline \hline
\end{tabular}
}}
\end{table*}
\clearpage


\begin{table*}
\centering
\vspace{0.1cm}
\caption{\Mt\ Cloudy-prediction for the tied \SiIII\ and \SiIV\ components of \sam\ and \san\ for
our fiducial KS18 UVB.}
\label{sitable}
\begin{tabular}{|c|c|c|c|c|c|c|c|r|r|c|}
\hline
 &  & \multicolumn{ 4}{c|}{Observed} & \multicolumn{ 4}{c|}{Photoionization Models} & \multicolumn{1}{l|}{} \\ \hline
\# & $z_{abs}$ & \multicolumn{ 2}{c|}{$log N (\SiIII)$} & \multicolumn{ 2}{c|}{$log N (\SiIV)$} & $log N (\SiIII)$ & $log N (\SiIV)$ & \multicolumn{1}{c|}{[C/H]} & \multicolumn{1}{c|}{[Si/H]} & \multicolumn{1}{l|}{Flag} \\ \hline
 &  &  & Error &  & Error &  &  & \multicolumn{1}{l|}{} & \multicolumn{1}{l|}{} & \multicolumn{1}{l|}{} \\ \hline
 &  & \multicolumn{ 6}{c|}{(in {\cmsq})} & \multicolumn{1}{l|}{} & \multicolumn{1}{l|}{} & \multicolumn{1}{l|}{} \\ \hline
\multicolumn{ 8}{|c|}{Sample \sam} & \multicolumn{1}{l|}{} & \multicolumn{1}{l|}{} & \multicolumn{1}{l|}{} \\ \hline
10 & 2.82656 & 11.45 & 0.08 & 12.02 & 0.02 & 11.49 & 12.02 & -0.8 & -0.69 & Detection \\ 
13 & 2.4499 & 12.40 & 0.00 & 12.22 & 0.03 & 11.66 & 12.22 & -0.94 & -0.32 & UL \\ 
18 & 2.46322 & 11.53 & 0.34 & 11.98 & 0.17 & 11.53 & 11.98 & -0.6 & -0.73 & Detection \\ 
19 & 2.46236 & 12.73 & 0.07 & 12.64 & 0.05 & 12.60 & 12.64 & -0.46 & -0.46 & Detection \\ 
20 & 2.46204 & 12.79 & 0.00 & 12.49 & 0.01 & 12.15 & 12.49 & -0.6 & -0.48 & UL \\ 
23 & 2.44411 & 12.29 & 0.01 & 12.16 & 0.02 & 12.23 & 12.16 & -0.23 & -0.25 & Detection \\ 
24 & 2.44264 & 12.81 & 0.00 & 12.63 & 0.01 & 12.59 & 12.63 & -0.38 & -0.21 & UL \\ 
29 & 2.39676 & 12.38 & 0.00 & 11.52 & 0.07 & 10.73 & 11.52 & -0.58 & -0.43 & UL \\ 
30 & 2.27756 & 12.92 & 0.00 & 11.73 & 0.11 & 11.21 & 11.73 & 0.08 & 0.09 & UL \\ \hline
\multicolumn{ 8}{|c|}{Sample \san} & \multicolumn{1}{l|}{} & \multicolumn{1}{l|}{} &  \\ \hline
1 & 2.45569 & 12.42 & 0.14 & 12.60 & 0.00 & 12.08 & 12.60 & -2.02 & -1.51 & UL \\ 
2 & 2.25106 & 12.02 & 0.41 & 12.38 & 0.00 & 11.83 & 12.38 & -0.55 & -0.83 & UL \\ \hline
\end{tabular}\\
\flushleft \text{UL: upper limits}
\end{table*}

\begin{table}
\caption{Linear regression fit coefficients for {\it L} as a function of \nc4\ and \nhi: log L = A + B log N(X) and associated Sphearman correlation coefficients.}
\begin{tabular}{c c c c c c}
\hline
Ion X & A & B & $\rho$ & $\rho/\sigma$ & Flag \\ \hline \hline
&& Sample \sam+\san \\ \hline 
\CIV &-5.20$\pm$2.43 & 0.38$\pm$0.19 & 0.19 & 1.7 & Full Sample \\ 
\CIV &-13.32$\pm$2.84 & 0.96$\pm$0.22 & 0.54 & 3.44 & For [C/H] $\geq$ -1.22 \\ 
\CIV &-8.99$\pm$2.34 & 0.73$\pm$0.19 & 0.5 & 3.16 & For [C/H] $<$ -1.22 \\ 
\HI&-11.97$\pm$1.45 & 0.80$\pm$0.10 & 0.59 & 5.27 & Full Sample \\ 
\HI&-10.84$\pm$2.62 & 0.71$\pm$0.19 & 0.51 & 3.22 & For [C/H] $\geq$ -1.22 \\ 
\HI&-7.35$\pm$1.97 & 0.50$\pm$0.13 & 0.41 & 2.59 & For [C/H] $<$ -1.22 \\
 \hline \hline
&&   Sample \sam \\ \hline  
\CIV&-3.65$\pm$4.62 & 0.26$\pm$0.36& -0.01& -0.01 & Full Sample \\ 
\CIV&-15.50$\pm$7.05&1.14$\pm$0.54&0.43&1.68 & For [C/H] $\geq$ -0.98 \\ 
\CIV&-12.47$\pm$3.13&1.01$\pm$0.25&0.57&2.22 & For [C/H] $<$ -0.98 \\ 
\HI& -9.95$\pm$2.94&0.67$\pm$0.20&0.30&1.65 & Full Sample \\ 
\HI& -6.56$\pm$4.91&0.41$\pm$0.35&0.26&1.02 & For [C/H] $\geq$ -0.98 \\ 
\HI& -7.99$\pm$2.80&0.56$\pm$0.19&0.32&1.24 & For [C/H] $<$ -0.98 \\  \hline 
\end{tabular}
\label{tablek21}
\end{table}

\begin{table}
\caption{Linear regression fit coefficients for [C/H] as a function of \nc4\ and \nhi: log [C/H] = A + B log N(X) and associated Sphearman correlation coefficients.}
\begin{tabular}{c c c c c c}
\hline
Ion X & A & B & $\rho$ & $\rho/\sigma$ & Flag \\ \hline \hline
&& Sample \sam+\san \\ \hline 
\CIV &-9.96$\pm$2.06 & 0.69$\pm$0.16 & 0.42 & 3.76 & Full Sample \\ 
\CIV &-2.55$\pm$1.93 & 0.16$\pm$0.15 & 0.15 & 0.94 & For [C/H] $\geq$ -1.22 \\ 
\CIV &-5.42$\pm$1.81 & 0.28$\pm$0.15 & 0.37 & 2.32 & For [C/H] $<$ -1.22 \\ 
\HI&9.66$\pm$1.31 & -0.75$\pm$0.09 & -0.65 & -5.82 & Full Sample \\ 
\HI&4.50$\pm$1.54 & -0.35$\pm$0.11 & -0.45 & -2.84 & For [C/H] $\geq$ -1.22 \\ 
\HI&2.50$\pm$1.42 & -0.29$\pm$0.09 & -0.36 & -2.3 & For [C/H] $<$ -1.22 \\ 
 \hline \hline
&&   Sample \sam \\ \hline 
\CIV&-11.24$\pm$3.63&0.80$\pm$0.29&0.52&2.90 & Full Sample \\ 
\CIV&0.10$\pm$3.64 & -0.04$\pm$0.28& -0.11 &-0.42 & For [C/H] $\geq$ -0.98 \\ 
\CIV& -3.80$\pm$3.47&0.16$\pm$0.28&0.36&1.40 & For [C/H] $<$ -0.98 \\ 
\HI&9.41$\pm$2.31&-0.73$\pm$0.16 &-0.56& -3.13 & Full Sample \\ 
\HI&4.09$\pm$1.99& -0.32$\pm$0.14& -0.35 &-1.35 & For [C/H] $\geq$ -0.98 \\ 
\HI&6.00$\pm$1.74& -0.53$\pm$0.12 &-0.81 &-3.13 & For [C/H] $<$ -0.98 \\ \hline
\end{tabular}
\label{tablek22}
\end{table}

\begin{table}
\caption{ Linear regression fit coefficients for {\it L} 
as a function of [C/H]: log L = A + B [C/H] and associated Sphearman correlation coefficients.}
\begin{tabular}{c c c c c c}
\hline
 A & B & $\rho$ & $\rho/\sigma$ & Flag \\ \hline \hline
&& Sample \sam+\san \\ \hline 
-1.27$\pm$0.12 & -0.78$\pm$0.09 & -0.68 & -6.09 & Full Sample \\ 
-1.32$\pm$0.16 & -1.02$\pm$0.23 & -0.61 & -3.83 & For [C/H] $\geq$ -1.22 \\ 
-1.27$\pm$0.39 & -0.76$\pm$0.20 & -0.43 & -2.72 & For [C/H] $<$ -1.22 \\ 
 \hline \hline
&&   Sample \sam \\ \hline 
-1.13$\pm$0.21 & -0.78$\pm$0.15 & -0.62 & -3.46 & Full Sample \\ 
-1.30$\pm$0.29&-1.31$\pm$0.48 & -0.64 & -2.49 & For [C/H] $\geq$ -0.98 \\ 
-0.75$\pm$0.58&-0.54$\pm$0.32 & -0.26 & -0.99& For [C/H] $<$ -0.98 \\ \hline

\end{tabular}
\label{tablek23}
\end{table}

\label{lastpage1}
\end{document}